\documentclass[12pt,bibliography=totoc,DIV=17]{scrartcl}
\pdfoutput=1
\usepackage[utf8]{inputenc}

\usepackage{amsmath,amsfonts,amssymb,dsfont,mathtools}
\usepackage{array}
\usepackage[symbol]{footmisc}
\usepackage{stackrel}
\let\refeq\undefined
\usepackage[mathscr]{euscript}
\usepackage{enumitem}
\usepackage{longtable}
\usepackage{booktabs}
\usepackage{multirow}
\usepackage{cite}
\usepackage{graphicx, ulem, hhline}
\usepackage[table]{xcolor}
\usepackage{xspace}
\usepackage{needspace}
\usepackage{siunitx}
\usepackage{soul}
\usepackage{wasysym}              
\usepackage{hyphenat}             

\usepackage{afterpage}
\usepackage{etoolbox,xstring,xspace,calc,xifthen}
\usepackage{centernot}

\usepackage{tikz}
\makeatletter
\newif\iftag@here
\newcommand*{\taghere}[1][0pt]
{\ifmeasuring@\else
  \global\tag@heretrue
  \tikz[remember picture,overlay]{\coordinate (taghere) at (0pt,#1);}%
\fi}

\def\place@tag{%
    \iftagsleft@
      \kern-\tagshift@
      \iftag@here
        \global\tag@herefalse
        \tikz[remember picture,overlay]%
          {\path (taghere) -| node[anchor=base]{\rlap{\boxz@}} (0pt,0pt);}%
      \else
        \if1\shift@tag\row@\relax
            \rlap{\vbox{%
                \normalbaselines
                \boxz@
                \vbox to\lineht@{}%
                \raise@tag
            }}%
        \else
            \rlap{\boxz@}%
        \fi
        \kern\displaywidth@
      \fi
    \else
      \kern-\tagshift@
      \iftag@here
        \global\tag@herefalse
        \tikz[remember picture,overlay]%
          {\path  (taghere) -|  node[anchor=base]{\llap{\boxz@}} (0pt,0pt);}%
      \else
        \if1\shift@tag\row@\relax
            \llap{\vtop{%
                \raise@tag
                \normalbaselines
                \setbox\@ne\null
                \dp\@ne\lineht@
                \box\@ne
                \boxz@
            }}%
        \else \llap{\boxz@}%
        \fi
      \fi
    \fi
}
\makeatother

\long\def\symbolfootnote[#1]#2{\begingroup%
\def\thefootnote{\fnsymbol{footnote}}\footnote[#1]{#2}\endgroup}

\makeatletter
\def\@fnsymbol#1{\ensuremath{\ifcase#1\or%
\ast\or \dagger\or \ddagger\or \mathsection\or \parallel\or \nparallel\or%
\mathparagraph\or \cap\or \cup\or \subset\or \supset\or%
\wedge\or \vee\or <\or >\or \diamond\or \circ\or%
\vartriangle\or \triangledown\or \triangleleft\or \triangleright\or%

\else\@ctrerr\fi}}
\makeatother

\makeatletter
\newlength{\fnhskip}
\renewcommand\@makefntext[1]{
  \settowidth{\fnhskip}{\@makefnmark}
  \leftskip=\fnhskip
  \hskip-\fnhskip
  \@makefnmark#1
}
\makeatother

\renewenvironment{subequations}[1][]{
  \refstepcounter{equation}%
  \setcounter{parentequation}{\value{equation}}
  \setcounter{equation}{0}
  \def\theequation{\theparentequation\alph{equation}}%
  \let\parentlabel\label
  \ifx\\#1\\\relax\else\label{#1}\fi
  \ignorespaces
}{%
  \setcounter{equation}{\value{parentequation}}
  \ignorespacesafterend
}

\newcommand*{\nextParentEquation}[1][]{
  \refstepcounter{parentequation}
  \setcounter{equation}{0}
  \ifx\\#1\\\relax\else\parentlabel{#1}\fi
}

\usepackage{setspace}

\usepackage[numbers,sort&compress]{natbib}
\makeatletter
\def\NAT@spacechar{\,}
\makeatother
\usepackage[
colorlinks=true
,urlcolor=blue
,anchorcolor=blue
,citecolor=blue
,filecolor=blue
,linkcolor=blue
,menucolor=blue
,linktoc=all
,unicode
,backref=page
]{hyperref}
\usepackage{hypernat}

\newrobustcmd*{\tocref}[1]{\hyperref[TOC]{\color{black}{#1}}}
\newcommand{\tocsection}[2][]{\section[\boldmath #2]{\boldmath\tocref{#2#1}}}
\newcommand{\tocsubsection}[2][]{\subsection[#2]{\boldmath\tocref{#2#1}}}

\usepackage[all]{hypcap}
\usepackage{footnotebackref}

\usepackage[format=plain, margin=0.5cm, font=footnotesize, labelfont=bf, labelsep=colon]{caption}

\renewcommand*{\backref}[1]{}
\renewcommand*{\backrefalt}[4]{%
  \ifcase #1%
  \or [p\,#2]%
  \else [pp\,#2]%
  \fi%
}

\newif\ifbackrefshowonlyfirst
\backrefshowonlyfirsttrue
\makeatletter
\let\BR@direct@old@hyper@natlinkstart\hyper@natlinkstart
\renewcommand*{\hyper@natlinkstart}{\phantomsection\BR@direct@old@hyper@natlinkstart}
\let\BR@direct@oldBR@citex\BR@citex
\renewcommand*{\BR@citex}{\phantomsection\BR@direct@oldBR@citex}%

\long\def\hyper@page@BR@direct@ref#1#2#3{\hyperlink{#3}{#1}}

\ifx\backrefxxx\hyper@page@backref
    \let\backrefxxx\hyper@page@BR@direct@ref
    \ifbackrefshowonlyfirst
    \fi
\else
    \ifbackrefshowonlyfirst
    \fi
\fi

\patchcmd{\Hy@backout}{Doc-Start}{\@currentHref}{}{\errmessage{I can't seem to patch backref}}
\makeatother

\let\theparentequation\theequation
\patchcmd{\theparentequation}{equation}{parentequation}{}{}

\apptocmd{\thebibliography}{\scriptsize}{}{}
\let\OLDthebibliography\thebibliography
\renewcommand\thebibliography[1]{
  \OLDthebibliography{#1}
  \setlength{\parskip}{1pt}
  \setlength{\itemsep}{1pt plus 0.3ex}
}

\addtokomafont{disposition}{\rmfamily}
\BeforeTOCHead[toc]{{\pdfbookmark[1]{\contentsname}{toc}}}

\newcommand{\unicodescriptO}{^^f0^^9d^^92^^aa}
\newcommand{\unicodesuptwo}{^^c2^^b2}

\newcommand{\unicodesubt}{^^e2^^82^^9c}
\newcommand{\unicodesubs}{^^e2^^82^^9b}
\newcommand{\unicodethinspace}{^^e2^^80^^89}

\newcommand{\unicodealpha}{^^f0^^9d^^9b^^bc}

\newcommand{\dotcirc}{%
  \pdfliteral{q 1 Tr .1 w [0.5 .5] 0d}%
  \textcolor{lightgray}{\rlap{\Circle}}%
  \pdfliteral{Q}%
  \hphantom{\Circle}%
}

\makeatletter
\patchcmd{\upbracefill}{\m@th}{\scriptstyle\m@th}{}{}
\patchcmd{\upbracefill}{$\braceld$}{$\scriptstyle\braceld$}{}{}
\patchcmd{\upbracefill}{\bracelu}{\bracelu\mkern-1mu}{}{}
\patchcmd{\upbracefill}{\hfill\braceru}{\hfill\mkern-1mu\braceru}{}{}
\DeclareOldFontCommand{\rm}{\normalfont\rmfamily}{\mathrm}
\DeclareOldFontCommand{\sf}{\normalfont\sffamily}{\mathsf}
\DeclareOldFontCommand{\tt}{\normalfont\ttfamily}{\mathtt}
\DeclareOldFontCommand{\bf}{\normalfont\bfseries}{\mathbf}
\DeclareOldFontCommand{\it}{\normalfont\itshape}{\mathit}
\makeatother

\newlength{\floatwidth}
\def\beq{\begin{equation}}
\def\eeq{\end{equation}}

\newcommand{\AtoB}[2]{\mbox{$#1\to #2$}}
\newcommand{\Amp}[4][\mathcal{A}]{#1^{\mbox{\tiny #2}}_{\mbox{\tiny #3}}\ifthenelse{\isempty{#4}}{}{{\left[#4\right]}}}

\def\twomat[#1,#2][#3,#4]{\left( \begin{array}{cc} #1 & #2 \\ #3 & #4 \end{array} \right)}
\def\threemat[#1,#2,#3][#4,#5,#6][#7,#8,#9]{\left( \begin{array}{ccc} #1 & #2 & #3\\ #4 & #5 & #6 \\ #7 & #8 & #9 \end{array} \right)}
\def\twovec[#1,#2]{\left( \begin{array}{c} #1  \\ #2 \end{array} \right)}
\def\thv[#1,#2,#3]{\left( \begin{array}{c} #1 \\ #2 \\ #3 \end{array} \right)}
\def\twv[#1,#2]{\left( \begin{array}{c} #1 \\ #2 \end{array} \right)}

\newcommand{\IE}{\textit{i.\,e.}\xspace}
\newcommand{\EG}{\textit{e.\,g.}\xspace}
\newcommand{\AP}{\textit{a~priori}\xspace}
\newcommand{\AH}{\textit{ad-hoc}\xspace}
\newcommand{\refeq}[1]{Eq.\,\eqref{#1}}

\newcommand{\sect}[1]{Sect.\,\ref{#1}}
\newcommand{\appx}[1]{Appx.\,\ref{#1}}
\newcommand{\fig}[1]{Fig.\,\ref{#1}}
\newcommand{\tab}[1]{Tab.\,\ref{#1}}
\newcommand{\citere}[1]{Ref.\,\cite{#1}}
\newcommand{\citeres}[1]{Refs.\,\cite{#1}}
\newcommand{\simord}{\mathord{\sim}\,}
\newcommand{\simeqord}{\mathord{\simeq}\,}

\newcommand{\yint}[3]{Y_{#1}^{[#2]}\ifthenelse{\isempty{#3}}{}{{\left(#3\right)}}}
\newcommand{\ysum}[3]{Y_{#1}^{#2}\ifthenelse{\isempty{#3}}{}{{\left(#3\right)}}}
\newcommand{\yonesum}[3]{{^0}Y_{#1}^{#2}\ifthenelse{\isempty{#3}}{}{{\left(#3\right)}}}
\newcommand{\tint}[2]{\ifthenelse{\isempty{#2}}{T_{#1}}{T_{#1}^{\left|#2\right.}}}

\newcommand{\toneint}[2]{\ifthenelse{\isempty{#2}}{{^0}T_{#1}}{^0T^{\left|#2\right.}_{#1}}}

\newcommand{\yoneint}[3]{{^0}Y_{#1}^{[#2]}\ifthenelse{\isempty{#3}}{}{{\left(#3\right)}}}
\newcommand{\bint}[1]{B_{0}\ifthenelse{\isempty{#1}}{}{{\left(#1\right)}}}
\newcommand{\aint}[1]{A_{0}\ifthenelse{\isempty{#1}}{}{{\left(#1\right)}}}
\newcommand{\binteps}[2]{\ifthenelse{\isempty{#2}}{B_{0}^{\left|#1\right.}}{B_{0}^{\left|#1\right.}{\!\left(#2\right)}}}

\newcommand{\CP}{\ensuremath{\mathcal{CP}}\xspace}
\newcommand{\MS}{\ensuremath{\overline{\text{MS}}}\xspace}
\newcommand{\DR}{\ensuremath{\overline{\text{DR}}}\xspace}

\newcommand{\Real}[1]{\Re\hspace{-1pt}\mathfrak{e}{\left[#1\right]}}

\newcounter{notecount}
\hypersetup{ pdfauthor={Florian Domingo and Sebastian Paßehr}
  ,pdftitle={Fighting off field dependence in MSSM Higgs-mass
    corrections of order
    \unicodealpha\unicodesubt\unicodethinspace\unicodealpha\unicodesubs
    \ and \unicodealpha\unicodesubt\unicodesuptwo} ,pdfsubject={}
  ,pdfkeywords={} }

\begin{document}

\newcommand*{\mytitle}[1]{%
  \parbox{\linewidth}{\setstretch{1.5}\centering\Large\textsc{\textbf{\boldmath #1}}}
}

\thispagestyle{empty}

\def\thefootnote{\fnsymbol{footnote}}

\begin{flushright}
  BONN-TH-2021-01\\
  TTK-21-15
\end{flushright}

\vfill

\begin{center}

\mytitle{Fighting off field dependence in\\ MSSM Higgs-mass corrections of order
  $\alpha_t\,\alpha_s$ and $\alpha_t^2$}

\vspace{1cm}

Florian Domingo$^{1}$\footnote{email: florian.domingo@csic.es}
and
Sebastian Pa{\ss}ehr$^{2}$\footnote{email: passehr@physik.rwth-aachen.de}

\vspace*{1cm}

\textsl{
$^1$Bethe Center for Theoretical Physics \&
Physikalisches Institut der Universit\"at Bonn,\\
Nu\ss allee 12, D--53115 Bonn, Germany
}

\medskip
$^2$\textsl{Institute for Theoretical Particle Physics and Cosmology,}\\
\textsl{RWTH Aachen University, Sommerfeldstra{\ss}e 16, 52074 Aachen, Germany.}

\end{center}

\vfill

\begin{abstract}{}
The connection between gauge and Higgs sectors makes supersymmetric
extensions of the Standard Model predictive frameworks for the
derivation of Higgs masses. In this paper, we study the contamination
of such predictions by field-renormalization constants, in the MSSM
with two-loop gaugeless corrections of
$\mathcal{O}\big(\alpha_{t,b}\,\alpha_s,\,\alpha_{t,b}^2\big)$ and
full momentum dependence, and demonstrate how strict perturbative
expansions \mbox{allow} to systematically neutralize the dependence on
such unphysical objects. On the other hand, the popular procedure
consisting in an iterative pole search remains explicitly dependent on
field counterterms. We then analyze the magnitude of the intrinsic
uncertainty that this feature implies for the iterative method, both
in non-degenerate and near-degenerate regimes, and conclude that this
strategy does not improve on the predictions of the more
straightforward expansion. We also discuss several features related to
the inclusion of the orders $\alpha_{t,b}\,\alpha_s$ and
$\alpha_{t,b}^2$ in the so-called `fixed-order' \mbox{approach}, such
as the resummation of UV-logarithms for heavy supersymmetric
\mbox{spectra}.
\end{abstract}

\vfill

\def\thefootnote{\arabic{footnote}}
\setcounter{page}{0}
\setcounter{footnote}{0}

\hypersetup{linkcolor=black}
\tableofcontents\label{TOC}
\hypersetup{linkcolor=blue}

\tocsection{Introduction}

The discovery of a Standard-Model (SM)-like Higgs boson at the
LHC\,\cite{Aad:2012tfa,Chatrchyan:2012ufa,Aad:2015zhl} and the
growingly precise measurements of its
characteristics\,\cite{Khachatryan:2016vau,Sirunyan:2018koj,Aad:2019mbh}
make the need for controlled theoretical predictions
in the Higgs sector of models of new physics quite clear, if one
wishes to exploit associated observables and constrain the parameter
space of beyond-the-SM (BSM) theories. In many
cases, such precision calculations are difficult to conduct in the
full model, \EG~due to the non-perturbative nature of the theory, and
can only be performed in an effective context, in the limit of
decouplingly heavy new physics. While the absence of conclusive signs
of BSM physics at the LHC seems
to justify the relevance of this type of approach anyway, the case of
supersymmetric~(SUSY) extensions of the
SM\,\cite{Nilles:1983ge,Haber:1984rc} is still remarkable in that it
provides a perturbative framework up to relatively high energies.

Discussions of radiative corrections in the SUSY Higgs sector have a
long-lasting history, of which the reader can find a summary in the
recent review\,\cite{Slavich:2020zjv}. In particular, the higher-order
contributions have a sizable impact on the mass of the SM-like
state---which is automatically constrained to take tree-level values
of electroweak (EW) magnitude and acquires sensitivity to new-physics
scales via ultraviolet (UV) logarithms. Recently, the KUTS initiative
has stimulated substantial advances in various related directions,
such as calculations at fixed
order~(FO)\,\cite{Borowka:2014wla,Degrassi:2014pfa,Borowka:2015ura,Borowka:2018anu,Hollik:2014wea,Hollik:2014bua,Passehr:2017ufr,Goodsell:2016udb,Goodsell:2014bna,Drechsel:2016jdg,Goodsell:2014pla,Muhlleitner:2014vsa,Goodsell:2015ira,Goodsell:2015yca,Braathen:2016mmb,Braathen:2016cqe,Braathen:2017izn,Biekotter:2017xmf,Stockinger:2018oxe,Bahl:2018ykj,Dao:2019qaz,Goodsell:2019zfs}
or in the context of low-energy Effective Field Theories
(EFTs)\,\cite{Bagnaschi:2014rsa,Lee:2015uza,Vega:2015fna,Athron:2016fuq,Bagnaschi:2017xid,Bahl:2018jom,Braathen:2018htl,Bagnaschi:2019esc,Murphy:2019qpm,Bahl:2020mjy,Bahl:2020jaq,Staub:2017jnp,Harlander:2018yhj,Bahl:2019wzx},
hybrid approaches combining~FO and~EFT
results\,\cite{Bahl:2016brp,Bahl:2017aev,Bahl:2019hmm,Bahl:2020tuq,Harlander:2019dge,Kwasnitza:2020wli},
or the development of public
codes\,\cite{Staub:2015aea,Hahn:2015gaa,Harlander:2017kuc,Drechsel:2016htw,Athron:2017fvs,Bahl:2018qog,Gabelmann:2018axh}. The
two-loop~(2L) corrections to the SM-like Higgs mass of the
MSSM---where one-loop (1L) contributions are mostly of
$\mathcal{O}(\alpha_t)$\footnote{By convention, we denote the
considered classes of Feynman diagrams by the type of the involved
couplings $\alpha_f=Y_f^2\big/(4\,\pi)$,
$\alpha_s=g_s^2\big/(4\,\pi)$, and
$\alpha=g_{\mathrm{EW}}^2\big/(4\,\pi)$, where $Y_f$ is the Yukawa
coupling of fermion $f$, $g_s$ is the strong gauge coupling, and
$g_{\mathrm{EW}}$ is any of the electroweak gauge couplings. In
addition, we introduce $\alpha_q\equiv\alpha_{t,b}$ when considering
corrections controlled by the Yukawa couplings of both heavy
quarks.}---are dominated by the contributions of order
$\alpha_t\,\alpha_s$\,\cite{Heinemeyer:1998jw,Heinemeyer:1998np,Heinemeyer:1999be,Heinemeyer:2007aq,Borowka:2014wla,Degrassi:2014pfa,Borowka:2015ura,Borowka:2018anu}
and
$\alpha_t^2$\,\cite{Martin:2004kr,Brignole:2001jy,Dedes:2003km,Hollik:2014bua,Hollik:2014wea,Passehr:2017ufr,Goodsell:2016udb}---this
does not necessarily apply to the other Higgs states of the model
however.

In \citere{Domingo:2020wiy}, we argued that terms violating the EW
symmetry in an uncontrolled fashion were introduced in the predictions
of masses and decays at the 1L order unless one carefully expands and
truncates the various contributions entering the amplitudes. In
particular, momentum-dependent corrections can lead to an explicit
dependence on field-renormalization constants, thus producing
UV-logarithms that have no quantitative meaning. While such artificial
effects are formally of a higher order compared to that of the
explicitly included self-energies, it is important to assess their
numerical impact in order to estimate the actual gain in predictivity
when incorporating radiative corrections in a fashion that explicitly
depends on the regularization procedure.\footnote{Here, we employ the
term `regularization' in reference to the procedure giving a meaning
to ill-defined self-energies away from their mass shell, not in
reference to the counting of UV-divergences in Feynman
amplitudes. Field-renormalization constants are the usual
regulators---and the only ones that we consider in this
paper---allowing such an extrapolation, and the ambiguity rests with
their arbitrary finite part, from which actual observables are
supposedly independent.} Alternatively, a strict expansion and
truncation order-after-order can limit the dependence on the field and
gauge-fixing regulators and provide results where spurious
symmetry-violating effects are systematically neutralized. The purpose
of the present paper is to explain how such a formalism can be
extended to address 2L mass corrections. We focus on the
orders~$\alpha_q\,\alpha_s$ and~$\alpha_q^2$, since these are fully
exploitable at~2L, while the generic EW~corrections miss some
contributions needed for a direct connection to observable
quantities. We build upon the existing results
of \citeres{Hollik:2014wea,Hollik:2014bua,Passehr:2017ufr,Borowka:2018anu},
but include full momentum dependence now, also making use of the
relations derived in \citere{Goodsell:2019zfs}. Below, we restrict
ourselves to the simplest model where the considered orders matter,
\IE~the Minimal Supersymmetric Standard Model~(MSSM), although the
method straightforwardly applies to extensions, \EG~the
Next-to-MSSM~(NMSSM)---yet more care is then needed in order to
consistently account for the gaugeless limit and process the more
motivated scenarios with large radiative Higgs mixing.

In parallel with the computations at~FO, where self-energies are
derived order after order in the full model considered in a given
renormalization scheme, the mass-gap suggested by the absence of
discovery of new-physics particles has encouraged the discussion of
SUSY Higgs sectors in the context of EFTs, allowing for a resummation
of the large UV-logarithms that develop between the SUSY and EW
scales; see \EG~the recent work
of \citeres{Bagnaschi:2014rsa,Lee:2015uza,Vega:2015fna,Athron:2016fuq,Bagnaschi:2017xid,Bahl:2018jom,Braathen:2018htl,Bagnaschi:2019esc,Murphy:2019qpm,Bahl:2020mjy,Bahl:2020jaq,Staub:2017jnp,Harlander:2018yhj,Bahl:2019wzx}
and a more complete list of references in the
review\,\cite{Slavich:2020zjv}. While the effects that we discussed
in \citere{Domingo:2020wiy} also formally apply in such EFTs, the
weight of regulators in the definition of mass observables is less
critical than in FO~calculations. Indeed, the direct variation of the
gauge-fixing parameters or the field counterterms in the EFT is
necessarily reduced in such a formalism where large logarithms are
forced to follow an $SU(2)\times U(1)$-conserving pattern. On the
other hand, such a variation represents only a partial picture in
testing the connection between observables in the EFT context, as the
MSSM--EFT matching procedure then plays a determining part in
establishing this relation. This connection between the predicted
Higgs masses and observable input is however at the center of what we
probe through variations of regulators in the FO~approach. We will not
attempt to address the question of this relation in EFTs here.

\needspace{3ex}
In \sect{sec:2Lmasses}, we analyze how to avoid or minimize the
dependence of 2L corrections to the Higgs masses on the choice of
field renormalization, taking this criterion as our guideline for the
inclusion of the 2L contributions of
$\mathcal{O}\big(\alpha_t\,\alpha_s\big)$ and
$\mathcal{O}\big(\alpha_t^2\big)$. We then numerically compare the
corresponding mass predictions with those obtained through an
iterative pole search, both in non-degenerate and near-degenerate
scenarios, checking how variations of the field counterterms affect
each in view of the magnitude of the 2L effects. We also probe how the
effective-potential approximation performs for SM and non-SM Higgs
states. This forms the content of \sect{sec:nondegeneratecase}
and \sect{sec:degeneratecase}. In \sect{sec:resum}, we perform a
resummation of UV-logarithms of
$\mathcal{O}\big(\alpha_q,\,\alpha_q\,\alpha_s,\,\alpha_q^2\big)$ in
the FO~context, as the increasing weight of these effects for large
SUSY scales otherwise limits the applicability of the FO~approach. A
brief summary is provided in \sect{sec:conclusion}.

\tocsection{\label{sec:2Lmasses}Inclusion of 2L corrections to the Higgs mass observables}

In this section, we present a brief description of the computation of
Higgs masses at higher order, with reference to more detailed
derivations in the appendix. We explicitly extract the dependence of
the radiative contributions on Higgs-field counterterms and analyze
the conditions for its cancellation at the level of Higgs-mass
predictions at the 2L order.

\tocsubsection{\label{sec:fieldinv}Invariance under field-renormalization}

As usual in Quantum Field Theory, the mass observables are calculated
from the two-point truncated and connected correlators of the
model. The complex zeroes of the characteristic polynomial in the
Higgs sector should indeed correspond to the poles describing the
Higgs resonances in particle scattering: a derivation is proposed
in \appx{ap:scattering}. The defining equation for these
poles~$\mathcal{M}^2_{k}$ thus reads:
\begin{equation}\label{eq:massbasic}
  \det{\left[
    \mathcal{M}^2_{k}\,\mathds{1} - M^2_{\text{tree}} + \hat{\Sigma}\big(\mathcal{M}^2_{k}\big)
  \right]} = 0\,.
\end{equation}
where $M^2_{\text{tree}}=\text{diag}\big[m_i^2\big]$ represents the
tree-level mass matrix and $\hat{\Sigma}\big(p^2\big)$ is the
renormalized self-energy matrix for external momentum $p$.

A problematic feature in \refeq{eq:massbasic} is that the renormalized
self-energies $\hat{\Sigma}_{ij}\big(p^2\big)$ are not well-defined
away from their mass-shell at
$p^2=\big(m_i^2+m_j^2\big)\big/2$. Indeed, absorption of the
UV-divergences in such objects requires the introduction of
field-renormalization constants,
\begin{align}\label{eq:dZdef}
  \phi_i &\to \left(\sqrt{Z}\right)_{ij} \phi_j \equiv
  \left(\sqrt{\mathds{1} + \delta Z^{(1)} + \delta Z^{(2)} + \ldots}\right)_{ij}
  \phi_j\,,
\end{align}
where the upper indices in parentheses refer to the loop order of the
renormalization constant. However, since quantum fields are not
measurable as such, these field counterterms should be strict
bookkeeping quantities, which drop out at the level of observables. An
iterative resolution of \refeq{eq:massbasic}, as often advocated in
the
literature\,\cite{Frank:2006yh,Degrassi:2009yq,Williams:2011bu,Graf:2012hh,Athron:2014yba,Goodsell:2014bna,Goodsell:2016udb,Drechsel:2016jdg,Domingo:2017rhb,Athron:2017fvs,Bahl:2018qog,Dao:2019nxi},
leads to an explicit dependence of the predicted masses on the field
counterterms. This dependence should ideally become negligibly small
when sufficiently high orders are included in the calculation and, in
the meanwhile, the variation of the field counterterms could be seen
as setting a lower bound on the higher-order uncertainty, by weighing
such terms of higher order that are introduced in the pole search.

\needspace{5ex}
In \citere{Domingo:2020wiy}, we argued against this iterative
approach, in particular because it generates symmetry-violating terms
that are not controlled by the vacuum-expectation
values \mbox{(v.e.v.-s)} of the Higgs fields, but appear as pure
artifacts of the formalism. Consequently, these partial higher-order,
gauge-dependent terms introduced by the pole search do not represent a
genuine estimate of higher-order corrections but amount to an
intrinsic error of the procedure. Instead, we preferred to employ an
explicit expansion and truncation where the dependence on field
counterterms and gauge-fixing parameters vanishes or is minimized
order-by-order. At 1L, this procedure leads us to systematically use
diagonal self-energies evaluated on their mass-shell, while
off-diagonal self-energies can be dismissed as contributions of higher
order. Only in near-degenerate scenarios are the replacement
$\hat{\Sigma}_{ij}\big(p^2\big)\stackrel[p^2\,\sim\,m_i^2]{}{\xrightarrow{\hspace*{1.2cm}}}\hat{\Sigma}_{ij}{\left(\big(m^2_i+m^2_j\big)\big/2\right)}$
and the inclusion of off-diagonal self-energies justified by the
counting $m_i^2-m_j^2=\mathcal{O}(\text{1L})$.

Below, we discuss how to extend this procedure to include 2L mass
corrections. As the considered orders do not involve EW gauge
corrections, we must take solely independence from
field-renormalization constants as our guiding criterion. This
condition can be formulated in two different fashions:
\begin{enumerate}
\item the observables should be invariant under any choice of field
  counterterm;
\item the observables are UV-finite without need of considering field
  counterterms.
\end{enumerate}
The second prescription proves to be a weaker requirement than the
first one, in particular because the UV-divergent part of
$d\Sigma_{ij}\big/dp^2$ is symmetric under the exchange
$i\leftrightarrow j$, so that amplitudes may in general depend on
antisymmetric finite contributions to the field counterterms while
these are not needed to achieve UV-finiteness. In practice, the first
criterion is satisfied in the non-degenerate case, while we must
content ourselves with the second one in the near-degenerate
scenario---at least in the strategy presented
in \citere{Domingo:2020wiy}. We will look upon the second condition as
being roughly equivalent to an invariance under symmetric choices of
the field counterterms---in fact a more constraining requirement.

\tocsubsection{Non-degenerate case}

For a non-degenerate state with tree-level mass $m_i$, it is possible
to extract the nearby pole~${\cal M}^2_i$ in the propagator matrix by
solving the recursive \refeq{eq:massbasic}; at the 2L order this
amounts to solving ${\cal
M}^2_i\stackrel[]{!}{=}\widetilde{M}^2_i\big({\cal M}^2_i\big)$, where
$\widetilde{M}^2_i$ is the quantity defined in
\refeq{eq:nondegrecur}---refer \EG~to the derivation in
\appx{ap:propmat}. The right-hand side of this equation can
be further expanded using the condition ${\cal
M}^2_i=\overline{p}^2_{ii}+\mathcal{O}(\text{2L})=m_i^2+\mathcal{O}(\text{1L})$
in the argument of the renormalized self-energies
\mbox{($\hat{\Sigma}^{(k)}$ represents the renormalized self-energy of
$k$-loop order):}
\begin{align}\label{eq:nondegmassexp}
  \mathcal{M}^2_i = m^2_i - \hat{\Sigma}^{(1)}_{ii}\big(m^2_i\big)
  - \left[\hat{\Sigma}^{(2)}_{ii}\big(m_i^2\big)
    + \big(\overline{p}_{ii}^2 - m^2_i\big)\,
    \frac{d\hat{\Sigma}^{(1)}_{ii}}{dp^2}\big(m^2_i\big)
    - \sum_{j\neq i}\frac{\hat{\Sigma}^{(1)}_{ij}\big(m_i^2\big)\,
      \hat{\Sigma}^{(1)}_{ji}\big(m_i^2\big)}{m_i^2 - m^2_j}\right]
  + \mathcal{O}(\text{3L})\,.\taghere[-3ex]
\end{align}
The leading correction
$\hat{\Sigma}^{(1)}_{ii}\big(\mathcal{M}^2_i\big)$ has been expanded
so that it gives a contribution that is independent from field
renormalization (and gauge-fixing) at strict $\mathcal{O}(\text{1L})$.

Similarly, invariance under variations of the 2L field counterterm
dictates the choice $p^2\stackrel[]{!}{=}m_i^2$ as the argument of
$\hat{\Sigma}^{(2)}_{ii}$ in \refeq{eq:nondegmassexp}. In addition,
restriction to the orders $\alpha_{q}\,\alpha_s$ and $\alpha_{q}^2$
entails a specific distinction between corrections of Yukawa- and
gauge-type since both are otherwise mixed by the EW
symmetry-breaking. A clean separation is only possible in the
gaugeless limit (\IE~$g_{\mathrm{EW}}\stackrel[]{!}{=}0$), where the
tree-level spectrum slightly differs from the original one: the mass
`$m_i$' entering the self-energy of 2L order computed in the gaugeless
limit and ensuring independence from the 2L field counterterm is not
the original tree-level mass, but its gaugeless
counterpart~$\breve{m}_{i}$. The difference with $m_i$ corresponds to
the order neglected in the gaugeless approximation. In the
non-degenerate scenario, the identification of the `original' Higgs
state with its gaugeless counterpart is fortunately straightforward.

The choice of momenta in the 1L$^2$ terms of \refeq{eq:nondegmassexp}
should allow the cancellation of the dependence on the field
counterterms of 1L order in the 2L contributions. While we already
expanded the momenta in the off-diagonal terms of
\refeq{eq:nondegmassexp} and set them to $m_i^2$, any value
$\overline{p}_{ij}^2= m^2_i+\mathcal{O}(\text{1L})$ is \AP as
legitimate as long as dependence on the field counterterms is not
considered. Let us therefore analyze the dependence of
$\hat{\Sigma}^{(2)}_{ii}$ on the 1L field counterterms. It originates
in two pieces: 1L diagrams with counterterm insertion (1L$\times$CT)
and tree-level graphs with counterterm squared insertion
(CT$^2$):\footnote{In this analysis of the dependence on 1L field
counterterms $\delta Z_{ij}$ (for commodity we omit the
superscript~$^{\text{(1)}}$), we do not display the 2L
term \mbox{$\big(\delta Z^{(2)}_{ii} - \sum_k\tfrac{1}{4}\,\delta
Z_{ik}^2\big)\,\big(p^2 - m_i^2\big)$} emerging simply from the
expansion of \refeq{eq:dZdef}, which separately cancels out for
$p^2\,\stackrel[]{!}{=}\,m_i^2$, as the vanishing condition for the
dependence on the 2L~field counter\-term~$\delta Z^{(2)}_{ii}$. This
restores agreement between our expansion and the standard conventions
for field counterterms---see \refeq{eq:dZdef}.}
\begin{align}\label{eq:2Lfield}
  \hat{\Sigma}^{(2)}_{ii}\big(p^2\big) &\ni
  \frac{1}{4}\,\big(p^2-\breve{m}_i^2\big)\,\delta Z_{ii}^2
  + \delta Z_{ii}\left[\breve{\Sigma}^{(1)}_{ii}\big(p^2\big)
    - \delta \breve{M}^2_{ii}\right]\notag\\
  &\quad+ \sum_{j\neq i}\left\{\frac{1}{4}\,\big(p^2-\breve{m}^2_j\big)\,
  \delta Z_{ji}^2 + \delta Z_{ji}\left[\breve{\Sigma}^{(1)}_{ij}\big(p^2\big)
    - \delta \breve{M}^2_{ij}\right]\right\};
\end{align}
the 1L$\times$CT and CT$^2$ contributions to the tadpoles cancel one
another. Here, $\Sigma_{ij}$ (or $\breve{\Sigma}_{ij}$) stands for the
unrenormalized self-energy; $\delta M_{ij}^2$ (or
$\delta \breve{M}^2_{ij}$) denote the mass and $\delta Z_{ji}$ the
field counterterms of 1L order. The $\breve{\dotcirc}$~notation
highlights that the corresponding quantities are obtained in the
approximations applied in the 2L calculation (\EG~the gaugeless
limit).

From the 1L$^2$ terms emerge the following contributions depending on
field renormalization:
\begin{multline}\label{eq:1L^2}
  \big(\overline{p}_{ii}^2 - m^2_i\big)\,
  \frac{d\hat{\Sigma}^{(1)}_{ii}}{dp^2}\big(m^2_i\big)
  - \sum_{j\neq i}\frac{\hat{\Sigma}^{(1)}_{ij}\big(\overline{p}_{ij}^2\big)\,
    \hat{\Sigma}^{(1)}_{ji}\big(\overline{p}_{ij}^2\big)}{\overline{p}_{ij}^2-m^2_j}
  \ni\\
  \begin{gathered}
    \big(\overline{p}_{ii}^2 - m^2_i\big)\,\delta Z_{ii}
    - \begin{aligned}[t]
      \sum_{j\neq i}\frac{1}{\overline{p}_{ij}^2 - m^2_j}\,\bigg\{
      & \frac{1}{4}\left[\delta Z_{ij}\,\big(\overline{p}_{ij}^2 - m^2_i\big)
        + \delta Z_{ji}\,\big(\overline{p}_{ij}^2 - m^2_j\big)\right]^2\\[-2ex]
      &{+}\left[\delta Z_{ij}\,\big(\overline{p}_{ij}^2 - m^2_i\big)
        + \delta Z_{ji}\,\big(\overline{p}_{ij}^2 - m^2_j\big)\right]\!
      \left(\Sigma^{(1)}_{ij}\big(\overline{p}_{ij}^2\big) - \delta M^2_{ij}\right)
      \!\!\bigg\}\,.
    \end{aligned}
  \end{gathered}
\end{multline}
The requirement for cancellation of the
field-renormalization dependence between \refeq{eq:2Lfield}
and \refeq{eq:1L^2} dictates the following conditions:
\begin{equation}
  \overline{p}_{ii}^2 - m^2_i \to
  -\left[\breve{\Sigma}^{(1)}_{ii}\big(\breve{m}_i^2\big)
    - \delta \breve{M}^2_{ii}\right],\quad
  \Sigma^{(1)}_{ij}\big(\overline{p}_{ij}^2\big) - \delta M^2_{ij}\to
  \breve{\Sigma}^{(1)}_{ij}\big(\breve{m}_i^2\big)
  - \delta \breve{M}^2_{ij}\,,\quad
  \overline{p}_{ij}^2\to \breve{m}^2_i\,.
\end{equation}
In other words, the self-energies employed in the 1L$^2$ terms should
be calculated in the same approximation as in the 2L calculation in
order to avoid the inclusion of arbitrary UV-logarithms in the
calculation.

\tocsubsection{Degenerate case\label{subsec:neardeg}}

We assume the existence of a degenerate sector (denoted by `$D$'). In
this case, one should consider the effective mass matrix
$\widetilde{M}^2_{D}(s)$ of \refeq{eq:neardegmassmat}, evaluated at a
pole
\begin{align}
  \mathcal{M}^2_I &= \overline{p}_{ij}^2+\mathcal{O}(\text{2L}) =
  m_{ij}^2+\mathcal{O}(\text{1L})
\end{align}
of the propagator---refer to \appx{ap:propmat} for a derivation. For
two field directions~$i,j\in D$, we define the
notation \mbox{$m^2_{ij}\equiv\big(m_i^2+m_j^2\big)\big/2$}.

Then, there exists a unitary matrix~$\mathbf{U}^I$ such
that
\begin{align}
  \big(\mathbf{U}^{I}\big)^*\!\cdot\!\left[\mathcal{M}^2_I\,\mathds{1}_D
    - \widetilde{M}^2_{D}\big(\mathcal{M}^2_I\big)\right]\!\cdot\!
  \big(\mathbf{U}^{I}\big)^{\dagger} = \text{diag}[\mathcal{D}_J]
\end{align}
with $\mathcal{D}_I\equiv0$ ($\mathcal{D}_J$, $J\neq I$ has no
particular relevance), hence
\begin{equation}
  \sum_{j\in D}\left[\widetilde{M}^2_{D}\big(\mathcal{M}^2_I\big)\right]_{ij}
  \big(U^{I}_{Ij}\big)^* = \mathcal{M}^2_I\,\big(U^{I}_{Ii}\big)^*\,,\quad
  \sum_{i,j\in D}\big(U^{I}_{Ii}\big)^*\,\big(U^{I}_{Ij}\big)^*\!
  \left[\mathcal{M}^2_I\,\delta_{ij}
    - \left[\widetilde{M}^2_{D}\big(\mathcal{M}^2_I\big)\right]_{ij}
  \right] = 0\,.
\end{equation}
Thus, ${\cal M}^2_I$ is an eigenvalue of
$\widetilde{M}^2_{D}\big({\cal M}^2_I\big)$ and its eigenvector
$\big(U^{I}_{Ij}\big)^*$ generates the kernel of \mbox{$\big[{\cal
M}^{2}_I\,\mathds{1}_D - \widetilde{M}^{2}_{D}\big({\cal
M}^2_I\big)\big]^{\dagger}\!\cdot\!\big[{\cal M}^{2}_I\,\mathds{1}_D
- \widetilde{M}^{2}_{D}\big({\cal M}^2_I\big)\big]$}. These are the
defining properties that we are going to exploit below after expanding
$\widetilde{M}^{2}_D\big({\cal M}^2_I\big)$.

As explained in \citere{Domingo:2020wiy}, it is possible, at the 1L
order, to expand $\widetilde{M}^2_{D}\big({\cal M}^2_I\big)$ in a
fashion that is invariant under field renormalization (in the `weak'
sense discussed in \sect{sec:fieldinv}):
\begin{align}\label{eq:1Leffmass}
  \widetilde{M}^2_{D}\big({\cal M}^2_I\big) &=
  \widetilde{M}^{2\,\text{(1)}}_{D}+\mathcal{O}(\text{2L})
  \quad\text{with}\quad
  \big(\widetilde{M}^{2\,\text{(1)}}_{D}\big)_{ij} \equiv
  m_i^2\,\delta_{ij} - \hat{\Sigma}^{\text{(1)}}_{ij}\big(m^2_{ij}\big)
\end{align}
(which now has lost its dependence on the value of the pole). We
denote the associated poles and eigenvectors of 1L order as ${\cal
M}^{2\,\text{(1)}}_I$
and \mbox{$S_{Ij}\equiv \big(U^{I\,\text{(1)}}_{Ij}\big)^*$}: the
poles are simply given by the eigenvalues of
$\widetilde{M}^{2\,\text{(1)}}_{D}$ while the matrix
$\mathbf{U}^{I\,\text{(1)}}$ is obtained from the diagonalization
of \mbox{$\big[{\cal M}^{2\,\text{(1)}}_I\,\mathds{1}_D
- \widetilde{M}^{2\,\text{(1)}}_{D}\big]^{\dagger}\!\cdot\!\big[{\cal
M}^{2\,\text{(1)}}_I\,\mathds{1}_D-\widetilde{M}^{2\,\text{(1)}}_{D}\big]$}.
There is a subtle difference in the definition of the eigenvectors
with respect to the procedure presented in \citere{Domingo:2020wiy}:
we ensure the exact cancellation of off-diagonal terms in
\mbox{$\big(\mathbf{U}^{I\,\text{(1)}}\big)^*\!\cdot\!\big[{\cal
      M}^{2\,\text{(1)}}_I\,\mathds{1}_D -
    \widetilde{M}^{2\,\text{(1)}}_{D}\big]\!\cdot\!\big(\mathbf{U}^{I\,\text{(1)}}\big)^{\dagger}$}
now, while they could still amount to $\mathcal{O}(\text{2L})$ in
\citere{Domingo:2020wiy}; the reason is that we now also want to put
2L effects under control.

Let us now consider the 2L order:
\begin{align}\label{eq:massmatdeg}
  \widetilde{M}^2_{D}\big({\cal M}^2_I\big) &=
  \begin{aligned}[t] \Bigg[
    & \big(\widetilde{M}^{2\,\text{(1)}}_{D}\big)_{ij}
    - \hat{\Sigma}^{(2)}_{ij}\big(m^2_{ij}\big)
    - \big(\overline{p}_{ij}^2 - m^2_{ij}\big)\,
    \frac{d\hat{\Sigma}^{(1)}_{ij}}{dp^2}\big(m^2_{ij}\big)\\
    &{+}\,\sum_{l\notin D}\frac{\big(m_{ij}^2 - m_l^2\big)\,
      \hat{\Sigma}^{(1)}_{il}\big(\overline{p}_{ijl}^2\big)\,
      \hat{\Sigma}^{(1)}_{jl}\big(\overline{p}_{jil}^2\big)}
         {\big(m_i^2 - m_l^2\big)\,\big(m_j^2-m_l^2\big)}\Bigg]_{i,j\in D}
         + \mathcal{O}(\text{3L})\taghere
  \end{aligned}
\end{align}
Once again, the dependence on field-renormalization constants of 2L
order is neutralized by setting
\mbox{$p^2\stackrel[]{!}{=}\breve{m}^2_{ij}={\cal
    M}^2_I+\mathcal{O}(\text{1L})$} in the argument of
$\hat{\Sigma}^{(2)}_{ij}$. We still need to determine the
momenta \mbox{$\overline{p}_{ij}^2={\cal
M}^2_I+\mathcal{O}(\text{2L})$} and $\overline{p}_{ijl}^2={\cal
M}^2_I+\mathcal{O}(\text{1L})$. To this end, we consider the
dependence on 1L$^2$ field counterterms, directly restricting
ourselves to the symmetric case $\delta Z_{ji}=\delta Z_{ij}$. Then,
instead of the diagonal matrix element of \refeq{eq:2Lfield} one has
to take the full self-energy matrix for the degenerate sector into
account, depending on 1L field counterterms as follows:
\begin{align}\label{eq:2Loffdiag}
  \hat{\Sigma}^{(2)}_{ij}\big(p^2\big) &\ni
  \sum_k \begin{aligned}[t]\bigg\{
  & \frac{1}{4}\,\big(p^2 - \breve{m}_k^2\big)\,\delta Z_{ki}\,\delta Z_{kj}\\
  &{+}\, \frac{1}{2} \left(\delta Z_{ki} \left[
    \breve{\Sigma}^{(1)}_{kj}\big(p^2\big) - \delta \breve{M}^2_{kj}\right]
  + \delta Z_{kj} \left[
    \breve{\Sigma}^{(1)}_{ki}\big(p^2\big) - \delta \breve{M}^2_{ki}\right]
  \right)\!\!\bigg\}\,.\taghere\end{aligned}
\end{align}
With the choice $\overline{p}_{ijl}^2\equiv m^2_{ij}
\equiv\overline{p}_{jil}^2$ one has
\begin{multline}
  \sum_{l\notin D}\frac{\big(m_{ij}^2 - m_l^2\big)\,
    \hat{\Sigma}^{(1)}_{il}\big(m^2_{ij}\big)\,
    \hat{\Sigma}^{(1)}_{jl}\big(m^2_{ij}\big)}
      {\big(m_i^2 - m_l^2\big)\,\big(m_j^2 - m_l^2\big)} \ni\\
  \begin{aligned}[b]
    \sum_{l\notin D}\!\bigg\{
    & \frac{1}{4}\,\big(m_{ij}^2 - m_l^2\big)\,\delta Z_{li}\,\delta Z_{lj}
    + \frac{1}{2}\,\delta Z_{li} \left[\Sigma^{(1)}_{lj}\big(m_{ij}^2\big)
      - \delta {M}^2_{lj}\right]
    \left[1 + \frac{1}{2}\,\frac{m_j^2 - m_i^2}{m_i^2 - m_l^2}\right]\\
    &{+}\,\frac{1}{2}\,\delta Z_{lj} \left[\Sigma^{(1)}_{li}\big(m_{ij}^2\big)
      - \delta {M}^2_{li}\right]
    \left[1 + \frac{1}{2}\,\frac{m_i^2 - m_j^2}{m_j^2 - m_l^2}\right]
  \!\bigg\}\,.\end{aligned}
\end{multline}
Then, the dependence on field-renormalization constants $\delta
Z_{li}$ with $i\in D$, $l\notin D$, cancels out against the
one of \refeq{eq:2Loffdiag}, provided \mbox{$\Sigma^{(1)} -
  \delta{M}^2\to\breve{\Sigma}^{(1)} - \delta\breve{M}^2$}
and up to terms of 3L order involving an additional mass suppression.

If we assume that the degeneracy is lifted at the 1L~order,
\IE~$\big\lvert{\cal M}_I^{2\,\text{(1)}} - {\cal
  M}_J^{2\,\text{(1)}}\big\rvert>\mathcal{O}(\text{2L})$ for~$I\neq
J$, then the off-diagonal elements of
$\big(\mathbf{U}^{I\,\text{(1)}}\big)^*\!\cdot\!\big[{\cal
M}^2_I\,\mathds{1}_D - \widetilde{M}^2_{D}\big({\cal
M}^2_I\big)\big]\!\cdot\!\big(\mathbf{U}^{I\,\text{(1)}}\big)^{\dagger}$
are negligible, since of 2L order when the diagonal splitting is of 1L
order. We can thus focus on the diagonal element
\begin{align}\label{eq:2Lmassexp}
  \widetilde{\mathcal{D}_I} &\equiv \begin{aligned}[t]
    \sum_{i,j\in D}S_{Ii}S_{Ij}\,\Bigg\{
    & \!\!\left({\cal M}^2_I - {\cal M}^{2\,\text{(1)}}_I\right) \delta_{ij}
      + \hat{\Sigma}^{(2)}_{ij}\big(m^2_{ij}\big)
        + \big(\overline{p}_{ij}^2 - m^2_{ij}\big)\,
        \frac{d\hat{\Sigma}^{(1)}_{ij}}{dp^2}\big(m^2_{ij}\big)\\
    &{-} \sum_{l\notin D}\frac{\big(m_{ij}^2 - m_l^2\big)\,
          \hat{\Sigma}^{(1)}_{il}\big(m^2_{ij}\big)\,
          \hat{\Sigma}^{(1)}_{jl}\big(m^2_{ij}\big)}
        {\big(m_i^2 - m_l^2\big)\,\big(m_j^2 - m_l^2\big)}\Bigg\}\,.\taghere
  \end{aligned}
\end{align}
We define
$\overline{p}_{ij}^{2}\equiv\big(\overline{p}_{ij}^{2\,(i)}+\overline{p}_{ij}^{2\,(j)}\big)\big/2$
and, from the 1L eigenstate equation (adding a convenient 2L~term):
\begin{equation}\label{eq:2Lmass1L}
  \overline{p}_{ij}^{2\,(i)} \equiv m_i^2
  - S_{Ii}^{-1} \sum_{k\in D} S_{Ik} \left[\hat{\Sigma}^{(1)}_{ik}\big(m^2_{jk}\big)
    + \frac{m^2_{ik} - m_j^2}{2}\,
    \frac{d\hat{\Sigma}^{(1)}_{ik}}{dp^2}\big(m^2_{ik}\big)\right] =
  {\cal M}^2_I + \mathcal{O}(\text{2L})\,.
\end{equation}
Then, we obtain $\widetilde{\mathcal{D}_I}= \sum\limits_{i,\,j\in
  D}{S_{Ii}\,S_{Ij}\,\Big[\big({\cal M}^2_I - {\cal M}^{2\,\text{(1)}}_I\big)\,
    \delta_{ij} + \mathcal{W}_{ij}\Big]}$ with
\begin{align}\label{eq:2Lderterm}
  {\cal W}_{ij} &\equiv \begin{aligned}[t]
    \hat{\Sigma}^{(2)}_{ij}\big(m_{ij}^2\big)
    & - \frac{1}{2}\sum_{k\in D}\left\{\left[
      \hat{\Sigma}^{(1)}_{ik}\big(m^2_{ij}\big)
      + \frac{m^2_{ij} - m_k^2}{2}\,
      \frac{d\hat{\Sigma}^{(1)}_{ik}}{dp^2}\big(m^2_{ik}\big)\right]
    \frac{d\hat{\Sigma}_{jk}^{(1)}}{dp^2}\big(m^2_{jk}\big)
    + (i\leftrightarrow j)\right\}\\
    & - \sum_{l\notin D}\frac{\big(m_{ij}^2 - m_l^2\big)\,
      \hat{\Sigma}^{(1)}_{il}\big(m^2_{ij}\big)\,
      \hat{\Sigma}^{(1)}_{jl}\big(m^2_{ij}\big)}
    {\big(m_i^2 - m_l^2\big)\,\big(m_j^2 - m_l^2\big)}\,.\taghere
  \end{aligned}
\end{align}
\needspace{3ex}
The dependence of ${\cal W}_{ij}$ on $\delta Z_{ij}$ with $i,j\in D$
almost cancels (provided all 1L$^2$ quantities are calculated in the
same approximations as the self-energies of 2L order, \EG~in the
gaugeless limit), up to a remainder of 3L order:
\begin{equation}\label{eq:FDremainder}
  \frac{m_i^2 - m^2_j}{8} \left[
    \delta Z_{ki}\,\frac{d\Sigma_{jk}}{dp^2}\big(m^2_{jk}\big)
    - \delta Z_{kj}\,\frac{d\Sigma_{ik}}{dp^2}\big(m^2_{ik}\big)\right].
\end{equation}
We find no obvious method to absorb this piece by adding a finite term
of 3L order to $\mathcal{W}_{ij}$; thus,
after applying the condition
$\widetilde{\mathcal{D}}_I\stackrel[]{!}{=}\mathcal{O}(\text{3L})$, a
subleading field-renormalization dependence remains in the
determination of the pole of 2L order by
\begin{equation}\label{eq:2Lpoledeg}
  {\cal M}^{2\,(2)}_I \equiv {\cal M}^{2\,\text{(1)}}_I
  - \left.\sum_{i,\,j\in D}S_{Ii}\,S_{Ij}\,\mathcal{W}_{ij}\middle/
  \sum_{k\in D}S_{Ik}^2\right..
\end{equation}
This persisting dependence on field-renormalization constants in the
degenerate case is intimately related to the form
of the dependence of the off-diagonal self-energy of 2L order on 1L
field counterterms---see \refeq{eq:2Loffdiag}. In the non-degenerate
case, this object only intervenes in the mass corrections of 3L order,
so that we can expect the field-renormalization dependence to be
tackled by the inclusion of the diagonal $\hat{\Sigma}^{(3)}_{ii}$
(which, however, goes far beyond our purpose and our current
means). The 2L terms of \refeq{eq:2Lmass1L} (contributing at 3L order)
have been chosen so as to avoid a quadratic dependence on the field
counterterms. Other choices are of course possible. While the full
cancellation of the dependence on field counterterms thus fails in the
degenerate scenario, we still expect an improvement in the expansion
method, as compared to the iterative pole search, due to the careful
pairing of 2L and 1L$^2$ effects.

In case the degeneracy is not lifted at 1L order, we can still define
${\cal M}^{2\,(2)}_I$ as an eigenvalue of the effective mass matrix
\begin{equation}\label{eq:effmass2L}
  \widetilde{M}^{2\,\text{(2)}}_{\text{eff}}\equiv
  \left[m_i^2\,\delta_{ij} - \hat{\Sigma}^{(1)}_{ij}\big(m^2_{ij}\big)
  - \mathcal{W}_{ij}\right],\quad i,j\in D\,.
\end{equation}
Indeed, similarly to the 1L piece in \refeq{eq:1Leffmass}, the 2L
corrections~$\mathcal{W}_{ij}$ of \refeq{eq:2Lderterm} are independent
from the chosen pole, the selection being ensured by the projection
via the eigenvectors $\mathbf{U}^I$. It is then convenient to directly
consider these eigenvectors, albeit slightly dependent on the choice
of field renormalization, as determining the mixing
matrix~$\mathbf{S}$ in the degenerate subspace at 2L order. In
addition, the normalization \mbox{$\sum_{k\in
D}S_{Ik}^2\stackrel[]{!}{=}1$} simplifies the application of this
object, as can be read from \refeq{eq:2Lpoledeg} or the derivation of
the Higgs-decay amplitudes in \appx{ap:vertexcorr}.

Finally, we stress that it is necessary to compute all the 1L$^2$
pieces in the approximation of the 2L calculation, \IE~all
self-energies, mass counterterms and tree-level masses in
$\mathcal{W}$ are replaced by $\breve{\Sigma}^{(1)}$,
$\delta\breve{\cal M}$ and $\breve{m}$. An apparent difficulty
accompanies the observation that the identification of `original' and
`gaugeless' states is no longer trivial in the degenerate
scenario. Yet, as the full~2L\,+\,1L$^2$ mass contribution is
collected within the block~$\mathcal{W}$, it is possible to compute
the latter in the gaugeless base and then rotate it to the `original'
base using the gauge eigenbase as reference. This \AH procedure is
however a sign that the combination of gaugeless 2L effects with a
full 1L calculation is not defined in a completely consistent fashion
in the near-degenerate case.

\tocsection{Field-dependence in the mass predictions for a non-degenerate scenario\label{sec:nondegeneratecase}}

We first investigate the numerical significance of radiative
corrections to the masses of MSSM Higgs bosons in the clearly defined
configuration where all states are non-degenerate, and assess the
weight of the dependence on field counterterms in various methods of
evaluation. The required Feynman diagrams are computed with the help
of \texttt{FeynArts}\,\cite{Kublbeck:1990xc},
\texttt{FormCalc}\,\cite{Hahn:2000kx,Hahn:1998yk},
\texttt{TwoCalc}\,\cite{Weiglein:1993hd} and
\texttt{TLDR}\,\cite{Goodsell:2019zfs}. The 1L integrals
are implemented analytically, while the 2L integrals
are numerically evaluated with the assistance
of \texttt{TSIL}\,\cite{Martin:2005qm}.

\tocsubsection{Preliminary considerations}

We focus on a `typical' MSSM scenario with squarks of third generation
($\tilde{Q}_3$) and gluinos ($\tilde{g}$) at the edge of the mass
region probed by the LHC ($m_{\tilde{Q}_3}\sim1.5$\,TeV and
$M_3\sim2$\,TeV; $A_t=2.3$\,TeV, $A_b=1$\,TeV). EW-only-interacting
SUSY particles are given a sub-TeV mass. The ratio of the doublet
Higgs v.e.v.-s, $t_{\beta}$, is set to $10$ and we vary the
charged-Higgs mass between $0.5$\,TeV and~$4$\,TeV. In these
conditions, the neutral SM-like Higgs ($h$), with mass at the EW
scale, and the heavy-doublet states ($H$, $A$), with masses comparable
to that of the charged Higgs ($H^{\pm}$), are clearly
non-degenerate. The \CP-even ($H$) and \CP-odd ($A$) heavy states do
not mix in the absence of \CP-violation, which we assume in this
section. One can thus safely employ the formalism corresponding to the
non-degenerate scenario. The lagrangian parameters in the Higgs sector
are renormalized in the same scheme as employed
in \citeres{Domingo:2018uim,Domingo:2020wiy},
\IE~with cancellation of the tadpoles, on-shell conditions for the EW
gauge-boson, SM-fermion and charged-Higgs masses, while~$t_{\beta}$ is
defined in the \DR scheme.

\begin{figure}[t!]
  \centering
  \includegraphics[width=\linewidth]{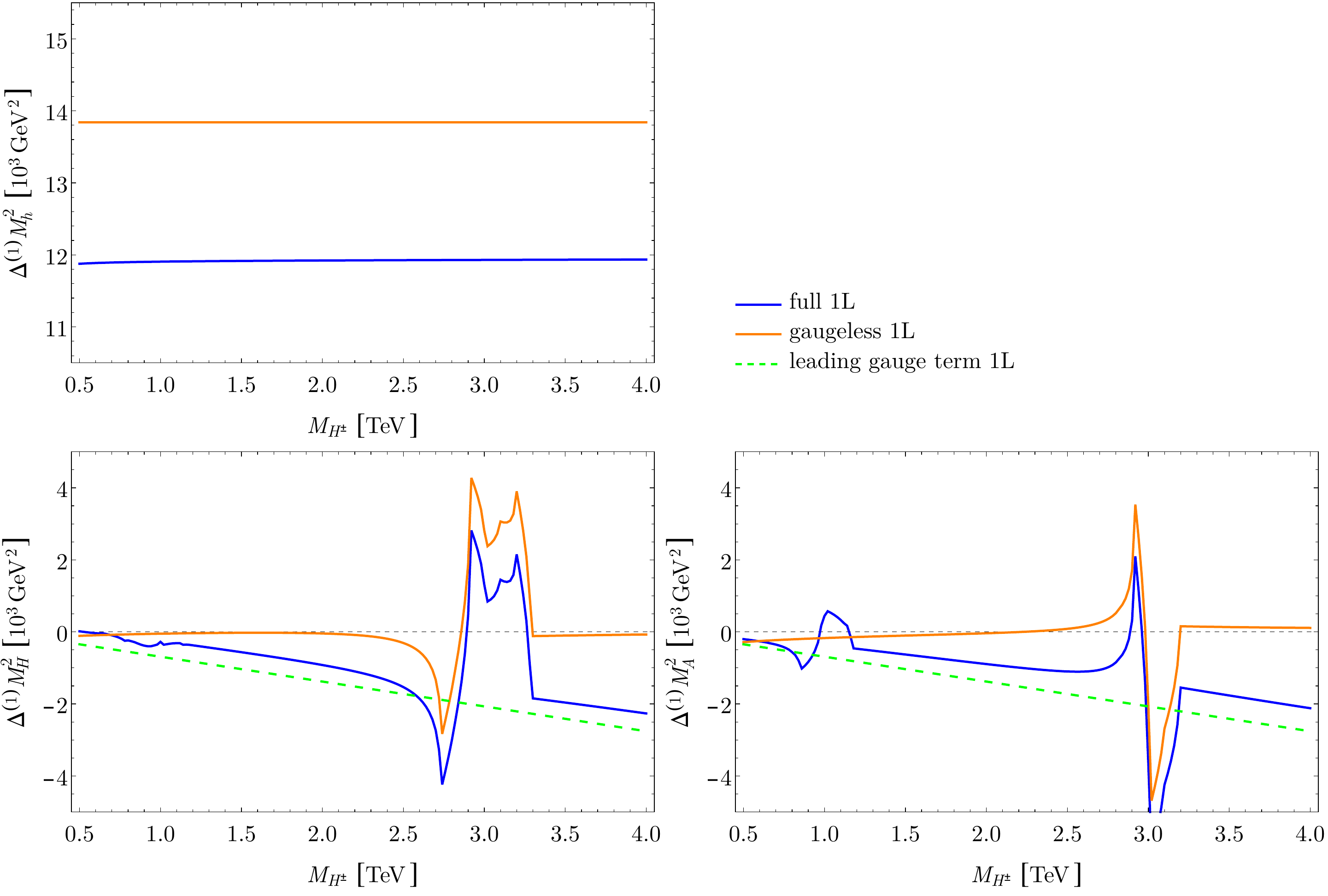}
  \caption{Mass shifts at 1L for the neutral Higgs states. The full 1L
    effect is shown in blue and the corresponding result in the
    gaugeless limit in orange. For the heavy states (lower row), the
    leading contribution from gauge loops, scaling linearly with the
    Higgs mass, is presented in dashed green.\label{fig:DM2H1L}}
\end{figure}

It is instructive to first consider the corrections of 1L order to the
tree-level masses $m_{h_i}$ of the neutral Higgs bosons
$h_i\in\{h,H,A\}$. In Fig.~\ref{fig:DM2H1L}, we show the corresponding
shifts in mass squared, $\Delta^{(1)}M^2_{h_i}\equiv
M^{2(1)}_{h_i}-m^2_{h_i}=-\Real{\hat{\Sigma}^{(1)}_{h_ih_i}\big(m^2_{h_i}\big)}$.\enlargethispage{.1ex}\footnote{In
the case of the heavy-doublet states, one could more properly consider
the radiative corrections to the mass-splitting between neutral and
charged states (otherwise, the squared tree-level mass $m^2_{h_i}$ is
not straightforwardly related to observable quantities): as the
charged Higgs is renormalized on-shell in our scheme, this definition
is equivalent.} The orange curves correspond to the gaugeless limit,
while the full 1L shift is shown in blue. The situation of the SM-like
Higgs is straightforward: the radiative corrections of Yukawa type
$\mathcal{O}(\alpha_{t,(b)})$, captured in the gaugeless approach,
indeed dominate the 1L shift. Contributions of EW type, beyond the
gaugeless description, amount to only $\simord15\%$. The gaugeless
limit can thus be seen as predictive for this state.\footnote{However,
contributions of $\mathcal{O}(\alpha_t)$ are reduced by large QCD
effects---which could be in part absorbed in the tree-level Yukawa
couplings in a convenient scheme. This ultimately increases the impact
of EW contributions lost in the gaugeless approximation.}

Nevertheless, the impact of the various contributions follows a
different pattern in the case of the heavy-doublet states. A first
identifiable feature corresponds to the `spikes' in the vicinity
of~$M_{H^{\pm}}=3$\,TeV: these are associated with threshold effects
in the loop integrals of the self-energies, when \mbox{$M_{H,A}\sim
m_X+m_Y$} for $X$ and $Y$ two particle species entering the loop
diagram ($m_X$ and $m_Y$, their masses). At $M_{H^{\pm}}\sim3$\,TeV,
these internal `on-shell' lines are squarks of the third generation,
contributing at~$\mathcal{O}(\alpha_{q})$. The sharp variation in the
mass shift is consequently a physical effect, although it is not
quantitatively described in the `free-particle' expansion---we will
make no attempt at addressing the threshold behavior by accounting for
squark interactions in this paper:
see \EG~\citere{Drees:1993uw}. Comparable features also appear close
to $M_{H^{\pm}}\approx1$\,TeV for the blue curve (full 1L corrections)
and correspond to electroweakino loops: these do not show in the
gaugeless limit as EW interactions are turned off.

Beyond this threshold behavior, the pattern of 1L corrections is
essentially flat in the gaugeless limit while a slope is definitely
identifiable in the presence of gauge effects. As explained in
Sect.\,3.2 of \citere{Domingo:2020wiy}, the radiative contributions of
gauge type to the squared mass-splitting between heavy-doublet states
generate a term scaling linearly with the Higgs mass. This leading
effect is shown as a dashed green line in \fig{fig:DM2H1L} and indeed
captures the slope of the full 1L result. The shift between the green
and blue curves is due to corrections scaling like
$M^2_{\text{EW}}\ln^{k}\!M^2_{H^{\pm}}\big/M^2_{\text{EW}}$, where
$k\in\{0,1\}$ and $M^2_{\text{EW}}\sim M_Z^2$ denotes the EW
scale. Therefore, radiative corrections of Yukawa type do not capture
the bulk of the contribution to the mass-splitting between heavy
states and corrections of orders $\alpha_{q}\,\alpha_s$ or
$\alpha_{q}^2$ are not expected to dominate the 2L corrections
either. While these orders are fully under control, their inclusion
does not improve the numerical precision of the mass predictions to
the heavy states as long as dominant corrections of EW gauge-type are
not considered. The latter are known as far as scalar self-energies
are concerned\,\cite{Goodsell:2019zfs}, but are currently not
exploitable as 2L contributions to the vector self-energies that are
needed for renormalization and the connection to observable input are
still missing. Consequently, when we discuss the known
$\mathcal{O}\big(\alpha_{q}\,\alpha_s,\,\alpha_{q}^2\big)$ at the
level of the heavy-doublet states below, it is purely for the sake of
testing the formalism described in \sect{sec:2Lmasses} on controlled
orders: the theoretical prediction is actually not improved as
compared to the strict 1L result.

\tocsubsection{Corrections of \texorpdfstring{$\mathcal{O}(\alpha_q\,\alpha_s)$}{\unicodescriptO(\unicodealpha\unicodesubt\unicodethinspace\unicodealpha\unicodesubs)}}

The (S)QCD corrections have a specific status in the contributions of
2L order to the Higgs masses, because no corresponding 1L$^2$ effects
are associated. Thus, this order does not entail the inclusion of any
off-diagonal self-energies---unless these are already needed at
1L, \IE~in the near-degenerate case. Given the large impact of QCD
corrections to effects of Yukawa-type, they are particularly relevant
for an accurate determination of the mass of the SM-like state.

\begin{figure}[b!]
  \centering
  \vspace{-1.4ex}
  \includegraphics[width=\linewidth]{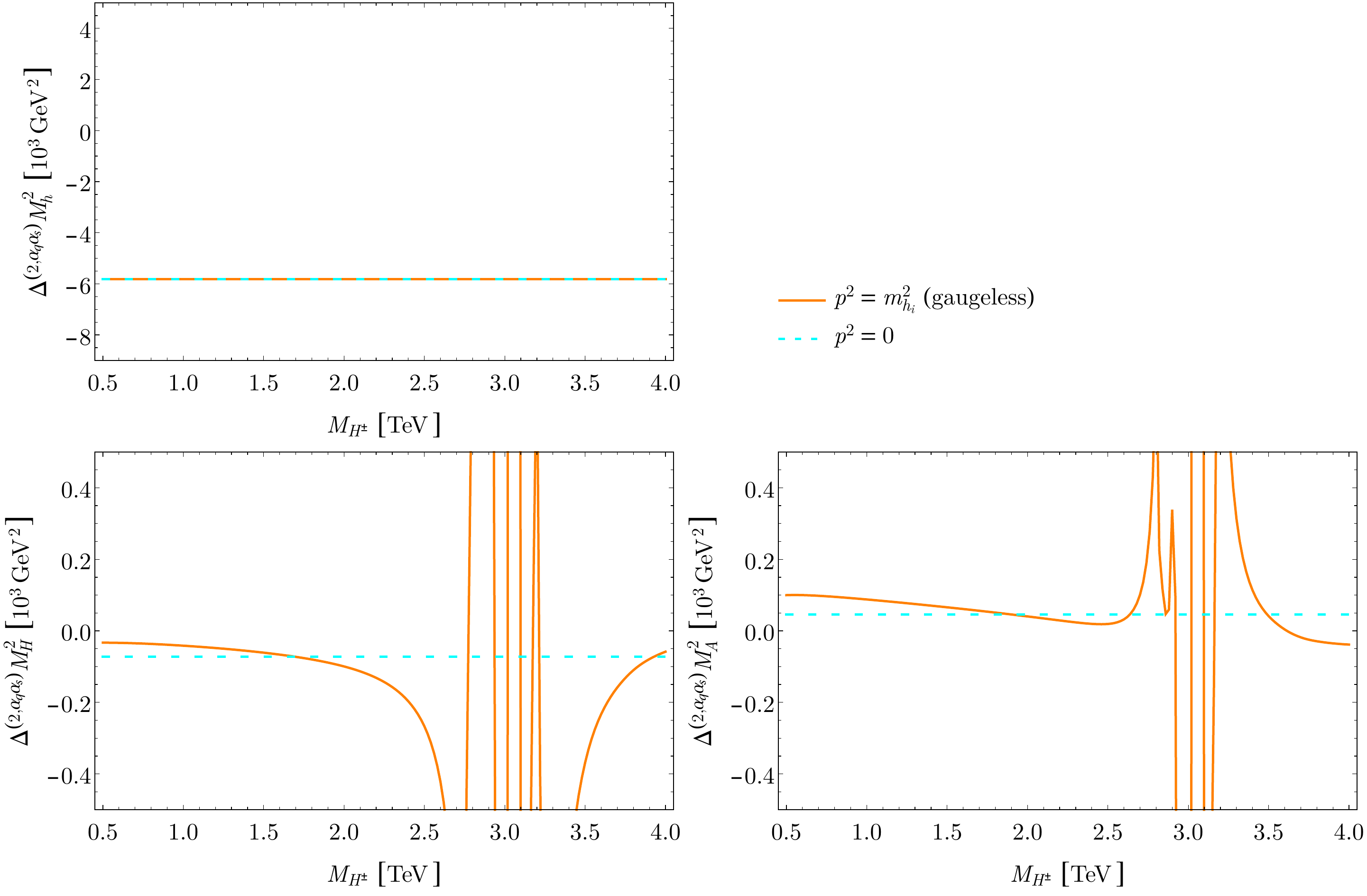}\vspace{-1ex}
  \caption{Mass shifts at $\mathcal{O}(\alpha_{q}\,\alpha_s)$ for
    the neutral Higgs states. The result in the gaugeless limit is
    shown in orange with momentum set to the (gaugeless) tree-level
    mass. The dashed cyan curve corresponds to the effective-potential
    approximation.\label{fig:DM2H2LQCD}}
  \vspace{-2ex}
\end{figure}

In \fig{fig:DM2H2LQCD}, we plot the shifts of the squared masses
obtained at the considered order, \IE
\begin{align}
  \Delta^{(2,\alpha_{q}\alpha_s)}M^2_{h_i} &\equiv
  M^{2(2,\alpha_{q}\alpha_s)}_{h_i} - M^{2(1)}_{h_i} =
  -\Real{\hat{\Sigma}^{(2,\alpha_{q}\alpha_s)}_{h_ih_i}\big(m^{2(\mathrm{gl})}_{h_i}\big)}\,,
\end{align}
with corrections obtained in the gaugeless (gl) limit. The solid
orange curves have momentum evaluated at the tree-level (gaugeless)
mass of the Higgs states. The cyan dashed lines are derived in the
effective-potential approximation, \IE~with momentum set to $0$. This
approximation is `exact' in the case of the SM-like Higgs, because the
corresponding tree-level mass in the gaugeless limit is indeed equal
to~$0$ (this is not necessarily true in extensions of the MSSM), so
that the orange and cyan curves overlap. For the heavy-doublet states,
threshold effects originating in squark loops are again prominent
(when considering momentum dependence). Even far from the threshold
region, the effective-potential approximation does not appear as a
particularly useful approach for the heavy states, as its predictions
for the order $\alpha_{q}\,\alpha_s$ are $\mathcal{O}(100\%)$ away
from the actual momentum-dependent contributions. This is not
surprising for Higgs states with mass differing significantly
from~$0$, even in the gaugeless limit. In this context, the
self-energies of order~$\alpha_{q}\,\alpha_s$ with vanishing momentum
can at best provide an estimate of the corresponding order in an
assessment of the uncertainties, but are not predictive.

In conformity with the discussion of \sect{sec:2Lmasses}, the mass
shifts of $\mathcal{O}(\alpha_{q}\,\alpha_s)$ presented in orange in
\fig{fig:DM2H2LQCD} (or those of 1L order in \fig{fig:DM2H1L}) are
independent from the field counterterms, because the self-energies
have been expanded for momenta in the vicinity of the tree-level
masses, and truncated at the strict desired order---meaning in
practice that the self-energies are evaluated at the tree-level Higgs
mass. For definiteness, we remind here the formal expression defining
the pole mass in this expansion-and-truncation approach for a
non-degenerate state $h_i$:
\begin{equation}\label{eq:asatexp}
  M^{2(2,\alpha_{q}\alpha_s)}_{h_i} = m^2_{h_i}
  - \Real{\hat{\Sigma}^{(1)}_{h_ih_i}\big(m^2_{h_i}\big)
    + \hat{\Sigma}^{(2,\alpha_{q}\alpha_s)}_{h_ih_i}\big(m^{2(\mathrm{gl})}_{h_i}\big)}\,.
\end{equation}

In competition with such an expansion, a popular approach consists in
iteratively replacing the momenta in the self-energies by the value
derived in the pole-mass determination. As off-diagonal effects do not
matter at $\mathcal{O}(\alpha_{q}\,\alpha_s)$ in a non-degenerate
scenario,\footnote{Off-diagonal contributions actually produce
EW-violating pieces of higher order, but numerically dominant in the
decoupling limit, see \citere{Domingo:2020wiy}.} we define such a pole
search of order $\alpha_{q}\,\alpha_s$ by simply considering the shift
of the diagonal element of the effective mass matrix. In other words,
the equation that we solve iteratively in this subsection reads
\begin{equation}\label{eq:2Lasatiter}
  \mathfrak{M}^{2(2,\alpha_{q}\alpha_s)}_{h_i} = m^2_{h_i}
  - \hat{\Sigma}^{(1)}_{h_ih_i}\Big(\mathfrak{M}^{2(2,\alpha_{q}\alpha_s)}_{h_i}\Big)
  - \hat{\Sigma}^{(2,\alpha_{q}\alpha_s)}_{h_ih_i}\Big(\mathfrak{M}^{2(2,\alpha_{q}\alpha_s)}_{h_i}\Big)
\end{equation}
with $\mathfrak{M}^{2(2,\alpha_{q}\alpha_s)}_{h_i}$ denoting the
(complex) pole mass---by an abuse of language, we will employ the same
notation for its real part in the discussion below. In order to
achieve UV-finite results for the renormalized self-energies
of \refeq{eq:2Lasatiter} away from $p^2=m^2_{h_i}$, the definition of
these objects requires the introduction of field counterterms, thereby
generating a dependence of higher order on field regulators in
$\mathfrak{M}^{2(2,\alpha_{q}\alpha_s)}_{h_i}$.

\needspace{3ex}
To understand the differences between the expansion and iteration
procedures, we investigate the mass shift between both through an
expansion, where we only keep leading terms:
\begin{align}\label{eq:itervsexp}
\mathfrak{M}^2_{h_i} - M^2_{h_i} &\approx
-\left(\mathfrak{M}^2_{h_i} - m^2_{h_i}\right)
\hat{\Sigma}^{\text{(1)}\prime}_{h_ih_i}
- \left(\mathfrak{M}^2_{h_i} - m^{2\text{(gl)}}_{h_i}\right)
\hat{\Sigma}^{\text{(2,gl)}\prime}_{h_ih_i}
- \hat{\Sigma}^{\text{(1,gl)}}_{h_ih_i}\,\hat{\Sigma}^{\text{(1,gl)}\prime}_{h_ih_i}\nonumber\\
&\approx \hat{\Sigma}^{\text{(2,gl)}}_{h_ih_i}\,
\hat{\Sigma}^{\text{(1,gl)}\prime}_{h_ih_i}
+ \left[m^{2\text{(gl)}}_{h_i} - m^2_{h_i} + \hat{\Sigma}^{\text{(1,gl)}}_{h_ih_i}
+ \hat{\Sigma}^{\text{(2,gl)}}_{h_ih_i}\right]
\hat{\Sigma}^{\text{(2,gl)}\prime}_{h_ih_i}\nonumber\\
&\quad + \hat{\Sigma}^{\text{(1,EW)}\prime}_{h_ih_i}\left[
\hat{\Sigma}^{\text{(1,gl)}}_{h_ih_i} + \hat{\Sigma}^{\text{(2,gl)}}_{h_ih_i}\right]
+ \hat{\Sigma}^{\text{(1,EW)}}_{h_ih_i}\left[\hat{\Sigma}^{\text{(1,gl)}\prime}_{h_ih_i}
+ \hat{\Sigma}^{\text{(2,gl)}\prime}_{h_ih_i}\right]
+ \hat{\Sigma}^{\text{(1,EW)}\prime}_{h_ih_i}\,\hat{\Sigma}^{\text{(1,EW)}}_{h_ih_i}
\end{align}
with all self-energies evaluated at the (gaugeless) tree-level mass,
$^{\prime}$ indicating differentiation with respect to the external
momentum squared, `gl' and `EW' referring to the gaugeless and
electroweak contributions respectively. We
assumed \mbox{$\hat{\Sigma}^{\text{(1)}}_{h_ih_i}\approx\hat{\Sigma}^{\text{(1,EW)}}_{h_ih_i}+\hat{\Sigma}^{\text{(1,gl)}}_{h_ih_i}$}
(which may be seen as a definition of the EW piece). The systematic
appearance of derivative self-energies makes evident the dependence on
field counterterms.

Below, we study the stability of Higgs-mass predictions of
$\mathcal{O}(\alpha_{q}\,\alpha_s)$ under variations of the
field-renormalization constants. For simplicity, we restrict ourselves
to a `minimal' form of the field counterterms, considering on-shell
(OS) or \DR field-renormalization constants. In the \DR approach, the
only loose parameter is the renormalization scale
$\mu_{\text{UV}}$; its variation between the
physical scales of the model, \IE~the EW and the SUSY scales, offers a
measurement of the arbitrariness introduced in the definition and
UV-regularization of physical observables, hence a lower bound on the
associated `error'.\footnote{It indeed corresponds to an `error'
introduced by the formalism rather than an `uncertainty', because the
partial higher orders contained in the field counterterms typically
violate the properties that are expected from actual higher-order
corrections; a genuine uncertainty estimate would have to
quantitatively assess the impact of higher orders (\EG~by estimating
genuine higher-order logarithms).} Yet, this very specific pattern
where distinct counterterms are correlated by a common regulator may
underestimate the actual theoretical uncertainty. In this picture, the
MSSM Higgs-field counterterms can be written as
\begin{equation}
  \delta Z_{ij}^{\text{\DR}} = \left(X^R_{id}\,X^R_{jd} + X^I_{id}\,X^I_{jd}\right)
  \delta Z_{H_d} + \left(X^R_{iu}\,X^R_{ju} + X^I_{iu}\,X^I_{ju}\right) \delta Z_{H_u}
\end{equation}
where $X^{R,I}_{if}$ encodes the decomposition of the tree-level
neutral Higgs state $h_i$ on the gauge-eigenbase:
$h_i=X^R_{id}\,h_d^0+X^R_{iu}\,h_u^0+X^I_{id}\,a_d^0+X^R_{iu}\,a_u^0$. At
the 1L order one has\footnote{We refer the reader
to \citeres{Sperling:2013eva,Sperling:2013xqa} and references therein
for calculations of the anomalous dimension of the Higgs fields. We
recover the same results at the one- and two-loop order by explicitly
computing derivatives of the Higgs self-energies and assessing their
UV-divergence.}
\begin{equation}\label{eq:1LfieldCT}
  \delta Z^{\text{(1)}}_{H_d} = -\frac{3\,\alpha_b + \alpha_{\tau}}{4\,\pi}
  \left[\overline{\Delta}^{-1}_{\text{UV}}
  + \ln\frac{\mu_{\text{ren}}^2}{\mu_{\text{UV}}^2}\right],\quad
  \delta Z_{H_u}^{\text{(1)}} = -\frac{3\,\alpha_t}{4\,\pi}
  \left[\overline{\Delta}^{-1}_{\text{UV}}
  + \ln\frac{\mu_{\text{ren}}^2}{\mu_{\text{UV}}^2}\right].
\end{equation}
Here, we neglect the Yukawa couplings of the first and second
generation. The symbol $\overline{\Delta}^{-1}_{\text{UV}}$ represents
the UV-divergence (including universal finite pieces). The finite
piece can be viewed as resetting the renormalization scale of the
fields from $\mu_{\text{ren}}$ to $\mu_{\text{UV}}$. Similarly, the 2L
field-renormalization constants of order $\alpha_{q}\,\alpha_s$ read
\begin{subequations}\label{eq:asatfieldCT}
\begin{align}
  \delta Z^{\text{(2,$\alpha_{q}\alpha_s$)}}_{H_d} &=
  \frac{\alpha_b\,\alpha_s}{2\,\pi^2}
  \left[\Big(\overline{\Delta}^{-1}_{\text{UV}}
  + \frac{\mu_{\text{ren}}^2}{\mu_{\text{UV}}^2}\Big)^2
    - \Big(\overline{\Delta}^{-1}_{\text{UV}}
    + \frac{\mu_{\text{ren}}^2}{\mu_{\text{UV}}^2}\Big)\right],\\
  \delta Z_{H_u}^{\text{(2,$\alpha_{q}\alpha_s$)}} &=
  \frac{\alpha_t\,\alpha_s}{2\,\pi^2}
  \left[\Big(\overline{\Delta}^{-1}_{\text{UV}}
  + \frac{\mu_{\text{ren}}^2}{\mu_{\text{UV}}^2}\Big)^2
    - \Big(\overline{\Delta}^{-1}_{\text{UV}}
    + \frac{\mu_{\text{ren}}^2}{\mu_{\text{UV}}^2}\Big)\right].
\end{align}
\end{subequations}
Only Yukawa and the QCD-gauge couplings appear in these expressions of
the \DR field counter\-terms. Since the EW corrections (gauge,
gauginos) generate a vanishing UV-divergence at 1L, the corresponding
pieces are insensitive to scale variations in the field
counterterms. Accordingly, the variations of $\mu_{\text{UV}}$ that we
consider below probe all terms except for the last
of \refeq{eq:itervsexp}. Thus, this type of scale variation is
meaningful as long as Yukawa effects dominate and the gaugeless
approximation holds. On the other hand, if corrections of EW type are
large, the partial higher order appearing in the last term of
\refeq{eq:itervsexp} is not tested and the uncertainty from scale
variation is only partial.

The approach with OS-renormalized fields allows to take into account
the impact of EW corrections to a certain extent, but it does not
allow for variations. In this case, we simply express the field
counterterms as cancelling the differentiated self-energies at the
corresponding tree-level (OS) mass:
\begin{equation}\label{eq:OSfieldCTs}
  \delta Z_{ij}^{\text{OS}}\equiv -\frac{d\Sigma_{ij}}{dp^2}\big(p^2=m^2_{ij}\big)\,,
\end{equation}
which is a symmetric, but (for off-diagonal field counterterms) not
fully conventional choice. We could define this object in the
gaugeless limit, in which case EW corrections would still be
overlooked. In practice, one then recovers results within the scope of
the \DR scale variation. Therefore we dismiss this choice. More
usefully, we can define the 1L field counterterms in the full
model, \IE~including EW effects. We can then compare the differences
between this procedure and the \DR evaluations. However, we do not
attempt to express the 2L field counterterms in this scheme, keeping
them \DR with $\mu_{\text{UV}}=m_t$, first, because the 2L
contributions are explicitly calculated in the gaugeless limit,
secondly because the evaluation of the differentiated 2L functions is
technically involved.

\begin{figure}[p!]
  \centering \includegraphics[width=\linewidth]{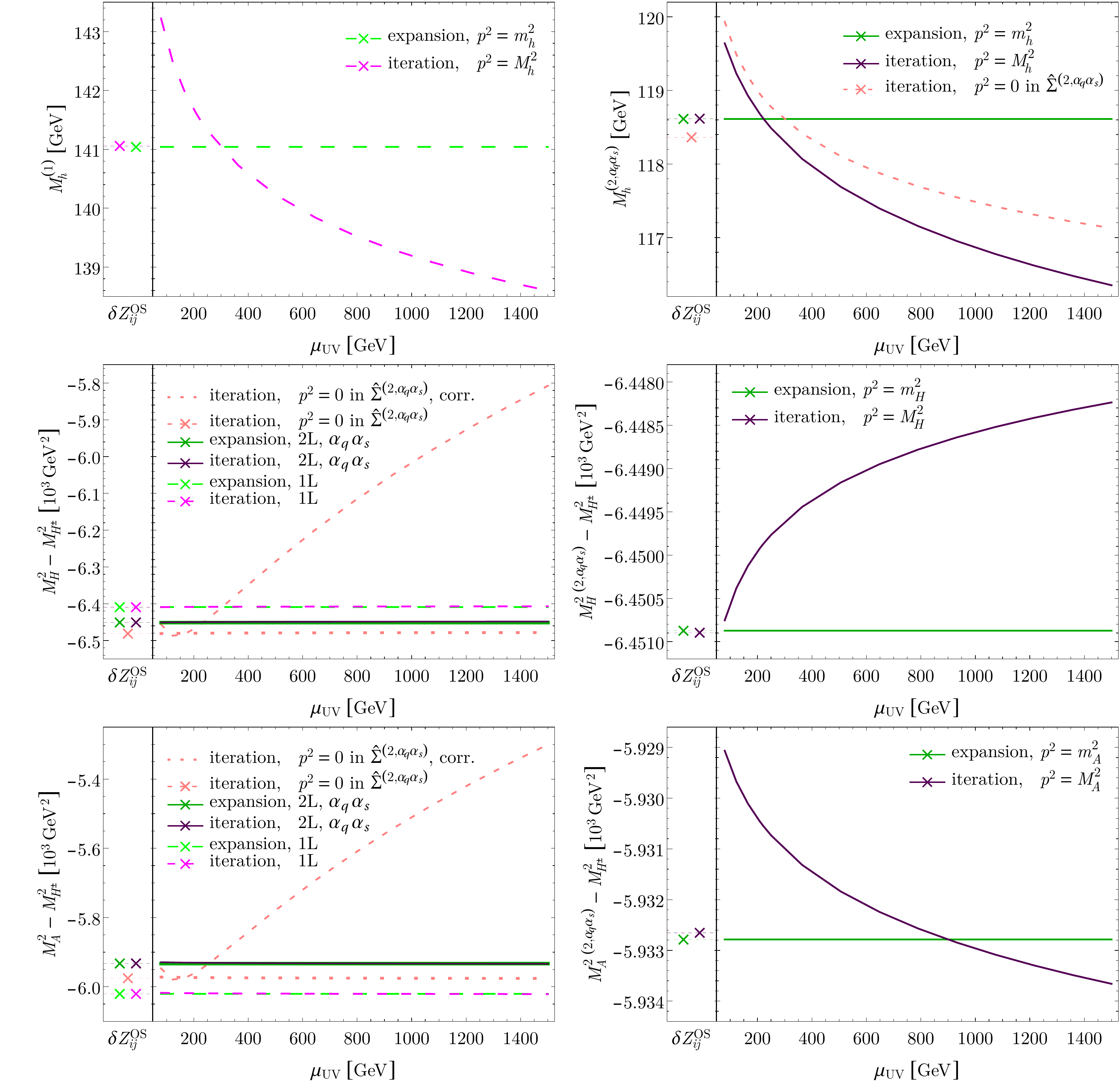}
  \caption{Dependence on the field counterterms at 1L order and
  $\mathcal{O}(\alpha_{q}\,\alpha_s)$ for $M_{H^{\pm}}=1$\,TeV. The
  off-diagonal effects are neglected in the iteration of the momenta
  injected in the self-energies. The long-dashed curves correspond to
  1L evaluations (green:~expansion; magenta:~iterative pole
  search). In the short-dashed and dotted pink curves, the
  $\mathcal{O}(\alpha_{q}\,\alpha_s)$ are included in the
  effective-potential approximation, but an iteration on the momentum
  of the 1L self-energy is included. In the short-dashed curves, the
  counterterms of the charged-Higgs fields are set to fixed values; in
  the dotted curves, they are correlated with those of the neutral
  fields. The solid lines represent calculations of
  $\mathcal{O}(\alpha_{q}\,\alpha_s)$ with full momentum dependence:
  the masses are derived via an expansion (dark green), or an
  iterative pole search (purple). The crosses in the columns on the
  left of the plots correspond to the evaluations with OS
  field counterterms.\newline {\em Up}: Mass of the light SM-like
  state at 1L (left plot; long-dashed curves) and at order
  $\alpha_{q}\,\alpha_s$ (right).\newline {\em Middle and Bottom}:
  Squared mass-splitting between the \CP-even (Middle) or \CP-odd
  (Bottom) heavy-doublet state and the charged Higgs; general
  perspective (left) and details of the 2L predictions (right).
\label{fig:M2Lasatiter}}
\end{figure}

\needspace{5ex}
In \fig{fig:M2Lasatiter}, we compare the Higgs-mass predictions of 1L
order and $\alpha_{q}\,\alpha_s$ obtained via an expansion---using
\refeq{eq:asatexp}---or an iterative pole search---using
\refeq{eq:2Lasatiter}---for $M_{H^{\pm}}=1$\,TeV. At the level of the
SM-like state (upper row), the mass shift between the 1L and
$\mathcal{O}(\alpha_{q}\,\alpha_s)$ predictions is sizable (in our
renormalization scheme), and much larger than the dispersion between
the expansion and iteration methods. The dependence of the iteration
method on the field-renormalization constants induces a variation with
the regulator $\mu_{\text{UV}}$, when evaluating the latter between
the EW and SUSY scales, which (in the considered scenario) amounts to
about $5$\,GeV at~1L (long-dashed magenta curve) and about $3$\,GeV at
$\mathcal{O}(\alpha_{q}\,\alpha_s)$ (solid purple curve). This
reduction of the scale dependence originates in the destructive
interplay between the two orders, not in the completion of the partial
order introduced by the pole search---which requires the terms of
$\mathcal{O}\big(\alpha_{q}^2\big)$ (see next subsection). Inclusion
of the $\mathcal{O}(\alpha_{q}\,\alpha_s)$ corrections in the
effective-potential approximation (short-dashed pink curve) leads to
very comparable results, confirming the adequacy of this approach for
the SM-like state: the 2L effects are in fact introduced at their
correct tree-level gaugeless mass value. The evaluation with OS field
counterterms, represented by a cross in the column on the left of each
plot, returns results very near that of the expansion.

In the second and third rows of \fig{fig:M2Lasatiter}, we consider the
shifts in squared mass for the heavy neutral states as compared to the
charged one. The plots on the left show the dispersion of the mass
predictions depending on the chosen order. Once again, the mass shift
associated with the inclusion of $\mathcal{O}(\alpha_{q}\,\alpha_s)$
corrections is larger than that induced by the choice of method
(expansion vs.\ iteration). However, contrarily to the case of the
SM-like state, the effective-potential approximation ostensibly falls
far away from the actual momentum-dependent result, implying that this
approach is not predictive for the heavy-doublet states. Here, we
should comment on the definition of this approach, since it is not
completely straightforward. Indeed, beyond the zero-momentum
assumption in the neutral Higgs self-energy of
$\mathcal{O}(\alpha_{q}\,\alpha_s)$, we also set the momentum equal
to~$0$ in the corresponding charged-Higgs counterterm.\footnote{One
could also evaluate the charged Higgs counterterm at
$p^2=M^2_{H^{\pm}}$, in which case one obtains a mass prediction very
close to the dashed pink line. However, if one employs the
effective-potential approximation in order to avoid the lengthy
evaluation of 2L momentum-dependent integrals, it appears more natural
to use the condition $p^2=0$ in all 2L self-energies.} However, the
charged Higgs self-energy at vanishing momentum requires the
introduction of charged-Higgs field counterterms as well in order to
obtain UV-finite results. Given that observables are supposed to be
independent from field counterterms, the latter could be chosen \AP
without connection to the neutral field counterterms: this is the
choice governing the short-dashed pink curves, where \DR conditions at
the scale $m_t$ are employed. Alternatively, one can correlate the
charged and neutral counterterms in the \DR~scheme, \IE~employ a
common scale $\mu_{\text{UV}}$: this is shown in dotted pink, with
much reduced variations along $\mu_{\text{UV}}$ due to
$SU(2)$-cancellations between neutral and charged
counterterms---though in no way suggestive of an improved reliability
of the effective-potential method (the results for full
momentum-dependent~$\mathcal{O}(\alpha_{q}\,\alpha_s)$ are
significantly away).\footnote{The charged-Higgs mass counterterm
appears neither in the diagonal nor off-diagonal self-energies
contributing to the mass of the lightest Higgs (in the gaugeless
limit); therefore, the distinction in the renormalization of the
charged-Higgs propagator is invisible in the first row
of \fig{fig:M2Lasatiter}.} We do not consider the OS field
counterterms for the charged Higgs in the effective-potential
approach, since these are infrared-divergent, which explains that only
one pink cross appears in each plot. The relative failure of the
effective-potential approximation to
capture the~$\mathcal{O}(\alpha_{q}\,\alpha_s)$ corrections is not
really surprising, as the corresponding Higgs masses are far
from~$p^2=0$, even in the gaugeless limit. Similarly, the large scale
dependence in the dashed pink line proceeds from the necessary
inclusion of a large correction proportional to the field counterterm
to absorb the UV-divergences of the self-energies at~$p^2=0$.

\needspace{5ex}
The plots on the right compare the momentum-dependent mass predictions
of $\mathcal{O}(\alpha_{q}\,\alpha_s)$ in the expansion (green) and
iteration (purple) methods. We observe that the scale variations are
not necessarily sufficient to allow both predictions to
overlap---though the size of this dispersion remains compatible with
the amplitude of the variations with $\mu_{\text{UV}}$. The origin of
this separation between the expansion and iteration method is
associated with the relevance of EW corrections at 1L for the
heavy-doublet states. The last term of
\refeq{eq:itervsexp} indeed becomes sizable but, as we explained
above, the scale variation does not probe the incompleteness of this
order. As it is, these partial $\mathcal{O}\big(\alpha^2\big)$ effects
are most certainly misleading and should not be interpreted as
meaningful contributions of the iterative pole search. In fact, the
evaluation with OS field counterterms typically pulls the iterative
pole search in the direction of the expansion.

\begin{figure}[t!]
  \centering
  \includegraphics[width=\linewidth]{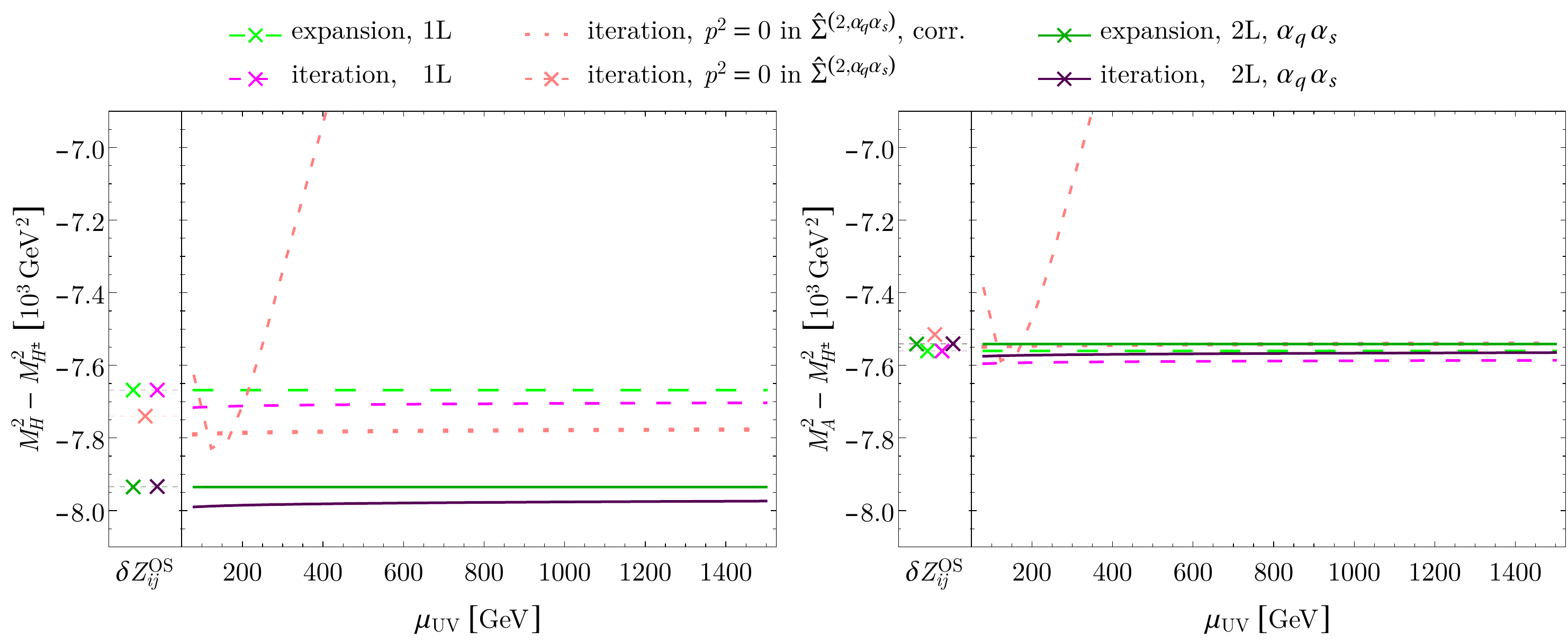}
  \caption{Dependence on the field counterterms at 1L order and
    $\mathcal{O}(\alpha_q\,\alpha_s)$ for
    $M_{H^{\pm}}=2.5$\,TeV. The conventions are similar to those of
    \fig{fig:M2Lasatiter}.
    \label{fig:M2Lasatiterbis}}
\end{figure}

We complete this discussion about the mass prediction at
$\mathcal{O}(\alpha_{q}\,\alpha_s)$ for the heavy-doublet states with
another example in \fig{fig:M2Lasatiterbis}: $M_{H^{\pm}}=2.5$\,TeV,
so that effects beyond the gaugeless approximation, both threshold and
EW, are more relevant. The main difference with respect to the
previous case is that the choice of procedure (expansion vs.\
iteration) leads to clearly separated predictions, not connected by
the scale variation. As explained above, this shift originates in the
last term of \refeq{eq:itervsexp}, which is not probed by the scale
variation, though it is of partial higher order, hence not
predictive. Considering the case of OS field counterterms---a
setup that is sensitive to this
term---extends the range of variation of the
iterative pole search to engulf the prediction of the expansion. Once
again, the effective-potential approximation provides no actual gain
in precision with respect to the strict 1L calculation.

\tocsubsection{Corrections of \texorpdfstring{$\mathcal{O}\big(\alpha_q^2\big)$}{\unicodescriptO(\unicodealpha\unicodesubt\unicodesuptwo)}}

The inclusion of order $\alpha_{q}^2$ brings about the interplay
between 2L and 1L$^2$ effects that we discussed in
\sect{sec:2Lmasses}. Off-diagonal self-energies indeed become
meaningful in the mass calculation even in the non-degenerate
case---while they led to non-decoupling $SU(2)$-violating effects of
higher order in a 1L calculation\,\cite{Domingo:2020wiy}. As explained
in this reference, it is crucial, in the case of the heavy-doublet
states, to properly include 1L$^2$ off-diagonal contributions to the
charged-Higgs mass in the 2L~counterterms.

\begin{figure}[t!]
  \centering
  \includegraphics[width=\linewidth]{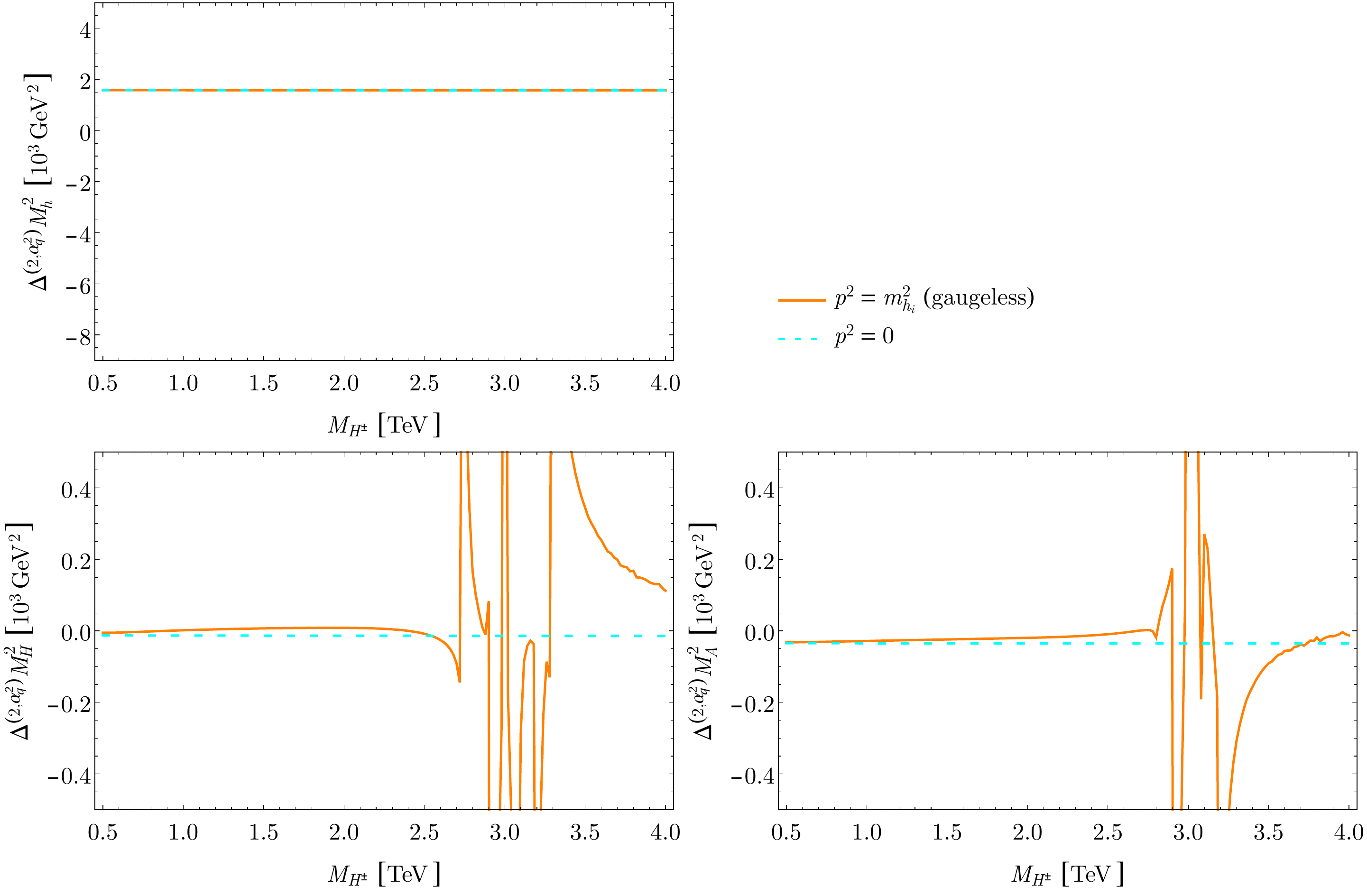}
  \caption{Mass shifts at $\mathcal{O}\big(\alpha_{q}^2\big)$ for the
    neutral Higgs states. The color-code follows the conventions
    of \fig{fig:DM2H2LQCD}. The wiggly lines at
    $M_{H^\pm}^2\gtrsim3.5$\,TeV reveal numerical instabilities in
    the evaluation of 2L integrals containing the tiny ratios of
    $m_b^2\big/M_{H^\pm}^2$.\label{fig:DM2H2LYuk}}
\end{figure}

\needspace{5ex}
Once again, we first investigate the mass shifts generated at this
order, in \fig{fig:DM2H2LYuk}. Here, with all objects defined in the
gaugeless limit, one has
\begin{align}\label{eq:2Lat2massexp}
  \Delta^{(2,\alpha_{q}^2)}M^2_{h_i} &\equiv
  M^{2(2,\alpha_{q}^2)}_{h_i} - M^{2(2,\alpha_{q}\alpha_s)}_{h_i}\nonumber\\
  &= -\Real{\hat{\Sigma}^{(2,\alpha_{q}^2)}_{h_ih_i}\big(m^2_{h_i}\big)
    - \hat{\Sigma}^{(1)}_{h_ih_i}\big(m^2_{h_i}\big)\,
    \frac{d\hat{\Sigma}^{(1)}_{h_ih_i}}{dp^2}\big(m^2_{h_i}\big)
    - \sum_{j\neq i}\frac{\hat{\Sigma}^{(1)}_{h_ih_j}\big(m^2_{h_i}\big)\,
      \hat{\Sigma}^{(1)}_{h_jh_i}\big(m^2_{h_i}\big)}{m^2_{h_i}-m^2_{h_j}}}\,.
\end{align}
As before, the effective-potential approximation is only meaningful
for the SM-like state, though it seems to work somewhat better for
heavy states at low values of $M_{H^{\pm}}$ than at
$\mathcal{O}(\alpha_{q}\,\alpha_s)$. Threshold effects again appear in
association with squark loops for $M_{H^{\pm}}\approx3$\,TeV, when the
momentum dependence is accounted for. The background far from
threshold contributions is essentially flat, as radiative corrections
of $\mathcal{O}\big(\alpha_{q}^2\big)$ typically scale like
$M^2_{\text{EW}}\ln^k\!M^2_{H^{\pm}}\big/M^2_{\text{EW}}$,
$k\in\{0,1,2\}$, at large~$M_{H^{\pm}}$ (provided the charged-Higgs
mass is renormalized on-shell).

The Higgs masses $M^{2(2,\alpha_{q}^2)}_{h_i}$ defined via the expansion in \refeq{eq:2Lat2massexp}
are by construction independent from field
renormalization. Alternatively, one can numerically solve
\refeq{eq:massbasic} with the momentum-dependent self-energy matrix
including contributions up to $\mathcal{O}\big(\alpha_{q}^2\big)$,
defining the poles~$\mathfrak{M}^{2(2,\alpha_{q}^2)}_{h_i}$. As
self-energies are evaluated away from their mass-shell, it is once
again necessary to call upon field renormalization to neutralize
UV-divergences and give a meaning to corresponding objects. In
addition to the 1L field counterterms of \refeq{eq:1LfieldCT}, one
should consider the 2L field counterterms of order $\alpha_{q}^2$,
reading (in the \DR scheme)
\begin{subequations}
\begin{align}
  \delta Z^{\text{(2,$\alpha_{q}^2$)}}_{H_d} &=
  -\frac{3\,\alpha_b\left(3\,\alpha_b + \alpha_t\right)}{32\,\pi^2}\left\{
  \left[\overline{\Delta}^{-1}_{\text{UV}}
  + \frac{\mu_{\text{ren}}^2}{\mu_{\text{UV}}^2}\right]^2
  - \left[\overline{\Delta}^{-1}_{\text{UV}}
  + \frac{\mu_{\text{ren}}^2}{\mu_{\text{UV}}^2}\right]\right\},\\
  \delta Z_{H_u}^{\text{(2,$\alpha_{q}^2$)}} &=
  -\frac{3\,\alpha_t\left(\alpha_b + 3\,\alpha_t\right)}{32\,\pi^2}\left\{
  \left[\overline{\Delta}^{-1}_{\text{UV}}
  + \frac{\mu_{\text{ren}}^2}{\mu_{\text{UV}}^2}\right]^2
  - \left[\overline{\Delta}^{-1}_{\text{UV}}
  + \frac{\mu_{\text{ren}}^2}{\mu_{\text{UV}}^2}\right]
  \right\}.
\end{align}
\end{subequations}
In addition to these 2L field counterterms, the self-energies of order
$\alpha_{q}^2$ depend on the 1L field counterterms of
\refeq{eq:1LfieldCT} according to \refeq{eq:2Loffdiag}---even when the
momentum is set to the tree-level Higgs mass. As explained in
\sect{sec:2Lmasses}, this dependence on field-renormalization
constants is mirrored by that of the 1L$^2$ terms resulting in a
cancellation for the full order. It is also worth noticing that the
dependence on the charged-Higgs field already vanishes separately
within the 2L\,+\,1L$^2$ terms forming the charged-Higgs mass
counterterm of $\mathcal{O}\big(\alpha_{q}^2\big)$.

\begin{figure}[t!]
  \centering
  \includegraphics[width=\linewidth]{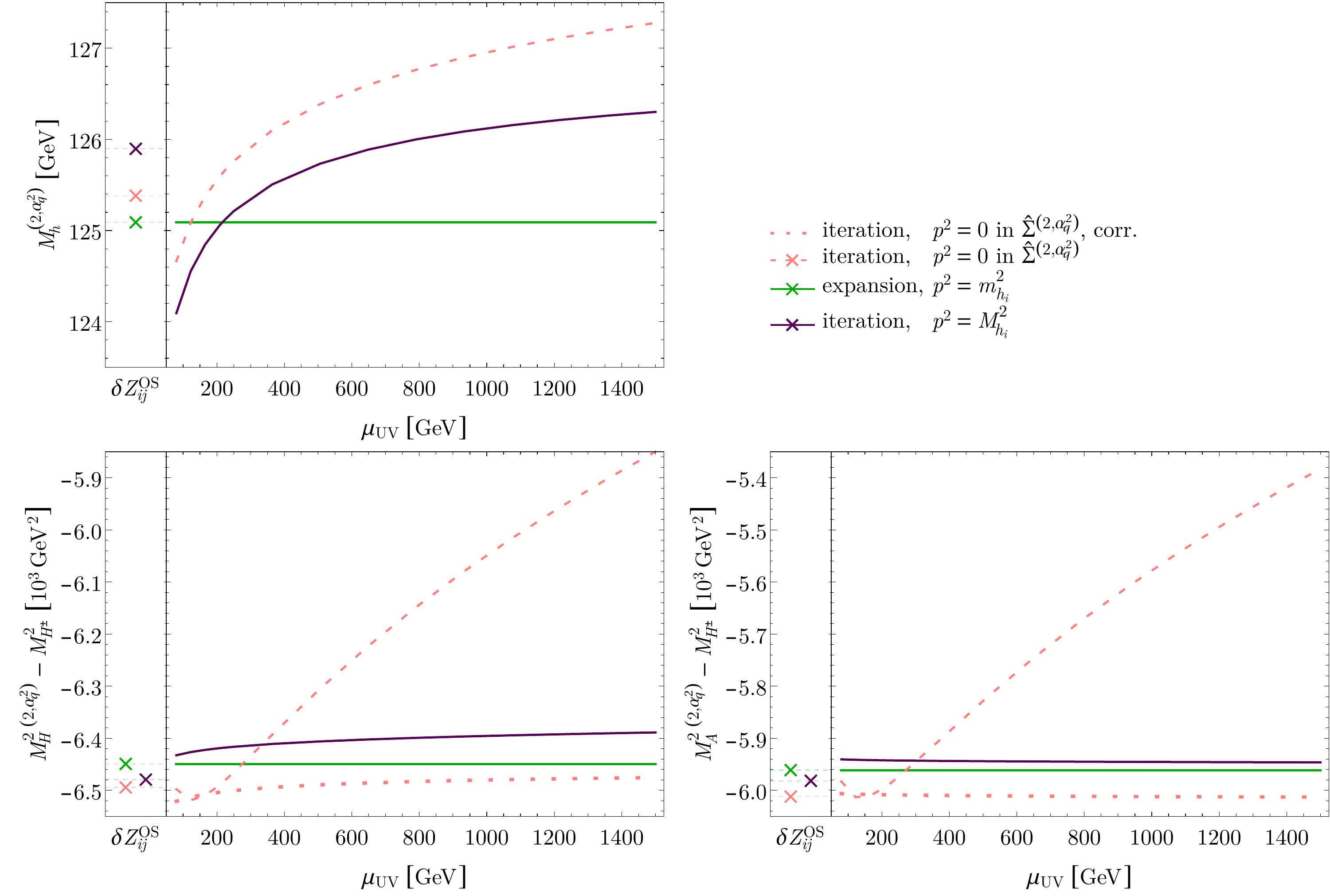}
  \caption{Dependence on the field-counterterms in mass predictions of
    order $\alpha_{q}^2$. The solid green line is derived with the
    expansion at strict order, the purple one, with an iterative pole
    search. The 2L self-energies are included in the effective
    potential approximation for the dashed pink
    curve.\label{fig:M2Lat2iter}}
\end{figure}

In \fig{fig:M2Lat2iter}, we show the field-counterterm dependence of
the mass prediction at~$\mathcal{O}\big(\alpha_{q}^2\big)$
for \mbox{$M_{H^{\pm}}=1$\,TeV}. Once again, we compare the strict
expansion---\refeq{eq:2Lat2massexp}---in solid green, with the
iterative pole search, in solid purple. In this last case, the full
\CP-even mass matrix of~$\mathcal{O}\big(\alpha_{q}^2\big)$ is included
in the pole search---with gaugeless approximation for the 2L
pieces. For the \CP-odd state, instead of considering the full matrix,
we directly add the off-diagonal mixing contribution with the
Goldstone boson
$\big[\hat{\Sigma}^{\text{(1,gl)}}_{A^0G^0}\big(p^2\big)\big]^2\big/m_A^2$
to the diagonal element---omission of this term would otherwise induce
an artificial $SU(2)$-breaking among the heavy-doublet
states, see \citere{Domingo:2020wiy}. Finally, the dashed
and the dotted pink curves include the 2L corrections of
$\mathcal{O}\big(\alpha_{q}\,\alpha_s,\,\alpha_{q}^2\big)$ in the
effective-potential approximation before performing the pole
search. There, we use two slightly different procedures for the
\CP-even and \CP-odd sectors, as we add a
$\big[\hat{\Sigma}^{\text{(1,gl)}}_{A^0G^0}(0)\big]^2\big/m_A^2$ term
in the latter case, instead of keeping a full momentum-dependent 1L
self-energy matrix. As before, the charged-Higgs field counterterms
are either set to a fixed value (dashed curves) or correlated with the
neutral sector (dotted).

At the level of the SM-like state, the corrections of
$\mathcal{O}\big(\alpha_{q}^2\big)$ are sufficiently significant to
make them necessary in any attempt at precision predictions for the
mass. Surprisingly, the variations with the field-renormalization
scale are barely reduced as compared to the order
$\alpha_{q}\,\alpha_s$---see~\fig{fig:M2Lasatiter}---for the iterative
approach. This situation originates in the inexact cancellation of
the~1L$^2$ field counterterms that are generated in the variation of
the 1L self-energy via the pole search, with those produced in the 2L
corrections and assuming the gaugeless limit. \mbox{A simple} estimate
actually suffices to recover the order of magnitude of these scale
variations. The leading contribution from the pole search is a term
$-\big(\mathfrak{M}^{2\,(2,\alpha^2_{q})}_h -
m^2_h\big)\,\hat{\Sigma}'_{hh}\big(m^2_h\big)$, generating a
field-renormalization dependence $\simord\delta
Z_{hh}\,\hat{\Sigma}^{(1,\alpha_{q}\alpha_s,\alpha_{q}^2)}_{hh}\big(m^2_h\big)$
because all orders are involved in the mass shift. On the side of 2L
self-energies, the dependence on field counterterms is dominated by
the corresponding term $-\delta
Z_{hh}\,\hat{\Sigma}^{\text{(1,gl)}}_{hh}\big(m^2_h\big)$---see
\refeq{eq:2Lfield}. The mismatch
$\hat{\Sigma}^{(1,\alpha_{q}\alpha_s,\alpha_{q}^2)}_{hh}\big(m^2_h\big)
- \hat{\Sigma}^{\text{(1,gl)}}_{hh}\big(m^2_h\big)$ is large---of the
order of~$100\%$ of the magnitude of radiative corrections to the mass
of the SM-like Higgs---while, dominated by~$\alpha_t$, $\delta
Z_{hh}\approx0.1$ for a variation between the EW and SUSY scales. In
the aftermath, an effect of $\simord2\%$ is assessed on the field
dependence at the level of the mass. Curiously enough, we can expect a
comparable uncertainty from missing EW orders due to the gaugeless
approximation. Indeed, as we observed in \fig{fig:DM2H1L}, the
gaugeless approximation works at~$\simord15\%$ at 1L for the SM-like
state while the squared-mass shift of
$\mathcal{O}\big(\alpha_{q}^2\big)$ is of order $2\cdot10^3$\,GeV$^2$;
combining the two numbers, we arrive at a FO~uncertainty of percent
level from uncontrolled 2L orders at the level of a $\simord125$\,GeV
Higgs mass---corresponding to the squared logarithms
of~$\mathcal{O}(\alpha_t\,\alpha)$. However, this coincidence does not
establish the variation of field counterterms as a realistic
uncertainty estimate for higher orders, as it strictly measures
partial higher-order effects associated with the regularization of the
pole-search procedure. We stress that the large dependence on field
counterterms is intimately related to the size of the radiative
corrections in the considered approach, \IE~to the FO~procedure: in an
EFT, contributions from the hierarchical spectrum would be absorbed
within the tree-level couplings. Similar behaviours had been observed
in \citere{Bahl:2018ykj}: see \EG~Fig.\,3 of this reference. In
addition, the mismatch between the total mass shift
($\hat{\Sigma}^{(1,\alpha_{q}\alpha_s,\alpha_{q}^2)}_{hh}\big(m^2_h\big)$)
and the gaugeless 1L~self-energy could certainly be minimized via an
adequate choice of renormalization conditions for the Yukawa
couplings, absorbing the impact of higher orders.

Concerning the heavy-doublet states, the field dependence induced by
the pole search is in fact increased in the \CP-even case with respect
to $\mathcal{O}(\alpha_{q}\,\alpha_s)$---see \fig{fig:M2Lasatiter}. In
contrast, the dependence on \DR~field-renormalization constants
remains mild in the \CP-odd case, which is a consequence of our adding
the off-diagonal self-energies in the gaugeless limit, hence
satisfying the cancellation of the field counterterms from
off-diagonal terms with the diagonal 2L contributions. Moreover, we
observe that even the enlarged field-scale variations of the \CP-even
mass predictions do not capture the full magnitude of the dispersion
between the expansion and iteration methods. Off-diagonal elements in
the pole search indeed add further EW terms insensitive to the scale
variation beyond that of the last term of \refeq{eq:itervsexp}; they
correspond to partial higher-order $\alpha\,\alpha_q$ and
$\alpha^2$~terms and cannot be interpreted as a genuine physical
effect, because the completion of the order $\alpha^2$ is likely to
sizably affect them. Comparing the \DR~regularization with that of OS
field counterterms further extends the magnitude of the variations
associated with the iterative approach, making it sensitive to EW
effects and filling the gap with the expansion procedure. Finally, for
2L self-energies derived in the effective-potential approximation
(dashed pink curves), the scale dependence caused by the approximation
$p^2=0$ in the 2L self-energies reaches a magnitude much larger than
the mass-squared shift induced by the inclusion of these contributions
with respect to the strict 1L order. The predicted masses in the case
of correlated field counterterms for the neutral- and charged-Higgs
sectors (dotted lines) also fall significantly far away from the
actual momentum-dependent 2L prediction---as compared to the shift
from predictions of 1L order. This approach is thus meaningless for
the heavy states.

To summarize, the inclusion of 2L corrections of orders
$\alpha_{q}\,\alpha_s$ and $\alpha_{q}^2$ is essential in a
precise prediction of the mass of the SM-like Higgs state and the
effective-potential approximation is reasonably predictive at this
level. However we have observed that these orders are of little
consequence for the heavy-doublet states in a non-degenerate scenario,
since one expects them to be superseded by EW corrections. In
addition, the effective-potential approximation works poorly for such
states, as the obtained mass shift does not quantitatively improve the
predictions as compared to $\mathcal{O}(\text{1L})$.
We also noted that the dependence on the field counterterm at the
level of the SM-like state remains large in an iterative pole search
including terms of $\mathcal{O}\big(\alpha_{q}^2\big)$ due to the
mismatch between the 1L$^2$ terms induced by the pole search and the
corresponding 2L~terms derived in the gaugeless limit: it thus appears
that the iterative pole search at FO~$\alpha_{q}^2$ contains an
intrinsic uncertainty of $\simord$GeV-size (depending on the hierarchy
between the EW and SUSY scales). Concerning the heavy-doublet states,
the dependence on \DR~field counterterms appears comparatively reduced
for a full calculation
of~$\mathcal{O}\big(\alpha_{q}^2\big)$. However, this simply indicates
that EW effects, not Yukawa, dominate the corrections to the
corresponding masses: the \DR~renormalization is insensitive to these
contributions, but the comparison with an OS regularization (as
defined in \refeq{eq:OSfieldCTs}) allows to probe them and proves that
the apparent gap between predictions from the expansion and the
iterative pole search is strictly artificial. We thus conclude that
there exists no advantage in employing the more costly iterative pole
search in a non-degenerate scenario, as compared to the more
straightforward expansion-and-truncation method. On the contrary, the
partial higher orders introduced in the pole search may be unphysical
and, in any case, they generate an `uncertainty' intrinsic to the
procedure and hardly representative of `genuine' higher-order
effects. Moreover, as the orders~$\alpha_{q}\,\alpha_s$ and
$\alpha_{q}^2$ are the leading 2L corrections only for the SM-like
Higgs, there is little significance---as long as 2L EW orders are not
under control---in maintaining them for the mass determination of
heavy-doublet states; in particular the inclusion in the
effective-potential approximation induces uncertainties that are
larger than the genuine mass shift.

\tocsection{Field-dependence in the mass predictions in near-degenerate scenarios\label{sec:degeneratecase}}

Higgs mixing in the near-degenerate scenario makes the situation more
subtle for the mass determination, and we study its practical
implementation in this section. As it is, the complications originate
less in the general formalism described in \sect{subsec:neardeg} than
in the features of the gaugeless approximation and its matching to the
full model.

\needspace{15ex}
\tocsubsection{\texorpdfstring{\CP}{CP}-violating mixing between heavy states}

In the MSSM, the phenomenologically most relevant scenario with mass
degeneracy involves \CP-violating mixing between the neutral
components of the heavy doublet. With the gaugeless description of
$\mathcal{O}\big(\alpha_q^2\big)$, it is actually not possible, even
for small mixing, to define loop-corrected masses according to the
strict expansion of \refeq{eq:nondegmassexp} since the \CP-even and
\CP-odd components are exactly degenerate at tree level in the
gaugeless limit.\footnote{For the SM-like Higgs, the mixing with the
Goldstone boson always vanishes, so that this issue does not appear.}
However, as the mass-splitting between the two neutral components is
of EW order, it can be regarded---consistently with the gaugeless
counting---as numerically comparable to 1L effects and the
near-degenerate formalism of \sect{subsec:neardeg} applies.

\begin{figure}[p!]
  \centering
  \includegraphics[width=\linewidth]{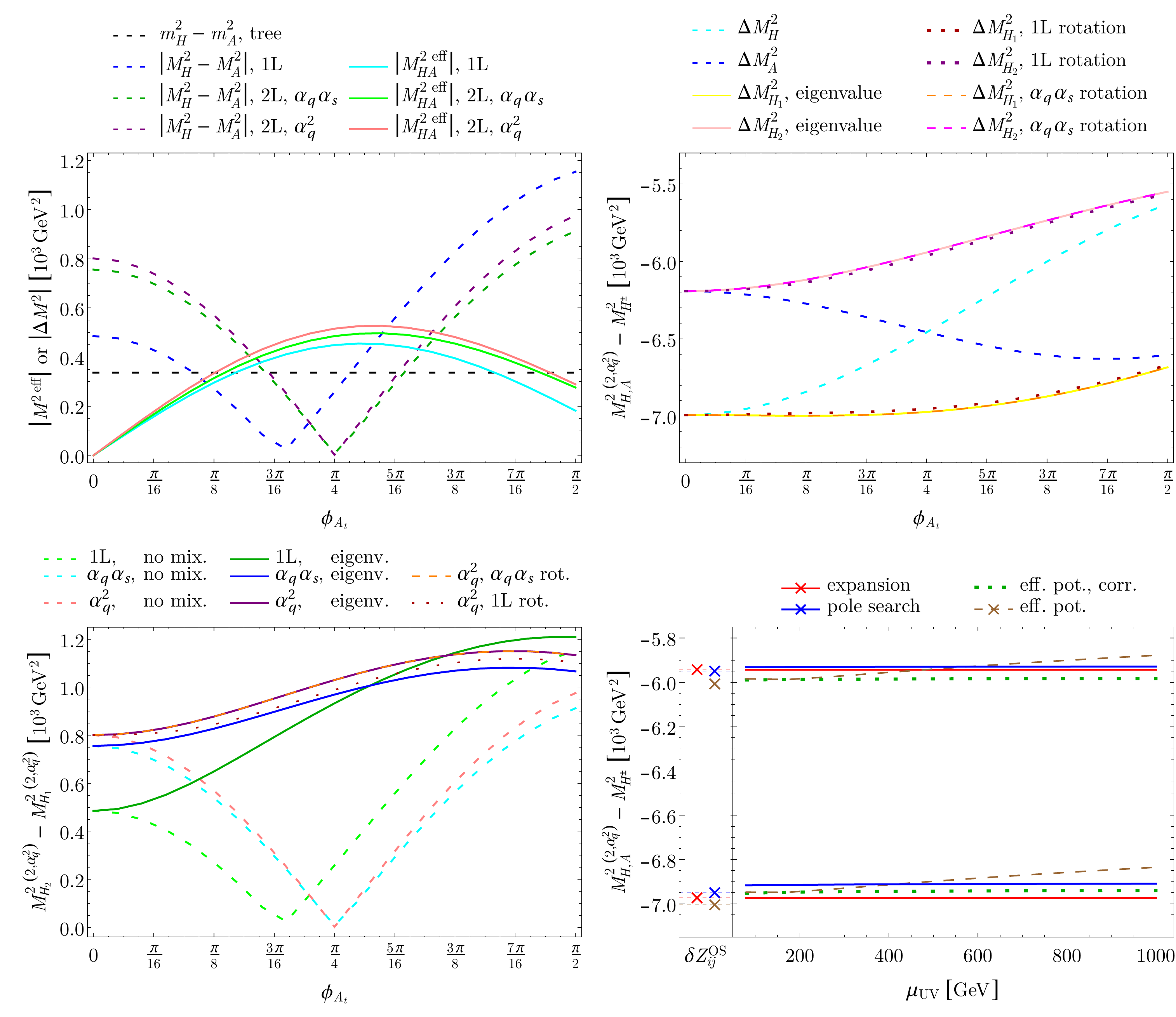}
  \caption{Mass predictions in the \CP-violating scenario at
  $\mathcal{O}\big(\alpha_q^2\big)$.\newline {\em Up Left}: The size
  of the elements of the effective matrix elements is plotted against
  $\phi_{A_t}$; more precisely, the mass-splitting between diagonal
  elements (dashed curves) is compared to the off-diagonal entry
  (solid curves) at various orders.\newline {\em Up Right}: Predicted
  mass-splitting between neutral and charged states at
  $\mathcal{O}\big(\alpha_q^2\big)$, without accounting for the mixing
  (blue/cyan dashed), or in various evaluations of the 2L
  mixing.  \newline {\em Down Left}: Predicted mass-splitting between
  neutral states at various orders, without accounting for the mixing
  or including it.  \newline {\em Down Right}: Field dependence of the
  predicted masses at $\mathcal{O}\big(\alpha_q^2\big)$ for
  $\phi_{A_t}=\frac{\pi}{4}$, in the expansion formalism (red), with
  an iterative pole search retaining full momentum dependence (blue),
  or in the effective-potential description with correlated (green) or
  uncorrelated (brown) charged field counterterms. The crosses on the
  left correspond to the predictions with OS counterterms, while the
  scale variation employs a \DR~regularization of the
  self-energies.  \label{fig:M2LCPVphase}}
\end{figure}

We turn to a scenario where \CP-violation is induced at the
loop-level---the tree-level MSSM Higgs sector is always
\CP-conserving---by the phase of the trilinear soft SUSY-breaking
coupling in the stop sector, $\phi_{A_t}$. In practice, generating a
sizable mixing (always scaling with an $SU(2)$-breaking v.e.v.) in
this fashion requires a rather large trilinear coupling $\lvert
A_t\rvert$ as compared to the diagonal soft SUSY-breaking stop masses
$m_{\tilde{Q}_3,\tilde{T}}$, which could potentially produce charge-
and color-breaking minima---see \EG~\citere{Hollik:2018wrr} for a
recent reference. As our discussion is meant to be strictly
illustrative of the mass calculation, we disregard this problem below,
and take $m_{\tilde{Q}_3,\tilde{T}}\approx1$\,TeV, $\lvert
A_t\rvert=3$\,TeV, $M_{H^{\pm}}=0.5$\,TeV, $t_{\beta}=10$.

In the upper left-hand quadrant of \fig{fig:M2LCPVphase}, we compare
the mixing entry of the effective mass matrix at 1L (solid cyan),
$\mathcal{O}(\alpha_q\,\alpha_s)$ (solid green),
$\mathcal{O}\big(\alpha_q^2\big)$ (solid magenta) with the diagonal
mass-splitting at tree level (dashed black), 1L (dashed blue),
$\mathcal{O}(\alpha_q\,\alpha_s)$ (dashed green),
$\mathcal{O}\big(\alpha_q^2\big)$ (dashed purple), for varying
$\phi_{A_t}$. We see that the diagonal mass-splitting at the radiative
level vanishes for $\phi_{A_t}\approx\frac{\pi}{4}$ while the
off-diagonal mass entry is comparatively large: we thus expect a
sizable mass-mixing in this region. The plot in the upper right-hand
corner shows the squared mass-splitting between neutral and charged
states at order $\alpha_q^2$ in the expansion formalism for degenerate
states. The blue and cyan dashed lines correspond to the diagonal
entries of the effective mass matrix of \refeq{eq:effmass2L}, crossing
at $\phi_{A_t}\approx\frac{\pi}{4}$. The other curves correspond to
various evaluations of the masses corrected with the \CP-violating
mixing: eigenvalues of the effective mass matrix in solid lines,
diagonal elements after rotation by the mixing matrix defined at order
$\alpha_q\,\alpha_s$ in dashed lines and diagonal elements after
rotation by the mixing matrix defined at 1L~order in dotted lines. We
see that these various definitions give very close results, even
though the maximal mixing is clearly shifted in phase at 1L. In the
lower left-hand quadrant, we plot the squared mass-splitting between
neutral states at 1L (green), $\mathcal{O}(\alpha_q\,\alpha_s)$ (blue)
and~$\mathcal{O}\big(\alpha_q^2\big)$ (purple, orange and red,
depending on the definition through eigenvalues or rotation of the
effective mass matrix) in the expansion formalism. The diagonal
splitting is shown in dashed lines. Once again, we observe the good
agreement among definitions at order $\alpha_q^2$.

Finally, in the lower right-hand panel of \fig{fig:M2LCPVphase}, we
consider the dependence on field counter\-terms for
$\phi_{A_t}=\frac{\pi}{4}$ in masses defined with the expansion
formalism (red), in an iterative pole search with full
momentum-dependence (blue) and in the effective-potential
approximation for 2L self-energies, with correlated
charged-Higgs \DR~field counterterms (dotted green) or independent
charged-Higgs field counterterms (dashed brown). In fact, the masses
obtained with the expansion method are fully independent from the
field counterterms, because the remainder of \refeq{eq:FDremainder}
exactly vanishes due to the exact degeneracy of the two tree-level
masses in the gaugeless approximation. For the iterative pole
searches, the situation is largely comparable to what we discussed in
the non-degenerate case. The predicted masses for the mixed (heavy)
states show only a mild dependence on field-scale variations: we
should stress here that the \CP-violating self-energy is UV-finite
without need of regularization by a field counterterm, so that the
mixing entry in the mass matrix is itself completely blind to
the \DR~regularization of the fields---adding to the already noted
insensitivity to EW effects. On the other hand, the OS~field
counterterms from \refeq{eq:OSfieldCTs} probe both effects: the
corresponding mass predictions, indicated by crosses in the column at
the left end of the plot, show a more significant deviation respective
to those using a scale variation in the \DR~renormalization, which
closes the apparent gap with the masses of the expansion
formalism. Once again, this difference appears as a strictly
artificial effect originating in the iterative method. Lastly, the
effective-potential approximation performs somewhat better than in the
non-degenerate scenario, which should be put in perspective with the
small value of~$M_{H^{\pm}}$. The dependence on field counterterms of
this description is made evident by the steeper variation with the
scale (when keeping the charged-field counterterms uncorrelated).

\tocsubsection{\texorpdfstring{\CP}{CP}-conserving mixing with the SM-like Higgs\label{sec:h0HHmix}}

As explained in the previous subsection, the \CP-violating mixing
between heavy-doublet states always falls in the degenerate limit,
even for small mixing, as long as 2L effects are considered in the
gaugeless approximation. Therefore, in order to study the transition
between degenerate and non-degenerate regimes, we consider the
scenario of degenerate \CP-even
states, \IE~\mbox{$M_{H^{\pm}}=\mathcal{O}(M_Z)$}, even though it is
now only marginally relevant from a phenomenological perspective due
to tight experimental limits---refer \EG~to the related work
in \citere{Bahl:2018zmf}. As the latter rather motivate weak- than
strong-mixing scenarios, we do not bother and try to accommodate a
Higgs boson with a mass of $125$\,GeV. This type of mixing also
differs from the \CP-violating mixing between heavy-doublet states in
that the remainder of \refeq{eq:FDremainder} no longer uniformally
vanishes, meaning that the expansion formalism retains some amount of
dependence on field counterterms, which we aim to quantify below.

In fact, the gaugeless approximation for 2L effects forbids an actual
`mass-crossing' in the off-diagonal 1L$^2$ term processed in the
non-degenerate expansion formalism of \refeq{eq:nondegmassexp},
because the SM-like state takes a zero-mass in this limit. Therefore,
this equation then returns a well-behaved mass-prediction, independent
from field counterterms. On the other hand, this approach meets a
first issue with the identification of gaugeless and `actual' Higgs
states, since the gaugeless state with mass equal to~$0$ is always
SM-like, while the $SU(2)$-partner of the Goldstone bosons in the
model with non-vanishing gauge couplings is distributed between both
tree-level states, according to an angle $\alpha-\beta$ ($\alpha$
denoting the
\CP-even tree-level mixing angle with respect to gauge eigenstates): as
$M_{H^{\pm}}$ narrows $M_Z$, $\alpha$ departs from~$\beta-\pi/2$ (for
$\beta>\tfrac{\pi}{4}$; while $\alpha$ is fixed to this value
in the gaugeless limit), underlining the irrelevance of a naive
identification. However, performing a rotation of $\alpha-\beta$ of
the total 2L effects obtained in the gaugeless limit (including
1L$^2$~contributions for controlled dependence on field counterterms)
is ill-defined outside of the degenerate regime. In addition, the
recourse to the near-degenerate formalism is originally motivated by
the need to consistently process off-diagonal contributions to the
Higgs masses intervening at 1L order, due to the near-degeneracy of
diagonal terms. On the other hand, it is misleading to maintain this
mixing-matrix description when the non-degenerate regime applies,
since it generates $SU(2)_{\mathrm{L}}$-violating pieces that are not
controlled by the EW-symmetry breaking,
see \citere{Domingo:2020wiy}. It is thus legitimate to worry about
defining the transition between both regimes.

\begin{figure}[p!]
  \centering
  \includegraphics[width=\linewidth]{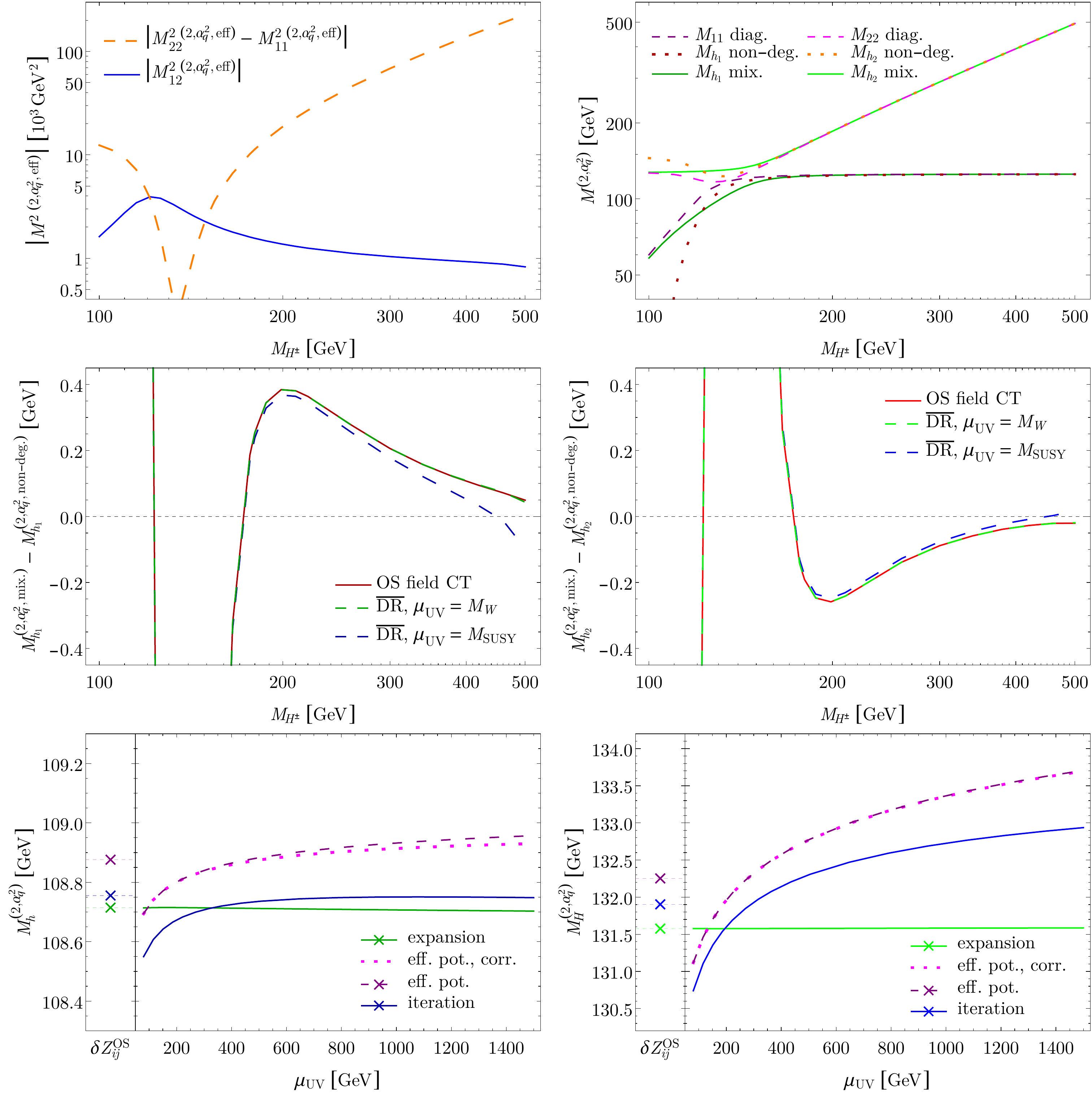}
  
  \caption{Mass predictions at $\mathcal{O}\big(\alpha_q^2\big)$ in
  the scenario with \CP-conserving mixing.  \newline{\em Up Left}:
  Magnitude of the elements of the effective mass matrix
  (with \DR~renormalization of the Higgs fields and
  $\mu_{\text{UV}}=m_t$) of $\mathcal{O}\big(\alpha_q^2\big)$,
  diagonal mass-splitting (dashed orange) vs.~off-diagonal entry
  (solid blue).  \newline{\em Up Right}: Mass predictions (obtained
  with the expansion formalism) for the \CP-even states at
  $\mathcal{O}\big(\alpha_q^2\big)$, in the non-degenerate description
  (dotted lines), from the diagonal elements of the effective mass
  matrix (dashed) and from its eigenvalues (solid)
  (with \DR~renormalization of the Higgs fields and
  $\mu_{\text{UV}}=m_t$).  \newline{\em Middle}: Difference between
  the mass predictions (obtained with the expansion formalism) in the
  near-degenerate and non-degenerate descriptions for the lightest
  (left) and heaviest (right) \CP-even states. Several regularizations
  of the self-energies with the Higgs field counterterms are
  considered: OS (solid red), \DR~(dashed) with $\mu_{\text{UV}}=M_W$
  (green) and $\mu_{\text{UV}}=M_{\text{SUSY}}$ (blue).  \newline{\em
  Down}: Dependence of the mass predictions at $M_{H^{\pm}}=140$\,GeV
  on the field regularization for the expansion formalism with
  degeneracy (green), in an effective-potential description (magenta
  and purple, depending on whether charged-Higgs field counterterms
  are correlated with the neutral ones) and in an iterative pole
  search retaining full momentum dependence (blue). The plot on the
  left (resp.\ right) corresponds to the lightest (resp.\
  heaviest) \CP-even state. The crosses on the left-hand side of the
  plots correspond to the OS~renormalization of the fields, and the
  curves to the \DR~renormalization with varying
  scale.  \label{fig:h0HHmix}}
\end{figure}

In \fig{fig:h0HHmix}, we consider the same region in parameter space
as in \sect{sec:nondegeneratecase}, but in the
range \mbox{$M_{H^{\pm}}\in[100,500]$}\,GeV. The first row of plots
shows the general perspective of mixing in the \CP-even sector. In the
plot on the left, we compare the off-diagonal mass-squared entry---we
use \DR~counterterms with $\mu_{\text{UV}}=m_t$---in the effective
mass matrix of \refeq{eq:effmass2L} (solid blue curve) to the diagonal
splitting (dashed orange curve): in the range
$M_{H^{\pm}}\approx120$--$150$\,GeV, the mass-squared splitting is
smaller than the off-diagonal self-energy, highlighting the need for a
near-degenerate formalism. We stress that we include the block of 2L
corrections obtained in the gaugeless limit after identification of
the gauge eigenbasis, \IE~after rotation by an angle
$\alpha-\beta$. On the right, we plot the masses of the \CP-even
states obtained in various versions of the expansion formalism. The
dashed purple and magenta lines correspond to the diagonal entries of
the effective mass matrix of \refeq{eq:effmass2L}, crossing at
$M_{H^{\pm}}\approx 130$\,GeV. The dotted orange and dark-red curves
represent the masses obtained in the non-degenerate
formalism, \IE~according to \refeq{eq:nondegmassexp}. These noticeably
depart from the diagonal entries of the mixing formalism at low
$M_{H^{\pm}}$, mainly due to the absence of corrections accounting for
the tree-level mixing (\IE~no $\alpha-\beta$ rotation), which are not
straightforward to include in a meaningful way in the non-degenerate
scenario.  Finally, the eigenvalues of the effective mass matrix are
plotted with solid green lines. Obviously, the predictions from the
mixing formalism barely differ from those of the non-degenerate one
above $M_{H^{\pm}}\sim160$\,GeV.

In the second row of plots in \fig{fig:h0HHmix}, we study the
difference between the degenerate and non-degenerate formalisms in the
higher range of $M_{H^{\pm}}$. The latter is obviously small---a
few~$100$\,MeV for $M_{H^{\pm}}\gtrsim160$\,GeV. In addition, the
dependence of the effective mass matrix on the field counter\-terms is
made obvious by the dispersion among the various choices (\DR with
$\mu_{\text{UV}}=M_W$, \DR~with~$\mu_{\text{UV}}=M_{\text{SUSY}}$,
OS). However, this dispersion remains within
$\mathcal{O}(10\,\text{MeV})$, although it tends to increase as
$M_{H^{\pm}}$ reaches $500$\,GeV, then outstretching somewhat the
regime of validity of the near-degenerate method. Given the good
agreement between the predictions obtained
with \refeq{eq:nondegmassexp} and \refeq{eq:effmass2L} in the range
$M_{H^{\pm}}\approx200$--$400$\,GeV (corresponding to a $10\%$--$1\%$
mixing), there is no difficulty to extrapolate between the
two. However, the combined results will still receive an uncertainty
of order $\mathcal{O}(100\,\text{MeV})$ from this extrapolation (in
addition to other sources of theoretical uncertainties). It is
actually unclear whether the non-degenerate regime should not be
altogether preferred in this intermediate regime. The difficulty here
consists in estimating the uncertainty from neglected off-diagonal EW
effects: the corresponding mixing entry in the effective mass matrix
indeed generates partial effects of EW 2L order, which are not
quantitatively reliable---in particular because they do not
necessarily satisfy the symmetries of the system,
see~\citere{Domingo:2020wiy}.

In the last row of \fig{fig:h0HHmix}, we compare the dependence on the
field-renormalization scale in the expansion, effective-potential and
iterative strategies for a point with near-maximal mixing
(\mbox{$M_{H^{\pm}}=140$}\,GeV). Expectedly, the field dependence in
the expansion approach, of the order of~$\simord10$\,MeV, is much
smaller than that obtained with the other approaches, of GeV~order. As
announced in \sect{sec:2Lmasses}, this is related to the careful
pairing of 2L and 1L$^2$ terms in the expansion, which limits the
contamination of the mass prediction by partial higher-order effects.

\begin{figure}[t!]
  \centering \includegraphics[width=.5\linewidth]{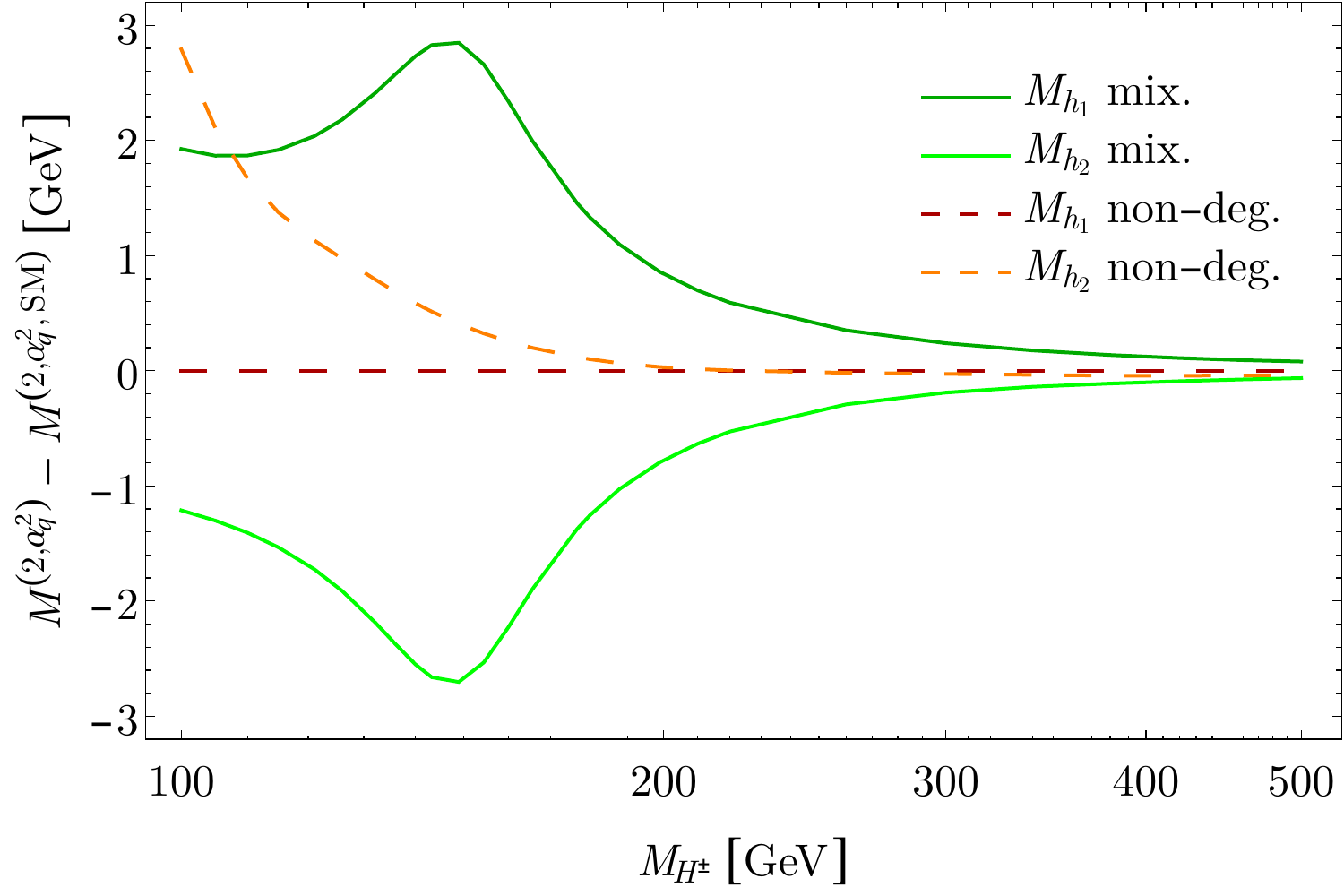}
  \caption{Difference between the predicted masses (in the expansion
  formalism with \DR~field counterterms at $\mu_{\text{UV}}=m_t$)
  including the $\mathcal{O}\big(\alpha_q\,\alpha_s,\,\alpha_q^2\big)$
  gaugeless corrections for all states ($M^{(\alpha_q^2)}$), and those
  obtained with 2L corrections only applied to the SM-like direction
  ($M^{(\alpha_q^2,\text{SM})}$).  \label{fig:h0HHmixwoHHgl}}
\end{figure}

Finally, we explained in \sect{sec:nondegeneratecase} that the 2L
corrections of $\mathcal{O}\big(\alpha_q\,\alpha_s,\,\alpha_q^2\big)$
are not really quantitatively meaningful for heavy-doublet states,
since one then expects larger EW 2L effects. On the other hand, these
orders are known to be dominant for the SM-like state. Therefore,
with~$M_{H^{\pm}}\sim M_Z$, it appears necessary to keep these
$\mathcal{O}\big(\alpha_q\,\alpha_s,\,\alpha_q^2\big)$ corrections for
the full effective mixing matrix. In \fig{fig:h0HHmixwoHHgl}, we
compare the masses $M^{(\alpha_q^2)}$ obtained with full
$\mathcal{O}\big(\alpha_q\,\alpha_s,\,\alpha_q^2\big)$ corrections (as
we always considered them till now) and those obtained in the
approximation where these corrections of order~$\alpha_q\,\alpha_s$
and~$\alpha_q^2$ are only applied in the SM-like
direction, denoted as~$M^{(\alpha_q^2,\text{SM})}$. The solid
green lines correspond to the masses derived in the degenerate
formalism, the dashed red and orange ones to those derived in the
non-degenerate formalism. Obviously, the introduction
of~$\mathcal{O}\big(\alpha_q\,\alpha_s,\,\alpha_q^2\big)$ for the
non-SM states and mixings has little impact for all states in the
parameter space above $M_{H^{\pm}}\sim300$~GeV, and may as well be
neglected since these orders are not quantitatively predictive. On the
other hand, they contribute significantly in the mixing regime,
justifying a complete inclusion at low~$M_{H^{\pm}}$.

\tocsubsection{Three-state mixing}

For completeness, we study a scenario involving three-state mixing,
although of little phenomenological relevance in the MSSM due to
strong experimental constraints on the properties of the observed
Higgs state. Such a setup has been considered in particular
in \citeres{Fuchs:2016swt,Fuchs:2017wkq}, in view of studying
interference effects close to degenerate scalar resonances, \EG~in the
`toy'-scattering~$b\bar{b}\to h_i\to\tau^+\tau^-$. The corresponding
cross-section is then dominated by the $S$-channel exchange of Higgs
bosons indeed. However, the formalism employed in these references
fully disregards the dependence on field counterterms, both in the
definition of the full propagator matrix and subsequent
approximations, implying a sizable `uncertainty' associated to
regulators, as we show below. Instead, we prefer to describe such
phenomena through the formalism derived
in \appx{ap:scattering}, \IE~in a fashion minimizing the dependence on
field counterterms (and linear gauge regulators).

For simplicity, we focus on the scattering $b\bar{b}\to
h_i\to\tau^+\tau^-$ via a trio of near-degenerate Higgs
states~$h_i$. A `naive' Feynman-diagrammatic calculation would
converge very slowly, due to the difference between the MSSM
tree-level masses and the actual poles. It is thus useful to directly
resum resonant effects. When performing this operation, one needs to
extrapolate the form of the effective propagator away from tree-level
Higgs masses. A direct resummation of self-energies as achieved
in \citere{Fuchs:2016swt} then explicitly contaminates the pole values
with gauge-dependent partial higher-order effects and field
regulators. Instead, the identification of poles, residues and
effective couplings via the expansion method, as proposed
in \appx{ap:scattering}, minimizes the dependence on field
counterterms and accordingly distributes the residues between pole and
effective couplings, leading to an \AP more predictive result.

Below, we consider a scenario similar to that of Sect.\,5.2
of \citere{Fuchs:2016swt}
with~$M_{H^{\pm}}=175$\,GeV, \mbox{$\tan\beta=50$},
$\phi_{A_{t}}=\pi/4$. A first formal difficulty in such a setup with
$M_{H^{\pm}}\sim M_{\text{EW}}$ originates in the mixing of (charged
and neutral) Higgs states with Goldstone and gauge bosons: in order to
avoid problems of consistency with the gaugeless limit, we assume that
this mixing can be processed in the perturbative (non-degenerate)
description. The latter is justified in the gaugeless evaluation
where \EG~$\big\lvert\hat{\Sigma}^{\text{(1,gl)}}_{H^+G^-}\big(M_{H^{\pm}}^2/2\big)\big\rvert\big/M_{H^{\pm}}^2\approx5\cdot10^{-4}$,
but also in the full model, with
$\big\lvert\hat{\Sigma}^{\text{(1)}}_{H^+G^-}\big((M_{H^{\pm}}^2+m^2_{G^{\pm}})/2\big)\big\rvert\big/M_{H^{\pm}}^2\approx1\cdot10^{-3}$---gauge
considerations justify the denominator~$M_{H^{\pm}}^2$
over~$M_{H^{\pm}}^2 - m^2_{G^{\pm}}$. We may then focus on the
three-state mixing in the neutral sector and consider the effective
mass matrix of \refeq{eq:effmass2L} or, alternatively, a mass
derivation through an iterative pole search. In fact, the mixing among
neutral states is not numerically large in this scenario and the
recourse to the degenerate formalism---although it is legitimized by
the proximity in mass---is only forced upon us by the need to connect
the gaugeless 2L corrections to the original Higgs states. The
(square-roots of the) poles are provided in \tab{tab:polval},
considering both the expansion and the iterative methods, and
employing three types of 1L field counter\-terms: OS, \DR
with~$\mu_{\text{UV}}=M_{\text{EW}}=M_W$ and \DR
with~$\mu_{\text{UV}}=M_{\text{SUSY}}=1$\,TeV. Similarly to the
previous examples, the poles derived in the expansion formalism hardly
depend on the choice of field renormalization (the variations are
actually at the level of $\simord1\cdot10^{-5}$), while those obtained
with an iterative pole search show fluctuations of GeV~order (with
varying hierarchies).

\begin{table}[t!]
  \centering
  \begin{tabular}{|c | c | c | c | c | c | c |}
  \hline
  poles & \multicolumn{3}{c|}{expansion} & \multicolumn{3}{c|}{iteration}\\
  \cline{2-7}
  [GeV] & OS & \vphantom{\Big|}\DR~EW & \DR~SUSY & OS & \DR~EW & \DR~SUSY\\
  \hline
  ${\cal M}_1$ & $123.3 - 6.9\,\imath$ & $123.3 - 6.9\,\imath$ &
  $123.3 - 6.9\,\imath$ & $124.0 - 5.8\,\imath$ & $124.7 - 0.8\,\imath$ &
  $123.5 - 7.8\,\imath$ \\
  \hline
  ${\cal M}_2$ & $123.8 - 7.6\,\imath$ & $123.8 - 7.6\,\imath$ &
  $123.8 - 7.6\,\imath$ & $124.6 - 6.6\,\imath$ & $125.0 - 6.9\,\imath$ &
  $124.3 - 8.4\,\imath$ \\
  \hline
  ${\cal M}_3$ & $125.8 - 0.7\,\imath$ & $125.8 - 0.7\,\imath$ &
  $125.8 - 0.7\,\imath$ & $126.0 - 0.8\,\imath$ & $125.3 - 7.7\,\imath$ &
  $127.5 - 0.5\,\imath$ \\
  \hline
\end{tabular}
\captionof{table}{Mass poles obtained in the scenario with three
near-degenerate neutral states. In the left trio of columns, the
masses are derived with the expansion method. In the righter one, the
iterative method is employed. Three schemes are considered for the 1L
Higgs fields: OS, \DR~with $\mu_{\text{UV}}=M_W$ (\DR~EW) and \DR~with
$\mu_{\text{UV}}=M_{\text{SUSY}}=1$\,TeV
(\DR~SUSY).\label{tab:polval}}
\end{table}

The effective couplings employed in the expansion approach are
obtained at 1L from \refeq{eq:effcoup}, which actually corresponds to
a decay amplitude for the (loop-corrected) Higgs states. Such objects
are independent on the choice of scheme for the fields. There exists
no particular inconsistency---only added uncertainty---in working with
different orders for the mass determination and the effective
couplings, except possibly in the identification of the correspondence
between poles and couplings. The latter offers no difficulty in the
scenario with weak mixing considered here, and can always be made
straightforward by using a mixing matrix defined at 2L order (at the
cost of introducing a small dependence on field counterterms). QCD
logarithms and $t_{\beta}$-enhanced corrections to the Higgs-bottom
couplings are factorized out and resummed. Turning to the poles
obtained in the iterative procedure, we adjoin to them effective
couplings that are derived according to the recipe
of \citere{Williams:2011bu} for decay amplitudes, objects that are
then also explicitly dependent on the choice of field
renormalization. We then combine these objects to define the
cross-section $\sigma[b\bar{b}\to H_{1,2,3}\to\tau^+\tau^-]$ in the
Breit--Wigner description. These quantities are also gauge-dependent,
see\,\citere{Domingo:2020wiy}, but we only consider the
't\,Hooft--Feynman gauge here.

\begin{figure}[t!]
  \centering
  \includegraphics[width=\linewidth]{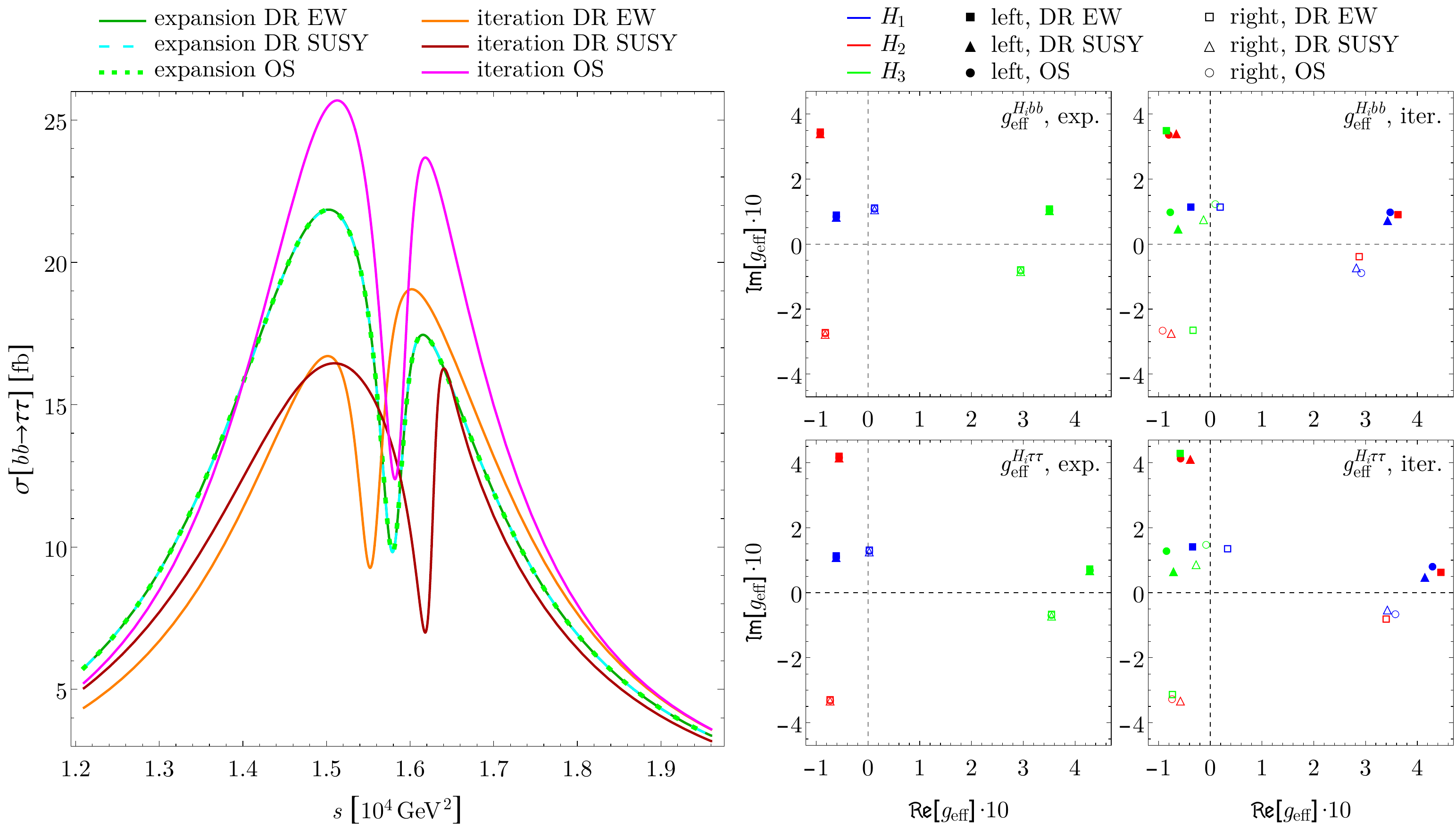}
  
  \captionof{figure}{Fermion scattering in the vicinity of three
  near-degenerate scalar resonances.  \newline{\em Left}:
  Cross-section $\sigma[b\bar{b}\to\tau^+\tau^-]$ obtained with poles
  and couplings in the expansion approach (dark-green, green and cyan,
  solid and dashed, depending on the scheme for field renormalization)
  and in the iterative method with off-shell momenta (solid orange,
  red and magenta in the \DR~EW, \DR~SUSY and OS schemes for fields,
  respectively). \newline{\em Right}: Effective Higgs couplings
  to bottom quarks (upper row)
  and $\tau$ leptons (lower row) obtained in the expansion (left
  column) and in the iterative approaches (right column). Left- and
  right-handed couplings are depicted in the complex plane with filled
  and empty symbols respectively.\label{fig:XSbbll}}
\end{figure}

The results for the $b\bar{b}\to\tau^+\tau^-$ scattering mediated by
neutral Higgs states are shown in the left panel
of \fig{fig:XSbbll}. The cross-sections obtained with the expansion
method, in green and cyan tones, are hardly distinguishable from one
another, illustrating the weak dependence of this description on field
renormalization. Conversely, the orange, red and magenta curves,
obtained with different prescriptions for the field counterterms,
demonstrate the strong dependence of the iterative method with
off-shell momenta on these regulators: the predicted cross-sections
then come with an in-built uncertainty of order $100\%$, strictly
induced by the field dependence of the would-be (pseudo-)\allowbreak
observables. Here, we note that this strong disparity among the
cross-sections is mostly driven by the imaginary parts of the poles
($\Gamma_{h_i}$): the amplitude of the resonances indeed scales like
$\Gamma_{h_i}^{-2}$, so that moderate fluctuations of these quantities
result in enhanced effects at the level of the scattering. The
effective Higgs couplings to SM fermions, $g_{\text{eff}}^{H_iff}$
with $f\in\{b,\tau\}$, are shown in the right-hand panel
of \fig{fig:XSbbll}. With the expansion method (left-most column), the
couplings of each Higgs resonance cluster at a definite point in the
complex plane, irrespectively of the choice of scheme for the field
renormalization. With the iteration method (right-most column), this
clustering is still perceptible but looser, again highlighting the
dependence on the choice of scheme. We note that the
property~$\big(g^{H_iff}_{\text{eff}}\big)_R=\big(g^{H_iff}_{\text{eff}}\big)_L^*$
cannot be maintained with a complex mixing matrix, as derived at
2L~order, which reveals the presence of partial 2L~pieces in the
couplings thus defined. However, at 1L~order, it would also be
possible to restore this property by employing a real mixing matrix,
as in \citere{Domingo:2020wiy}.

To summarize on these mass predictions in near degenerate scenarios,
we have seen how, in many cases, this formalism was rather forced upon
us by the gaugeless approximation for 2L~effects, than by the actual
size of the Higgs mixing. The exact degeneracy among the $SU(2)_{\text{L}}$~partners of the charged
Higgs at the tree level in the gaugeless limit thus forbids
the description of 1L$^2$~\CP-violating mixing terms in the formalism
of \refeq{eq:nondegmassexp}. Similarly, the tree-level mixing
angle~$\alpha-\beta$ of the full MSSM (including gauge terms),
relevant for $M_{H^{\pm}}\approx M_{\text{EW}}$, can only be accounted
for in an \AH fashion in the gaugeless description, requiring a
mixing-matrix formalism. This situation is not surprising as it
naturally emerges from the neglected EW effects in the gaugeless
description and can only be properly addressed by the inclusion of
full EW 2L corrections. As it is, we observed that the expansion
formalism for near-degenerate states, represented by the effective
mass matrix of \refeq{eq:effmass2L}, provides predictive mass (pole)
observables, showing little or no dependence on the choice of field
renormalization, when the converse defect limits the usefulness of the
iteration method.

\tocsection{Resummation of UV-logarithms of
\texorpdfstring{$\mathcal{O}\big(\alpha_q,\,\alpha_q\,\alpha_s,\,\alpha_q^2\big)$}{\unicodescriptO(\unicodealpha\unicodesubt,\unicodealpha\unicodesubt\unicodethinspace\unicodealpha\unicodesubs,\unicodealpha\unicodesubt\unicodesuptwo)} and field dependence\label{sec:resum}}

In the previous sections, we have derived the Higgs masses and
discussed the field dependence in a strict expansion
at~FO. Nevertheless, the large size of the radiative corrections
of~$\mathcal{O}(\text{2L})$ to the mass of the SM-like state in such
an approach points at the slow convergence of the perturbative series
in the presence of heavy (multi-TeV) SUSY particles. As is well-known,
UV-logarithms of the type
$\ln^{k}\!M^2_{\text{SUSY}}\big/M^2_{\text{EW}}$---with
$M_{\text{SUSY}}$ typically corresponding to the mass of the squarks
of third generation---are responsible for these large effects and
should be resummed for numerically meaningful predictions. One usually
turns to the EFT
framework---see \EG~\citeres{Bagnaschi:2014rsa,Lee:2015uza,Vega:2015fna,Athron:2016fuq,Bagnaschi:2017xid,Bahl:2018jom,Braathen:2018htl,Bagnaschi:2019esc,Murphy:2019qpm,Bahl:2020mjy,Bahl:2020jaq,Staub:2017jnp,Harlander:2018yhj,Bahl:2019wzx}---in
order to implement this resummation. However, for the corrections
of~$\mathcal{O}\big(\alpha_q,\,\alpha_q\,\alpha_s,\,\alpha_q^2\big)$
that we discuss here, this operation can be performed directly in the
FO~context: all the relevant parameter input---strong gauge- and
Yukawa couplings---is indeed accessible at low energy from SM
observables, so that a matching at high scale is superfluous. Then,
the (simple and squared) UV-logarithms contained in the FO~expansion
can be explicitly extracted and re-molded according to the flow of the
Callan--Symanzik equations applying to the SM-like Higgs mass (or the
mass-splitting between heavy states). While the method is thus
formally distinct from that of an~EFT, it is very similar at the
technical level to the hybrid procedure described
in \citere{Hahn:2013ria}, since both subtract UV-logarithms from the
FO~calculation to re-inject them in resummed form.

The relevant Renormalization-Group Equations (RGEs) are those obtained
after screening-off the heavy fields---due to the large mass that cut
these off in the emergence of UV-logarithms at the level of the loop
functions: in other words, they match the field content of the SM, or
of the Two-Higgs Doublet Model (THDM) for intermediate scales
(if~$M_{H^{\pm}}\ll M_{\text{SUSY}}$). Further thresholds could be
considered, depending on the relative scales of the gluino and the
squarks of third generation, but we will not discuss them here. For
the considered orders, the RGEs are entirely determined by the running
of the quartic Higgs couplings, the EW v.e.v., the strong gauge- and
Yukawa couplings. For definiteness, we collect the resummation
formulae for the SM-like state in the gaugeless limit below:
\begin{subequations}\label{eq:SMRGE}
\allowdisplaybreaks\begin{align}
\hat{\Sigma}_{hh}^{\text{(gl,resum)}} &=
  \int\limits_{\mathclap{\ln M_{\text{EW}}^2}}^{\mathclap{\ln M_{\text{SUSY}}^2}}d\ln\mu^2\,
  \bigg\{\begin{aligned}[t]
  & {-}\,12\left[\alpha_t^2(\mu)\,s_{\beta}^4 + \alpha_b^2(\mu)\,c_{\beta}^4\right]
  \left(1 + \frac{16}{3}\,\frac{\alpha_s(\mu)}{4\,\pi}\right)
  + \beta_{\lambda}^{\alpha_q^2}(\mu)\\
  &{+}\,\frac{6\,(\delta\lambda)^{(\alpha_q)}}{4\,\pi}
  \left[\alpha_t(\mu)\,s_{\beta}^2 + \alpha_b(\mu)\,c_{\beta}^2\right]
  \!\!\bigg\}\,v^2(\mu)\,,\taghere\end{aligned}\\
  \beta_{\lambda}^{\alpha_q^2}(\mu) &\equiv \frac{12}{4\,\pi}
  \left\{\begin{aligned}
  & 5\,\big[\alpha_t^3(\mu)\,s_{\beta}^4 + \alpha_b^3(\mu)\,c_{\beta}^4\big]
  + \alpha_t(\mu)\,\alpha_b(\mu)\,
  \big[\alpha_t(\mu)\,s_{\beta}^2 + \alpha_b(\mu)\,c_{\beta}^2\big],
  & \mu>M_{H^{\pm}}\,,\\
  & 5\,\big[\alpha_t^3(\mu)\,s_{\beta}^6 + \alpha_b^3(\mu)\,c_{\beta}^6\big]
  - \alpha_t(\mu)\,s_{\beta}^2\,\alpha_b(\mu)\,c_{\beta}^2\,
  \big[\alpha_t(\mu)\,s_{\beta}^2 + \alpha_b(\mu)\,c_{\beta}^2\big],
  & \mu<M_{H^{\pm}}\,,
  \end{aligned}\right.\\
  \frac{d\,\alpha_s}{d\ln\mu^2} &\equiv -7\,\frac{\alpha_s^2}{4\,\pi}
  + \mathcal{O}(\alpha_s^3,\alpha_s^2\,\alpha_q)\,,\\
  \frac{d\,\alpha_t}{d\ln\mu^2} &\equiv \begin{cases}
  \frac{1}{4\,\pi}\,\alpha_t\left[\frac{9}{2}\,\alpha_t + \frac{1}{2}\,\alpha_b
  - 8\,\alpha_s\right] + \mathcal{O}(\alpha_q\,\alpha,\alpha_s^2\,\alpha_q,\alpha_s\,\alpha_q^2,\alpha_q^3)\,, & \mu>M_{H^{\pm}}\,,\\
  \frac{1}{4\,\pi}\,\alpha_t\left[\frac{9}{2}\,\alpha_t\,s_{\beta}^2
  + \frac{3}{2}\,\alpha_b\,c_{\beta}^2 - 8\,\alpha_s\right] + \mathcal{O}(\alpha_q\,\alpha,\alpha_s^2\,\alpha_q,\alpha_s\,\alpha_q^2,\alpha_q^3)\,,
  & \mu<M_{H^{\pm}}\,,\end{cases}\\
  \frac{d\,\alpha_b}{d\ln\mu^2} &\equiv \begin{cases}
  \frac{1}{4\,\pi}\,\alpha_b\left[\frac{9}{2}\,\alpha_b + \frac{1}{2}\,\alpha_t
  - 8\,\alpha_s\right] + \mathcal{O}(\alpha_q\,\alpha,\alpha_s^2\,\alpha_q,\alpha_s\,\alpha_q^2,\alpha_q^3)\,, & \mu>M_{H^{\pm}}\,,\\
  \frac{1}{4\,\pi}\,\alpha_b\left[\frac{9}{2}\,\alpha_b\,c_{\beta}^2
  + \frac{3}{2}\,\alpha_t\,s_{\beta}^2 - 8\,\alpha_s\right] + \mathcal{O}(\alpha_q\,\alpha,\alpha_s^2\,\alpha_q,\alpha_s\,\alpha_q^2,\alpha_q^3)\,,
  & \mu<M_{H^{\pm}}\,,
  \end{cases}\\
  \frac{d\,v^2}{d\ln\mu^2} &\equiv
  -\frac{3\,v^2}{4\,\pi}\left[\alpha_t\,s_{\beta}^2 + \alpha_b\,c_{\beta}^2\right]
  + \mathcal{O}(\alpha,\alpha_s\,\alpha_q,\alpha_q^2)\,,\\
  (\delta\lambda)^{(\alpha_q)} &\equiv
  \left[(\delta\lambda_2)^{(\alpha_q)}\,s_{\beta}^4
  + (\delta\lambda_1)^{(\alpha_q)}\,c_{\beta}^4\right]\!\cdot\!
  \left[\alpha_t(\mu)\,s_{\beta}^2 + \alpha_b(\mu)\,c_{\beta}^2\right]^{-1},\quad
  \mu>M_{H^{\pm}}\,,\\
 &\quad\; \frac{d(\delta\lambda_1)^{(\alpha_q)}}{d\ln\mu^2} \equiv
 -6\,\alpha_b^2(\mu) + \mathcal{O}\big(\alpha_q^2\,\alpha_s,\,\alpha_q^3\big)\,,
 \quad \frac{d(\delta\lambda_2)^{(\alpha_q)}}{d\ln\mu^2} \equiv
 -6\,\alpha_t^2(\mu) + \mathcal{O}\big(\alpha_q^2\,\alpha_s,\,\alpha_q^3\big)\,,
 \nonumber\\
 \frac{d(\delta\lambda)^{(\alpha_q)}}{d\ln\mu^2} &\equiv
 -6\left[\alpha_t^2(\mu)\,s_{\beta}^4 + \alpha_b^2(\mu)\,c_{\beta}^4\right]
 + \mathcal{O}\big(\alpha_q^2\,\alpha_s,\,\alpha_q^3\big)\,,\quad
 \mu<M_{H^{\pm}}\,.
\end{align}\end{subequations}
The resummation of gaugeless orders has little significance for the
heavy-doublet masses, where EW corrections are dominant, and we thus
omit the corresponding formulae. Contrarily to the EFT description
where the EFT parameters are run down from a UV-matching
scale to the Higgs (EW) scale, the RGEs of the
FO~description are run up from the SM (or THDM) input
scale~$M_{\text{EW}}$---set equal to~$m_t$ in practice---towards the
scale corresponding to the heavy screened-off particles. The
conversion of observable input (\EG~fermion and gauge pole masses) to
running (\MS) parameters generates further next-to-leading logarithms.

The RGEs summarized in \refeq{eq:SMRGE} incorporate all the
logarithmically-enhanced corrections that one can derive in the
gaugeless limit at 2L order. We thus explicitly restrict ourselves to
the leading order~(LO) and next-to-LO~(NLO) that are explicitly
contained in the diagrammatic calculation of Higgs self-energies
at~2L~FO. However, RGEs of higher order could also be employed---in
that case, without subtracting the higher-order logarithms in expanded
form, since these have no equivalent in the diagrammatic calculation
of~$\mathcal{O}(\text{2L})$. Furthermore, a few subtleties associated
with the Higgs self-coupling parameter~$(\delta\lambda)^{(\alpha_q)}$
that are generated by loop effects of~$\mathcal{O}(\alpha_q)$ may be
worth discussing. Indeed, the traditional (EFT) counting would
associate to this object the boundary
condition~$(\delta\lambda)^{(\alpha_q)}(M_{\text{SUSY}})\stackrel[]{!}{=}0$,
resulting in the low-energy
boundary~$(\delta\lambda)^{(\alpha_q)}(M_{\text{EW}})\approx6\,\Big[\alpha_t^2\,s_{\beta}^4
+ \alpha_b^2\,c_{\beta}^4\Big]\ln
M^2_{\text{SUSY}}\big/M^2_{\text{EW}}$. This is also the choice that
allows to reproduce the explicit logarithmic expansion of the
FO~calculation. Nevertheless, the
choice~$(\delta\lambda)^{(\alpha_q)}(M_{\text{EW}})\stackrel[]{!}{=}0$
is also a perfectly legitimate condition due to the fundamental
ambiguities in order counting---$\alpha_q\sim\alpha$
and~$\alpha_q\,\ln M^2_{\text{SUSY}}\big/M^2_{\text{EW}}\sim1$---which
allow the transmutation of the 2L~quantity
$\alpha_q^2\,\ln^2\!M^2_{\text{SUSY}}\big/M^2_{\text{EW}}$ into the
1L~object $\alpha\,\ln M^2_{\text{SUSY}}\big/M^2_{\text{EW}}$: instead
of being resummed as logarithms of Yukawa type, the squared logarithms
thus overlooked would be left as simple EW logarithms---and possibly
resummed as such when absorbed within the quartic Higgs
coupling~$\lambda(M_{\text{EW}})$. As we do not perform this
EW~resummation here, however, the numerical difference between the two
procedures can be sizable at large squark masses. To keep the
comparisons simple, we therefore adopt the traditional counting by
default, \IE~set
\begin{align}
  (\delta\lambda)^{(\alpha_q)}(M_{\text{EW}}) &\stackrel[]{!}{=}
  \int\limits_{\ln M^2_{\text{EW}}}^{\ln M^2_{\text{SUSY}}}d\ln\mu^2\,6
  \left[\alpha_t^2\,s_{\beta}^4 + \alpha_b^2\,c_{\beta}^4\right].
\end{align}

With LO and NLO logarithms
of~$\mathcal{O}\big(\alpha_q,\,\alpha_q\,\alpha_s,\,\alpha_q^2\big)$
properly resummed (up to another caveat that we discuss below), the
higher-order uncertainty, controlled by
terms~$\simord\alpha_q\,\alpha_s^2\ln^3\!M_{\text{SUSY}}^2\big/M^2_{\text{EW}}$
($\simeqord10$--$30\%$, at the level of the squared mass, for
$M_{\text{SUSY}}=1.5$--$10$\,TeV) in the strict expansion, is now
pushed back---as far as gaugeless orders are
concerned---to~$\simord\alpha_q\,\alpha_s^2\ln
M_{\text{SUSY}}^2\big/M^2_{\text{EW}}$ ($<1\%$). In fact, the
higher-order uncertainty is now controlled by
EW~2L~effects~$\simord \alpha_q\,\alpha\ln^2\!M_{\text{SUSY}}^2\big/M^2_{\text{EW}}$
($\simord0.2$--$2\%$ for
$M_{\text{SUSY}}=1.5$--$10$\,TeV). Furthermore, another term of
gaugeless order appears in the RGEs, originating in Higgs
self-interactions:
\begin{equation}\label{eq:Bglresum}
  \hat{\Sigma}_{hh}^{(\text{gl},\alpha_q^3)} =
  \int\limits_{\mathclap{\ln M_{\text{EW}}^2}}^{\mathclap{\ln M_{\text{SUSY}}^2}}d\ln\mu^2\,
  \frac{12}{16\,\pi^2}\left[(\delta\lambda)^{(\alpha_q)}(\mu)\right]^2v(\mu)^2
  = \mathcal{O}\big(\alpha_q^3\ln^2\!M^2_{\text{SUSY}}\big/M^2_{\text{EW}}\big)\,.
\end{equation}
\needspace{3ex}
\noindent
As this is a contribution of 3L order, it does not appear in the 2L
calculation at~FO. In order to resum all NLO~logarithms of
Yukawa-type, we add this term to our calculation nonetheless, instead
of leaving it as a contribution of~$\mathcal{O}(\alpha)$. Finally, it
is fair to mention in this short uncertainty estimate that not only
logarithmically-enhanced contributions may require the recourse to
resummation techniques: see \EG~\citere{Kwasnitza:2021idg} for a
resummation of squark-mixing terms.

EW logarithms of LO could also be resummed, as is routinely performed
in the EFT description. Our method with input at the low-energy
boundary continues to apply, without need of a matching at high
scale. EW gauge couplings are indeed well-defined at the low-energy
end. As to the quartic Higgs couplings, they can be obtained from the
observables (Higgs masses and decays) that are predicted in the
FO~calculation, without or with partial UV-resummation. This is a
straightforward recipe for a low-energy effective SM---only the
(preliminary) mass prediction for the SM-like state is then
needed---but accounting for a THDM threshold becomes more intricate as
a larger basis of observables is then needed in order to identify the
more numerous parameters of the Higgs potential; yet, the situation is
no different at this level in an EFT~approach, in principle. In fact,
the somewhat less robust boundary condition for the Higgs
self-interactions is the only drawback as compared to the EFT
procedure: we anticipate it to amount to little at the numerical
level, albeit a detailed comparison (beyond our current scope) would
be needed in order to ascertain this claim. In this paper, we choose
not to carry out the resummation of EW logarithms and go through the
intricacies of a THDM threshold, first because this would exceed the
bounds of gaugeless orders that we meant to discuss here, then because
these effects remain comparatively small (and not necessarily related
to the scale of the sole gluino and squarks of third generation),
finally because the actual focus of the paper concerns the dependence
on field counterterms and that the gaugeless orders are quite
sufficient to measure its impact in a resummed mass prediction for the
SM-like Higgs.

\begin{figure}[p!]
  \centering
  \includegraphics[width=\linewidth]{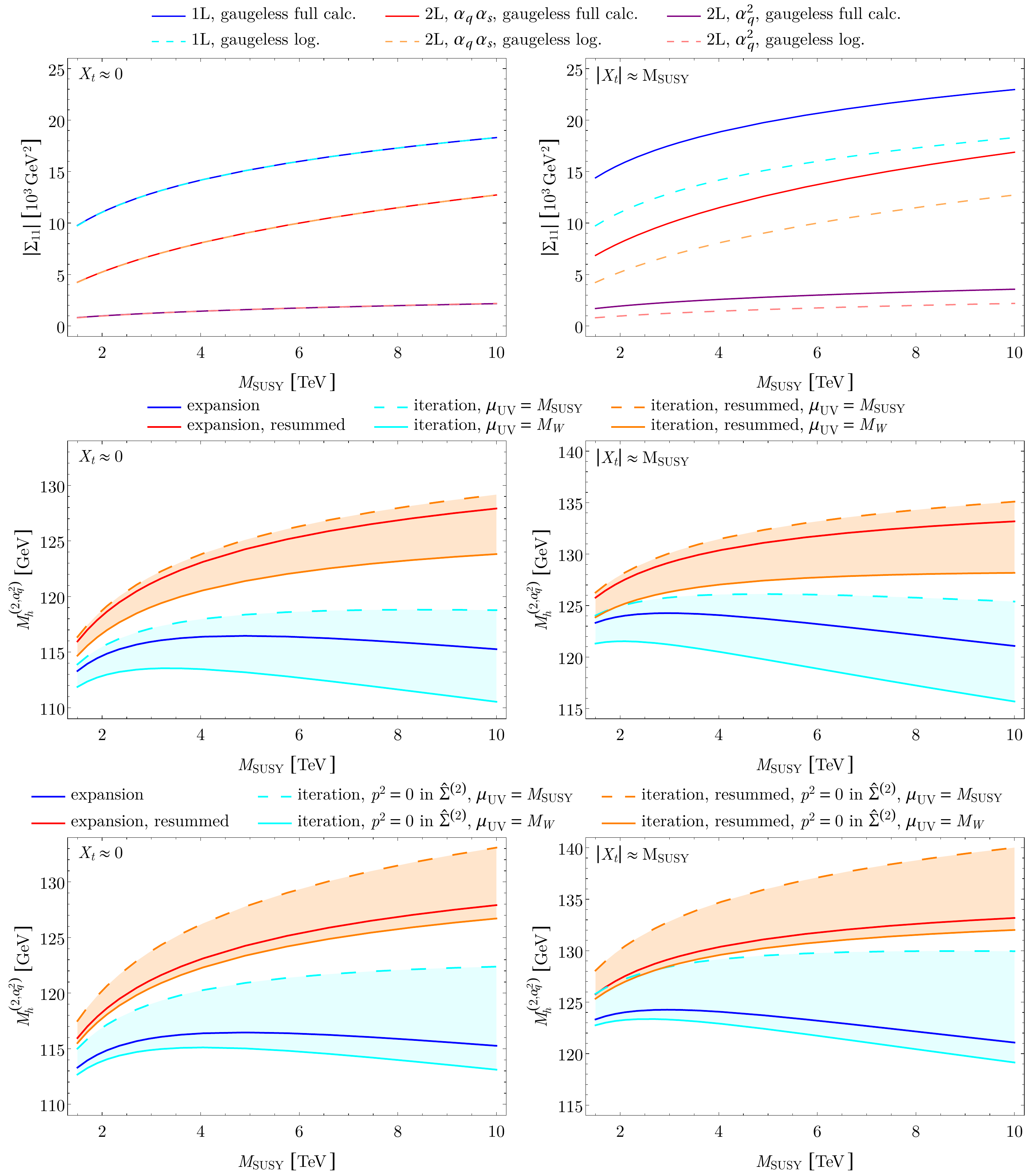}
  \caption{Impact of the resummation of UV-logarithms of
  $\mathcal{O}\big(\alpha_q,\,\alpha_q\,\alpha_s,\,\alpha_q^2)$ for
  the Higgs-mass prediction and the field dependence. \newline{\em
  Up}: Higgs self-energies in the gaugeless limit (solid curves)
  compared to the logarithmic expansion (dashed curves). \newline{\em
  Middle}: Mass predictions with resummed UV-logarithms (red and
  orange) compared to the strict FO~expansion (blue and cyan). Both
  the expansion formalism (red and blue) and the iterative pole search
  with \DR~field counterterms (orange and cyan; solid:
  $\mu_{\text{UV}}=M_W$; dashed: $\mu_{\text{UV}}=M_{\text{SUSY}}$)
  are considered. \newline{\em Down}: Same as in the middle row, but
  with two-loop self-energies evaluated at $p^2=0$. \newline{\em
  Left}: The stop mixing is set to negligible values. \newline{\em
  Right}: The stop mixing is kept at the SUSY scale.\label{fig:Resum}}
\end{figure}

In the upper row of \fig{fig:Resum}, we compare the explicit
calculation of Higgs self-energies at~FO in the gaugeless limit and
the corresponding expansion in
$\ln^k\!M_{\text{SUSY}}^2\big/M^2_{\text{EW}}$. This latter expansion
is obtained from the integration of the RGEs of \refeq{eq:SMRGE} in a
linear way, but---as mentioned above---additional logarithms of~NLO
intervene from the conversion of parameters in our scheme to \MS~ones
at the
scale~$M_{\text{EW}}$---\EG~$m_t^{\text{OS}}=m_t^{\text{\MS}}(m_t)\left(1+\frac{16}{3}\,\frac{\alpha_s}{4\,\pi}+\mathcal{O}(\alpha_q)\right)$. The
plots in the middle and lower rows show the impact of the resummation
of the UV-logarithms for the prediction of the SM-like mass. Here, the
UV-resummation is achieved by subtracting the identified UV-logarithms
and substituting a resummed version where \refeq{eq:SMRGE} is
integrated numerically. The corresponding evaluations are shown in red
and orange lines, while the blue and cyan lines correspond to the
predictions without resummation. These calculations are conducted for
both the expansion formalism (blue and red) and for the iterative pole
search (cyan and orange) with \DR~regularization of the Higgs fields
($\mu_{\text{UV}}=M_W$ in solid and $\mu_{\text{UV}}=M_{\text{SUSY}}$
in dashed lines). In the lower row, the orange and cyan curves
correspond to the approximation~\mbox{$p^2=0$} in the contributions of
2L~order processed in an iterative fashion: their colors match those
of the corresponding curves in the middle row, accounting for full
momentum dependence.

On the left-hand side of \fig{fig:Resum}, we consider an MSSM scenario
with decoupling squarks and gluinos ($m_{\tilde{Q}_3,\tilde{T}}\approx
M_3\stackrel[]{!}{=} M_{\text{SUSY}}\gg m_t$), keeping the mixing in
the stop sector, $X_t\equiv A_t-\mu/t_{\beta}$, minimal, and setting
$M_{H^{\pm}}=1$\,TeV, $t_{\beta}=10$ for the THDM sector. This is the
ideal setup for checking the agreement between the logarithmic
expansion and the FO~calculations, as can be observed in the very
narrow matching of the various self-energies. Turning to the mass
predictions, the resummation of UV-logarithms accounts for a shift of
already~$\simord1.5$\,GeV at~$M_{\text{SUSY}}\sim1.5$\,TeV and
over~$10$\,GeV at~$10$\,TeV. As already noted in earlier
works---see \EG~\citere{Slavich:2020zjv}---the resummation counteracts
the tendency of mass predictions of~$\mathcal{O}\big(\alpha_q^2\big)$
to fall at large~$M_{\text{SUSY}}$---due to
large~$\ln^2\!M^2_{\text{SUSY}}\big/M^2_{\text{EW}}$ added linearly in
the strict FO~expansion.

On the right-hand side of \fig{fig:Resum}, we maintain the stop mixing
at~$\lvert X_t\rvert\approx M_{\text{SUSY}}$, which generates sizable
shifts between the logarithmic expansion and the FO~calculation
(without endangering the relevance of the resummation of
UV-logarithms). At least the leading orders from squark mixing are
properly included within the non-logarithmic terms of the
FO~calculation.

\begin{figure}[t!]
  \centering
  \includegraphics[width=\linewidth]{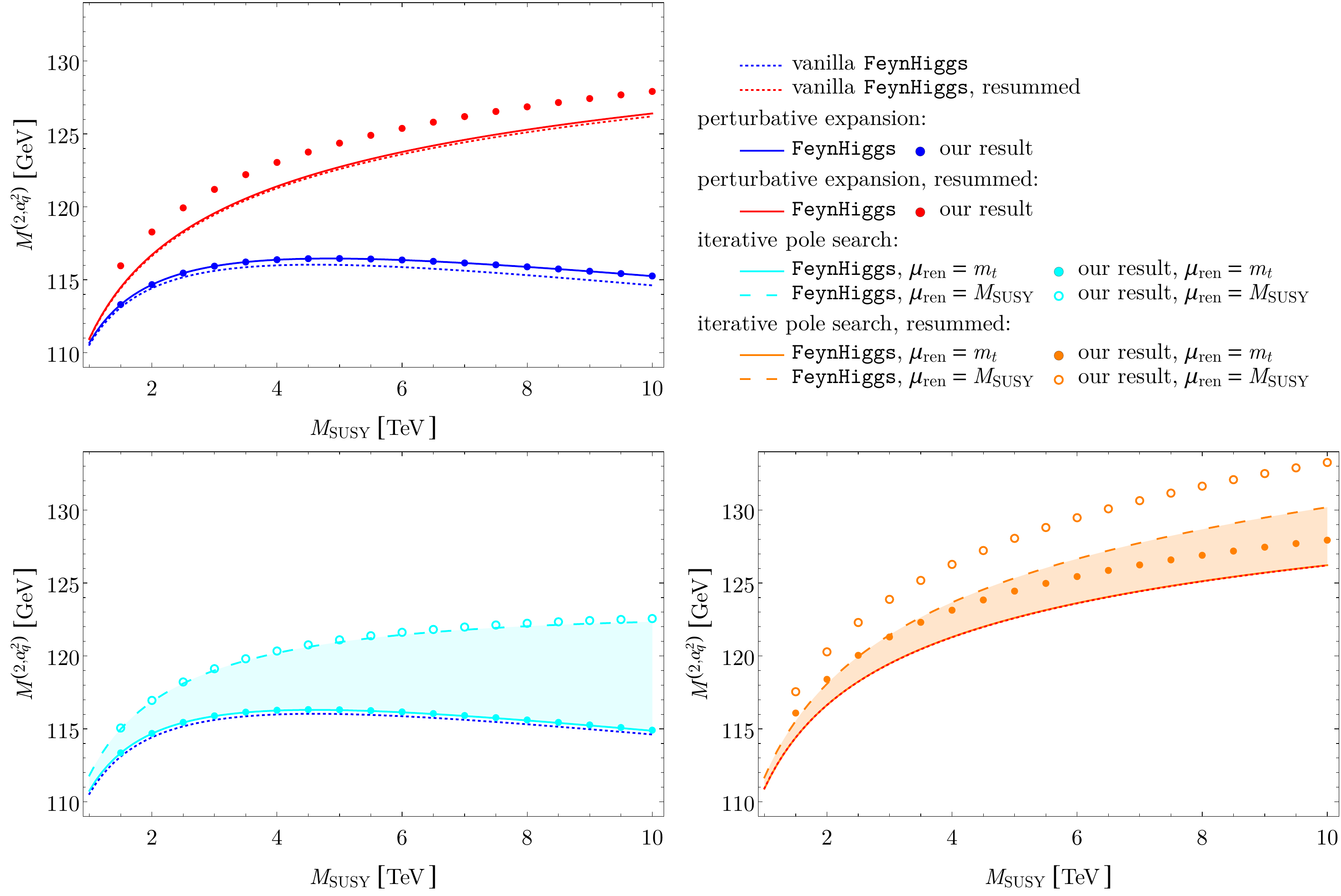}
  \caption{\label{fig:FHcomp} Comparison of our results with those
  of \texttt{FeynHiggs}. Blue/cyan curves and symbols correspond to
  the mass predictions with strict FO~expansion, while the red/orange
  ones include a resummation of UV-logarithms. Actual lines refer to
  masses determined with \texttt{FeynHiggs} whereas the disks and
  circles proceed from our own calculations. The default output
  of \texttt{FeynHiggs}, `vanilla \texttt{FeynHiggs}', is plotted with
  short-dashed lines in all plots to serve as a reference. All other
  curves test different settings. \newline {\em Up Left}: The partial
  expansion available in \texttt{FeynHiggs} (solid lines) is compared
  with our own (disks). \newline {\em Bottom Left}: Results of the
  iterative pole search without resumming UV-logarithms. The
  calculations are performed with two different renormalization scales
  and the \DR~input is correspondingly adjusted. \newline {\em Bottom
  Right}: Similar to bottom left, but now including the resummation of
  UV-logarithms.}
\end{figure}

As to the dependence of masses on field counterterms, in case an
iterative pole search is employed, it is obvious that the resummation
affects it only marginally, as the momenta evaluated in self-energies
are still shifted with respect to the tree-level mass (the latter
choice ensuring invariance in the expansion method). Correspondingly,
the uncertainty associated with field variations steeply grows with
increasing~$M_{\text{SUSY}}$ in the FO~approach with iterative pole
search, reaching~$\mathcal{O}(10\%)$
at~$M_{\text{SUSY}}\sim10$\,TeV---hence becoming larger than the
actual higher-order uncertainty after resummation of the
UV-logarithms. The approximation~$p^2=0$ in pieces of 2L~order of the
iterative pole search (lower row) tends to systematically
over-estimate the mass as compared to the prediction with full
momentum dependence, but does not reduce the magnitude of the
dependence on field-renormalization constants. Given the similarity of
the procedure employed in the hybrid calculation
of \citere{Hahn:2013ria}, then refined
in \citeres{Bahl:2016brp,Bahl:2017aev,Bahl:2018jom,Bahl:2018ykj,Bahl:2019hmm,Bahl:2020mjy,Bahl:2020tuq},
the corresponding predictions of the public
code \texttt{FeynHiggs}\,\cite{Heinemeyer:1998yj,Heinemeyer:1998np,Degrassi:2002fi,Frank:2006yh,Hahn:2013ria,Bahl:2016brp,Bahl:2017aev,Bahl:2018qog}
are thus \AP subject to the large error associated with the field
dependence. Yet, an \AH choice of field counterterms, derived by
comparison of the logarithms with those of an EFT
(see \citere{Bahl:2018ykj}), then restores a more predictive
behavior---largely compatible with that of the more straightforward
expansion approach. We aim at a more detailed comparison
in \fig{fig:FHcomp} for the same scenario as in the left-hand column
of plots of \fig{fig:Resum} ($X_t\approx0$).

\needspace{3ex}
\texttt{FeynHiggs-2.18.0} delivers Higgs-mass predictions in the MSSM.
The 2L corrections of order~$\alpha_q\,\alpha_s$ and~$\alpha_q^2$ are
included in the effective-potential approximation, which is a relevant
choice for the SM-like state, as already discussed. UV-logarithms can
optionally be resummed and included in the hybrid approach described
in \citeres{Hahn:2013ria,Bahl:2016brp,Bahl:2017aev,Bahl:2019hmm,Bahl:2020tuq}.
The default setup (labelled `vanilla \texttt{FeynHiggs}' below)
employs an iterative pole-search algorithm to find the loop-corrected
Higgs masses. In order to neutralize the large unphysical scale
dependence that is thus introduced by the field-renormalization
constants, a special scheme has been devised in \citere{Bahl:2018ykj};
it includes all finite SUSY contributions in the field counterterms,
thereby yielding the same logarithmic dependence as that of an EFT
calculation. The predictions of `vanilla \texttt{FeynHiggs}' are
depicted as reference in all plots of \fig{fig:FHcomp} (blue and red
short-dashed lines).

For consistency checks, \texttt{FeynHiggs} also provides some hidden
flags that can be set through environment variables. Below, we make
use of the following settings: 
\begin{itemize}
\item\texttt{FHFOPOLEEQ=1} switches the determination of the
loop-corrected masses to a partial perturbative expansion---only
corrections for the diagonal self-energies are included; in the
considered scenario with hierarchical Higgs sector, this is expected
to agree relatively well with our calculation using the expansion
formalism;
\item with \texttt{FHFINFIELDREN=0} the
finite terms of the field-renormalization constants are set to zero,
hence reverting the field counterterms to a simple \DR~form;
\item \texttt{FHTBSCALE=\#} interprets the input value for~$\tan\beta$
at the scale \texttt{\#}\,GeV, which can largely be seen as resetting
the renormalization scale~$\mu_{\text{ren}}$ for this
parameter.
\end{itemize}

In the upper-left plot of \fig{fig:FHcomp}, we compare the mass
predictions of the partial perturbative expansion available
in \texttt{FeynHiggs} (solid curves) with the results using our own
expansion formalism (disks). The plain FO~determination is shown in
blue, while a resummation of logarithms is included in the red
curves. In the case of \texttt{FeynHiggs}, the
setting \texttt{loglevel=2} is employed, which corresponds to the
resummation of next-to-leading logarithms~(NLL) for the case where all
SUSY masses are large.\footnote{A setting for higher precision is
possible and performs a next-to-NLL resummation, but such effects are
not considered in our calculation (though they could be added at a
later stage), hence appear of limited use for the comparison that we
conduct here and which focusses on field-dependence aspects.} We
observe an almost exact agreement of \texttt{FeynHiggs} and our
formalism for this expansion approach; small differences at the level
of predictions with resummed UV-logarithms are of the expected
percent-order magnitude for unresummed EW logarithms (in our setup)
and can thus be attributed to additional classes of logarithms that
are included in the resummation of \texttt{FeynHiggs}. In addition,
the resummation of EW~logarithms may not be completely suited to our
scenario where electroweakino and slepton masses are not varied
with~$M_{\text{SUSY}}$.

\needspace{3ex}
In the lower row of \fig{fig:FHcomp}, we compare our predictions
(disks and circles) to the corresponding ones obtained
with \texttt{FeynHiggs} (solid and dashed curves) when using an
iterative pole-search algorithm with \DR-renormalized fields. In order
to resemble the setup of \texttt{FeynHiggs}, our predictions include
the 2L~corrections in the effective-potential approximation. The plot
on the left-hand side shows the FO~results, while the curves on the
right-hand side include resummed logarithms. In both plots we display
predictions at the renormalization scales~$\mu_{\text{ren}}=m_t$
(solid curves or disks) and~$\mu_{\text{ren}}=M_{\text{SUSY}}$ (dashed
curves or circles). The input value for the \DR~parameter~$\tan\beta$
is interpreted at the scale~$m_t$ before being run to the chosen
renormalization scale and fed as input: we thus make sure that we are
comparing the same points in parameter space. The running of the
additional \DR~parameters~$m_b$ and~$A_b$ has a negligible impact in
the chosen scenario. This whole procedure is meant to emulate the
direct variation of finite field counterterms, as we considered it
before, since an independent variation is not straightforwardly
accessible in \texttt{FeynHiggs}. Again, we are able to recover the
predictions of \texttt{FeynHiggs} at~FO (left plot) to a good
approximation, which confirms our observations as to the large
inherent uncertainty associated with an iterative approach to the pole
determination. A somewhat larger deviation is visible in the resummed
prediction at the SUSY scale: we could not completely understand the
origin of this discrepancy, but this may not matter much since the
dependence on the choice of field renormalization in such a
description spoils the logarithmic behavior anyway. In any case, it is
obvious that the introduction of the UV-resummation does not
neutralize the intrinsic uncertainty originating in the pole search.
We note that the results obtained with $\mu_{\text{UV}}=m_t$ are very
close to the\allowbreak---in our opinion more reliable---predictions
of the expansion procedure. This good performance is not completely
mysterious in the considered scenario. Indeed, with the external
momentum set to~$0$ in the 2L self-energies and a sizable hierarchy
between~$M_{\text{EW}}$ and~$M_{H^{\pm}}$, the difference between the
iteration and expansion procedures for the determination of the mass
of the SM-like Higgs can be estimated as $-\big(\mathfrak{M}^2_h -
m_h^2\big)\,\hat{\Sigma}^{\text{(1)}\,\prime}_{hh}\big(m_h^2\big)
+ \hat{\Sigma}^{\text{(1,gl)}}_{hh}(0)\,\hat{\Sigma}^{\text{(1,gl)}\,\prime}_{hh}(0)$
(similarly to \refeq{eq:itervsexp}). Then, in the gaugeless
approximation,
$\hat{\Sigma}^{\text{(1,gl)}\,\prime}_{hh}\approx-\tfrac{3\,\alpha_t}{4\,\pi}\,s_{\beta}^2\,\Big[\ln\mu_{\text{UV}}^2\big/m_t^2
- \tfrac{2}{3}
+ \mathcal{O}\big(X_t^2\big/M^2_{\text{SUSY}}\big)\Big]$, which
vanishes for~$\mu_{\text{UV}}=\mathcal{O}(m_t)$ at small~$X_t$. The
inclusion of EW orders does not spoil this picture, as long as large
logarithms do not develop from the electroweakino loops. Therefore,
the apparent predictivity of the choice~$\mu_{\text{UV}}=m_t$ is very
specific to this scenario and not reliable on fundamental grounds.

Finally, we turn to the predictions by
`vanilla \texttt{FeynHiggs}'. These appear to be relatively close to
our results obtained with the perturbative expansion (upper
plot). Again, this comparative proximity is not really mysterious, as
the injected counterterms in \texttt{FeynHiggs} have been designed
such that one recovers the correct logarithmic behavior. Such a choice
is comparable to our OS~scheme for the field counterterms, for which
we also observed comparative agreement with the expansion in similar
setups: see the pink cross in the upper plot
of \fig{fig:M2Lat2iter}. Therefore, although \texttt{FeynHiggs}
employs an iterative pole search by default, it escapes the large
uncertainty associated with this procedure through a judicious choice
of field counterterms. Yet, a discrepancy
reaching~$\mathcal{O}(1\,\text{GeV})$ at~$M_{\text{SUSY}}=10$\,TeV is
visible in the strict FO~approach---blue curves; both calculations are
comparable in that they consider exactly the same
orders. Unquestionably, this difference originates in the choice of
procedure (expansion vs.~iteration), since the simplified expansion
available in \texttt{FeynHiggs} agrees with our method. This is
therefore the magnitude of the error---again, we stress that this
shift has no predictive value---contained in this choice for the mass
determination, which should be included as an irreducible uncertainty
to the predictions (beyond estimated higher-order effects). This
contribution is somewhat reduced after inclusion of the UV-resummation
(in this scenario), because $\mathfrak{M}^2_h$ then comes closer
to~$m_h^2-\hat{\Sigma}_{hh}^{(1,gl)}(0)$ (${\approx}\,162^2$\,GeV$^2$
at~$M_{\text{SUSY}}=10$\,TeV)---the difference between these two
quantities controls the leading dependence on the field counterterm,
proportional to~$\delta Z_{hh}$.

In addition, the use of the technically more involved iteration method
remains a choice of questionable efficiency since the predictivity of
the FO~calculation, directly accessible with the simple expansion
procedure, is first wasted, then restored through the cross-reference
of the logarithms with the EFT method. Finally, it is unlikely that
this method can simultaneously produce predictive results at the level
of the mass-splitting among heavy states. We therefore recommend the
use of the more robust expansion and truncation procedure that we
described in Sect.\,\ref{sec:2Lmasses} after the principle of
independence of observables from the choice of scheme for field
renormalization.

\needspace{20ex}
\tocsection{Conclusions\label{sec:conclusion}}

In this paper, we investigated the dependence of MSSM Higgs masses on
field counterterms in a FO~approach at~2L. This dependence on
regulators originates in the arbitrary regularization of Higgs
self-energies away from their (tree-level) mass-shell, and is
exacerbated when processing~2L and~1L$^2$~corrections in independent
fashions. This situation is further complicated by the fact that---due
to missing EW~corrections in 2L~vector self-energies---only effects of
$\mathcal{O}\big(\alpha_q\,\alpha_s,\,\alpha_q^2\big)$ are fully
exploitable at~2L, rendering an evaluation in the gaugeless
approximation necessary at the technical level. Masses derived in the
strict expansion formalism evade these difficulties through a careful
pairing of field-dependent pieces and the neutralization of field
counterterms. On the other hand, the popular mass determination via an
iterative pole search, computationally more costly, retains an
explicit dependence on field regulators, which generates an
irreducible uncertainty inherent to the method. In the case of the
SM-like Higgs state, we have seen that this `error' already amounts to
a few~GeV for a SUSY~sector at the TeV~scale. Concerning the
heavy-doublet states, unlike the SM-like one, the known
orders~$\alpha_q\,\alpha_s$ and~$\alpha_q^2$ are of limited relevance
since EW~corrections are expected to dominate. We still considered
these contributions in order to test the impact of the regularization
of Higgs self-energies away from their mass-shell. Then, a scale
variation with \DR~counterterms is insufficient to capture the full
extent of the field dependence, which is driven by leading EW~effects,
and we also introduced an OS~regularization for comparison. In all
cases, variation of the field regulators demonstrated that the
mass-shifts generated by the iterative pole search with respect to the
expansion approach are purely artificial in nature, and we thus
believe it justified to prefer the simpler---and field
independent---method. This argument adds to the one of the symmetry
\mbox{considerations that we raised at 1L~order
in \citere{Domingo:2020wiy}}.

In the presence of mass degeneracies, the expansion formalism can be
extended to account for mixing effects in a fashion keeping the
dependence on field regulators to a minimum. We studied several
scenarios involving large mixing effects as well as the transition
with the non-degenerate regime. The inclusion of 2L effects in the
gaugeless approximation has various consequences at this level, such
as imposing the degeneracy of heavy Higgs states in the presence of
CP-violating mixing or complicating the connection between the
gaugeless and `full' tree-level Higgs states. At a technical level, it
would thus seem desirable to put full 2L EW corrections under control,
which would allow to escape the constraints of the gaugeless
approximation and limit the use of a near-degenerate formalism
strictly to scenarios with large mixing. As far as the dependence on
field counterterms is concerned, the situation is very similar,
however, in the non-degenerate and near-degenerate scenarios.

In addition, we observed that the effective-potential approximation
for 2L~self-energies leads to quantitatively reliable results only
when applied to the mass of the SM-like state. Considering the
numerical cost of evaluating 2L~integrals at non-vanishing momentum,
as well as the irrelevance (in the absence of EW~2L~corrections) of
the orders~$\alpha_q\,\alpha_s$ and~$\alpha_q^2$ for heavy-doublet
states, it makes limited sense, in the non-degenerate case, to
consider 2L~gaugeless self-energies for any other external state than
the SM-like one. Furthermore, the limited pertinence of this
approximation for non-vanishing tree-level masses questions as to the
applicability of corresponding calculations to extensions of the MSSM,
away from an MSSM-like regime, since tree-level masses do not
necessarily vanish then, even in the gaugeless limit.

Given that large UV-logarithms develop with increasingly heavy
SUSY~spectrum, the strict FO~formalism suffers from a slow convergence
of the perturbative series. This issue can be evaded through an
explicit resummation of logarithmic effects. This resummation can be
directly included in the context of the FO~calculation, without
resorting to a matching scale, simply by exploiting low-energy
observables as input, and we explicitly performed this operation for
the orders~$\alpha_q$, $\alpha_q\,\alpha_s$ and~$\alpha_q^2$. As this
resummation does not modify the problem of the regularization of Higgs
self-energies away from their mass-shell, it does not affect our
conclusions concerning the dependence of mass predictions at~FO on
field counterterms. This situation contrasts with the concurrent
computation method through~EFTs, where artificial field-dependent
terms cannot receive large logarithmic enhancement---due to the very
structure of the~EFT that embeds UV-logarithms within effective
tree-level couplings---but not with \EG~the hybrid approach
of \citere{Hahn:2013ria}. Nevertheless, we have also shown how the
judicious choice of field renormalization devised
in \citere{Bahl:2018ykj} for the public code \texttt{FeynHiggs}
largely shields the latter from excessive uncertainties associated
with the iterative mass determination.

\section*{\tocref{Acknowledgments}}

We thank Henning Bahl for discussions about properties of
\texttt{FeynHiggs}. S.\,P. acknowledges \mbox{support} by the BMBF
Grant No.\,05H18PACC2. F.\,D.~acknowledges support of the BMBF
Verbund-Projekt 05H2018 and the DFG grant SFB CRC-110 {\em Symmetries
  and the Emergence of Structure in QCD}.

\appendix
\tocsection{\label{ap:scattering}Scattering by scalar
  resonances and propagator matrix}

\tocsubsection{\label{ap:scatgeneral}General considerations}

For definiteness, we consider the 2\,$\to$\,2 scattering process
\AtoB{b\bar{b}}{\tau^+\tau^-} mediated by scalar (neutral Higgs)
resonances, with a center-of-mass energy $\sqrt{s}$. The amplitude can
be formally written as
\begin{equation}
  \Amp{}{}{\AtoB{b\bar{b}}{\tau^+\tau^-}}{} =
  \left[\bar{v}_b(\bar{p}_b)\,\imath\,\check{V}^{\alpha}_{Sbb}\,u_{b}(p_b)\right]
  \imath\,\check{P}_{\alpha\beta}^S\left[
    \bar{v}_{\tau}(\bar{p}_{\tau})\,\imath\,\check{V}^{\beta}_{S\tau\tau}\,
    u_{\tau}(p_{\tau})\right].
\end{equation}
In this appendix, the $\check{\dotcirc}$~notation represents vectors
and matrices in scalar space---corresponding to the Greek indices,
$\alpha$, $\beta$, which are implicitly summed over. The symbols have
the following meaning:
\begin{itemize}
\item The propagator $\check{P}^S$ is a symmetric matrix in scalar
  space, depending on the external momentum squared
  $s\equiv(p_b+\bar{p}_b)^2$. As is customary in particle physics, we
  assume that this object can be decomposed into single poles and a
  continuum:
  \begin{align}
    \check{P}^S(s) &= \sum\limits_H\frac{\check{R}_H}{s-{\cal M}_H^2}
    + \check{C}(s)
  \end{align}
  where $\check{C}$ is a `smooth' function
  of $s$.

  Without loss of generality,
  \begin{align}
    \check{R}_H &= \sum_m r_{H_m}\,\check{E}_{H_m}\,\check{E}_{H_m}^T\,,
  \end{align}
  where $\check{E}_{H_m}$ are vectors generating the subspace
  associated with the pole ${\cal M}_H^2$, while $r_{H_m}$ is a
  normalization (`residue' if $\check{E}_{H_m}$ is normalized).
\item The vertex operator $\check{V}_{Sff}$ is a `column'-vector in
  scalar space and---since it corresponds to interactions with scalar
  resonances---can be decomposed
  as \EG \mbox{$\check{V}_{Sff}=\check{\cal
  V}^L_{Sff}\,P_L+\check{\cal V}^R_{Sff}\,P_R$} in spinor space, where
  $P_{L,R}$ represent the chiral projectors while $\check{\cal
  V}^{L,R}_{Sff}$ have no spinor indices left. \textit{A priori},
  $\check{\cal V}^{L,R}_{Sff}$ are functions of $p_f$ and $\bar{p}_f$,
  which we can replace by $m_f^2$ (kept implicit) and
  $s=\big(p_f+\bar{p}_f\big)^2$ due to Lorentz invariance. By
  `physical' hypothesis, these functions of~$s$ are `smooth', so that,
  near a pole ${\cal M}^2_H$ of the
  propagator,
  \begin{align}
    \check{\cal V}^{L,R}_{Sff}(s)
    &\stackrel[\mathclap{\quad\,s\sim {\cal M}_H^2}]{}{=}\quad
    \check{\cal V}^{L,R}_{Sff}\big({\cal M}_H^2\big)
    + \left(s - {\cal M}_H^2\right)
    \frac{d\check{\cal V}^{L,R}_{Sff}}{ds}\big({\cal M}_H^2\big) + \cdots\,.
  \end{align}
  We write the scalar product
  $r_{H_m}^{1/2}\big(\check{E}^T_{H_m}\,\check{\cal
  V}^{L,R}_{Sff}({\cal M}_H^2)\big)$ as $g^{L,R}_{H_mff}$\,.
\end{itemize}

From the analysis above, the scattering amplitude may thus be written
as follows:
\begin{equation}\label{eq:scatamp}
  \Amp{}{}{\AtoB{b\bar{b}}{\tau^+\tau^-}}{} =
    f(s) + \sum_{H,m}\!\left[\bar{v}_b(\bar{p}_b)\,\imath
      \left(g^{L,R}_{H_mbb}\,P_{L,R}\right) u_{b}(p_b)\right]
    \frac{\imath}{s - {\cal M}^2_H}\left[\bar{v}_{\tau}(\bar{p}_{\tau})\,\imath
      \left(g^{L,R}_{H_m\tau\tau}\,P_{L,R}\right) u_{\tau}(p_{\tau})\right]
\end{equation}
where $f$ is a smooth function. Perturbative QFT, assuming it is in
its regime of validity, should offer a predictive framework for the
calculation of the objects defined above, in particular the (complex)
poles ${\cal M}^2_H$ and the effective couplings $g^{L,R}_{H_mff}$.

\tocsubsection{\label{ap:propmat}Propagator matrix in perturbative QFT
  at 2L order}
  
The propagator matrix is defined as the inverse of the two-point
function in the scalar sector:
$\check{P}^S(s)=\big[s\,\check{\mathds{1}} -
  \check{M}_S^2(s)\big]^{-1}$, where
$\check{M}^2_S(s)=\check{M}^{2}_{\text{tree}} - \hat{\Sigma}(s)$ with
$\check{M}^{2}_{\text{tree}}$ denoting the tree-level mass matrix and
$\hat{\Sigma}(s)=\hat{\Sigma}^{(1)}(s) + \hat{\Sigma}^{(2)}(s) +
\mathcal{O}(\text{3L})$ denoting the renormalized self-energy
matrix. We work in the basis of the tree-level mass eigenstates,
\IE~$\check{M}^{2}_{\text{tree}}=\text{diag}\big[m_i^2\big]$.

\paragraph{Non-degenerate case:}
Let us consider $s=m_i^2+\mathcal{O}(\text{1L})$, with $m_i^2$
representing a non-degenerate tree-level mass, and define the matrix
$\check{\mathbf{\Omega}}^i(s)$ from its elements ($j,k\neq i$)
\begin{subequations}
\begin{align}
  \check{\Omega}^i_{ii}(s) &\equiv 1 - \frac{1}{2}\sum_{j\neq i}\left[
    \frac{\hat{\Sigma}^{\text{(1)}}_{ij}(s)}{m_i^2-m_j^2}\right]^2,\quad
  \check{\Omega}^i_{jk}(s) \equiv \delta_{jk} - \frac{\hat{\Sigma}^{\text{(1)}}_{ij}(s)\,
    \hat{\Sigma}^{\text{(1)}}_{ik}(s)}
        {2\left(m_i^2 - m_j^2\right)\left(m_i^2 - m_k^2\right)}\,,\\
  \check{\Omega}^i_{ij}(s) &= -\check{\Omega}^i_{ji}(s) \equiv -\frac{1}{m^2_i - m^2_j}\left[
    \hat{\Sigma}^{\text{(1)}}_{ij}(s) + \hat{\Sigma}^{\text{(2)}}_{ij}(s)
    + \frac{\hat{\Sigma}^{\text{(1)}}_{ii}(s)\,
      \hat{\Sigma}^{\text{(1)}}_{ij}(s)}{m_i^2 - m_j^2}
    - \sum_{k\neq i}\frac{\hat{\Sigma}^{\text{(1)}}_{ik}(s)\,
      \hat{\Sigma}^{\text{(1)}}_{jk}(s)}{m_i^2 - m_k^2}\right].
\end{align}
\end{subequations}
It satisfies
$\check{\mathbf{\Omega}}^i(s)\!\cdot\check{\mathbf{\Omega}}^{i\,T}(s)=\check{\mathds{1}}+\mathcal{O}(\text{3L})$
as well as
\begin{align}
  \left[s\,\check{\mathds{1}} - \check{M}_S^2(s)\right]_{jk} &=
  \left(s - \widetilde{M}_i^2(s)\right) \check{\Omega}^{i}_{ij}(s)\,\check{\Omega}^{i}_{ik}(s)
  + \smash{\sum\limits_{m,n\neq i}}\check{\Omega}^{i}_{mj}(s)\,\check{\Omega}^{i}_{nk}(s)
  \left(s\,\delta_{jk} - \widetilde{M}_{jk}^2(s)\right)
  + \mathcal{O}(\text{3L})\,,
\end{align}
\IE~the $\check{\mathbf{\Omega}}^i(s)$ `rotation' isolates the direction $i$
up to terms of 3L order, with the diagonal element (`eigenvalue') reading
\begin{equation}\label{eq:nondegrecur}
  \widetilde{M}_i^2(s) = m_i^2 - \hat{\Sigma}^{\text{(1)}}_{ii}(s)
  - \hat{\Sigma}^{\text{(2)}}_{ii}(s)
  + \sum_{j\neq i}\frac{\hat{\Sigma}^{\text{(1)}}_{ij}(s)^2}{m_i^2 - m^2_j}
  + \mathcal{O}(\text{3L})\,.
\end{equation}
Coming back to the propagator matrix, we then have
\begin{align}
  \check{P}^S_{jk}(s) &\stackrel[\mathclap{\ \ s\sim m^2_i}]{}{=}\quad
  \frac{1}{s - \widetilde{M}_i^2(s)}\,\check{\Omega}^{i}_{ij}(s)\,\check{\Omega}^{i}_{ik}(s)
  + \cdots\,,
\end{align}
where the ellipses represent non-singular pieces in the vicinity of
$s\sim m^2_i$. We thus obtain a pole~${\cal M}_i^2$ defined by the
recursive condition ${\cal M}_i^2=\widetilde{M}^2_i({\cal M}_i^2)$,
together with a residue
\begin{align}
  r_i^{-1} &= 1 + \frac{d\hat{\Sigma}^{\text{(1)}}_{ii}}{ds}\big({\cal M}_i^2\big)
  + \mathcal{O}(\text{2L})
\end{align}
and the associated `eigenvector' $\check{\Omega}^i_{ij}\big({\cal
  M}_i^2\big)$. Further perturbative expansion of the argument in the
self-energies, \mbox{${\cal
    M}_i^2=m_i^2-\hat{\Sigma}^{\text{(1)}}_{ii}\big(m_i^2\big)+\mathcal{O}(\text{2L})$},
leads to the expression of \refeq{eq:nondegmassexp} with the residue
\mbox{$r_i^{-1}=1+\frac{d\hat{\Sigma}^{\text{(1)}}_{ii}}{ds}\big(m_i^2\big)+\mathcal{O}(\text{2L})$}---as
we restrict ourselves to vertex corrections of 1L order below, we do
not attempt to control the residue beyond 1L order.

\paragraph{Near-degenerate case:}
Now, let us consider a degenerate sector $D$. We can still isolate it
via a `rotation' $\check{\mathbf{\Omega}}^D(s)$ satisfying
$\check{\mathbf{\Omega}}^D(s)\!\cdot\check{\mathbf{\Omega}}^{D\,T}(s)=\check{\mathds{1}}
+ \mathcal{O}(\text{3L})$. In this case, one may choose ($i,j\in D$,
$k,l\notin D$)
\begin{subequations}
\begin{align}
  \check{\Omega}^D_{ij}(s) &\equiv \delta_{ij}
  - \sum_{k\notin D}\frac{\hat{\Sigma}^{\text{(1)}}_{ik}(s)\,
      \hat{\Sigma}^{\text{(1)}}_{jk}(s)}
    {2\left(m_i^2 - m_k^2\right)\left(m_j^2 - m_k^2\right)}\,,\quad
  \check{\Omega}^D_{kl}(s) \equiv \delta_{kl}
  - \sum_{i\in D}\frac{\hat{\Sigma}^{\text{(1)}}_{ik}(s)\,
    \hat{\Sigma}^{\text{(1)}}_{il}(s)}
  {2\left(m_i^2 - m_k^2\right)\left(m_i^2 - m_l^2\right)}\,,\taghere\\
  \check{\Omega}^D_{ik}(s) &= -\Omega^i_{ki}(s) \equiv -\frac{1}{m^2_i - m^2_k}\left[
    \hat{\Sigma}^{\text{(1)}}_{ik}(s) + \hat{\Sigma}^{\text{(2)}}_{ik}(s)
    + \sum_{j\in D}\frac{\hat{\Sigma}^{\text{(1)}}_{ij}(s)\,
      \hat{\Sigma}^{\text{(1)}}_{jk}(s)}{m_j^2 - m_k^2}
    - \sum_{l\notin D}\frac{\hat{\Sigma}^{\text{(1)}}_{il}(s)\,
      \hat{\Sigma}^{\text{(1)}}_{kl}(s)}{m_i^2 - m_l^2}\right].
\end{align}
\end{subequations}
Then, the object
$\check{\mathcal{P}}^{-1}(s)\equiv\check{\mathbf{\Omega}}^D(s)\!\cdot\!\left[s\,\check{\mathds{1}}
  - \check{\cal M}^{2}(s)\right]\!\cdot\check{\mathbf{\Omega}}^{D\,T}(s)$
satisfies
$\check{\mathcal{P}}^{-1}_{ik}(s)=\mathcal{O}(\text{3L})=\check{\mathcal{P}}^{-1}_{ki}(s)$,
for $i\in D$ and $k\notin D$, while
$\displaystyle\check{\mathcal{P}}^{-1}_{ij}(s)=s\,\delta_{ij}-\widetilde{M}^2_{D\,ij}(s)$
for $i,j\in D$, with
\begin{equation}\label{eq:neardegmassmat}
  \widetilde{M}^2_{D\,ij}(s) \equiv m_i^2\,\delta_{ij}
  - \hat{\Sigma}^{\text{(1)}}_{ij}(s) - \hat{\Sigma}^{\text{(2)}}_{ij}(s)
  + \sum_{k\notin D}\frac{\left(m_i^2 + m_j^2 - 2\,m_k^2\right)
    \hat{\Sigma}^{\text{(1)}}_{ik}(s)\,\hat{\Sigma}^{\text{(1)}}_{jk}(s)}
  {2\left(m_i^2 - m_k^2\right)\left(m_j^2 - m_k^2\right)}
  + \mathcal{O}(\text{3L})\,.
\end{equation}
$\big(\check{\mathcal{P}}^{-1}_{ij}(s)\big)_{i,j\in D}$ is still a
symmetric matrix, hence can be written as
$\check{\mathbf{U}}^{D\,T}(s)\!\cdot\check{\mathcal{D}}(s)\!\cdot\check{\mathbf{U}}^D(s)$,
with a diagonal matrix $\check{\mathcal{D}}(s)$ and a unitary matrix
$\check{\mathbf{U}}^D(s)$. The equation
$\text{det}[\check{\mathcal{D}}(s)]\stackrel[]{!}{=}0$, defining the
zeroes of the inverse propagator matrix in the degenerate sector,
implies
\mbox{$\text{det}[s\,\check{\mathds{1}}_D-\widetilde{M}^2_D(s)]\stackrel[]{!}{=}0$},
since $\text{det}[\check{\mathbf{U}}^D(s)]\neq0$. Thus, the zeroes ${\cal
  M}_I^2$ in the subspace $D$ still satisfy an implicit eigenvalue
  condition
\begin{align}
  \det{\left[{\cal M}_I^2\,\check{\mathds{1}}_D
      - \widetilde{M}^2_D\big({\cal M}_I^2\big)\right]} &= 0\,.
\end{align}

At the 2L order, it is no longer sufficient to simply derive
$\check{\mathbf{U}}^D$ from the eigenvectors of $\widetilde{M}^2_D$, because
the orthogonality property is not necessarily satisfied by the
diagonalizing matrices in the complex case. Instead, one should
determine $\check{\mathbf{U}}^D({\cal M}_I^2)$ through the diagonalization of
$\big[{\cal M}_I^2\,\check{\mathds{1}}_D-\widetilde{M}^2_D\big({\cal
    M}_I^2\big)\big]^{\dagger}\!\cdot\!\big[{\cal
    M}_I^2\,\check{\mathds{1}}_D-\widetilde{M}^2_D\big({\cal
    M}_I^2\big)\big]$. Then, close to the pole, the propagator matrix
looks like
\begin{subequations}
\begin{align}
  \check{P}^S(s) &\stackrel[\mathclap{\quad s\sim {\cal M}^2_I}]{}{=}\quad
  \frac{r_I\left(S_{Ii}\,S_{Ij}\right)_{i,j\in D}}{s - {\cal M}_I^2} + \cdots\,,
  \quad
  S_{Ii} \equiv \sum_{j\in D}\left(\check{U}^{D}_{Ij}\big({\cal M}_I^2\big)\right)^*
  \check{\Omega}^D_{ji}\big({\cal M}_I^2\big)\,,\\
  r_I^{-1} &\equiv \sum_{i,j\in D}\left(\check{U}^{D}_{Ii}\big({\cal M}_I^2\big)\right)^*
  \left(\check{U}^{D}_{Ij}\big({\cal M}_I^2\big)\right)^*\left[
    \delta_{ij} + \frac{d\hat{\Sigma}^{\text{(1)}}_{ij}}{ds}\big({\cal M}_I^2\big)
  \right] + \mathcal{O}(\text{2L})\,.
\end{align}
\end{subequations}
We note that for $i\in D$ and $k\notin D$ one has
\begin{align}
  S_{Ii} &= \left(\check{U}^{D}_{Ii}\big({\cal M}_I^2\big)\right)^*
  + \mathcal{O}(\text{2L})\,,\quad
  S_{Ik} = -\sum_{j\in D}\left(\check{U}^{D}_{Ij}\big({\cal M}_I^2\big)\right)^*
  \frac{\hat{\Sigma}^{\text{(1)}}_{jk}\big({\cal M}_I^2\big)}{m^2_j - m^2_k}
  + \mathcal{O}(\text{2L})\,.
\end{align}
In addition, from
$\big(\check{\mathbf{U}}^{D}(s)\big)^*\!\cdot\!\big[s\,\check{\mathds{1}}_D -
  \widetilde{M}^2_D(s)\big]\!\cdot\!\big(\check{\mathbf{U}}^{D}(s)\big)^\dagger=\check{\mathcal{D}}(s)$
and $\check{\mathcal{D}}_{IJ}\big({\cal M}_I^2\big)=0$, we obtain
\begin{align}
  \sum_{j\in D}\left(\widetilde{M}^2_D\big({\cal M}_I^2\big)\right)_{ij}
    \left(\check{U}^{D}_{Ij}\big({\cal M}_I^2\big)\right)^* &=
         {\cal M}_I^2\left(\check{U}^{D}_{Ii}\big({\cal M}_I^2\big)\right)^*,
\end{align}
\IE~$\big(\check{U}^{D}_{Ii}\big({\cal M}_I^2\big)\big)^*$ is still an
eigenvector of $\widetilde{M}^2_D\big({\cal M}_I^2\big)$ for the
eigenvalue ${\cal M}_I^2$.

\tocsubsection{Vertex corrections in perturbative QFT at 1L order\label{ap:vertexcorr}}

Having extracted the poles from the propagator matrix, we may now turn
to the effective couplings in \refeq{eq:scatamp}. We restrict
ourselves to an analysis of strict 1L order.

\paragraph{Non-degenerate case:}
The effective couplings of a resonance associated with a pole ${\cal
  M}_i^2$ are straightforwardly derived from their definition in
\sect{ap:scatgeneral} and those of the residue and of the `rotation'
matrix from \sect{ap:propmat}:
\begin{align}
  g^{L,R}_{iff} &\equiv r_i^{1/2}\sum_j\check{\Omega}^i_{ij}\big({\cal M}_i^2\big)
  \left(\check{\mathcal{V}}_{Sff}^{L,R}\big({\cal M}_i^2\big)\right)_j\notag\\
  &= \!\left[
    1 - \frac{1}{2}\,\frac{d\hat{\Sigma}^{\text{(1)}}_{ii}}{ds}\big(m_i^2\big)
    \right]\!\left(\check{\mathcal{V}}_{Sff}^{L,R\,\text{(tree)}}\right)_i
  - \sum_{j\neq i}\frac{\hat{\Sigma}^{\text{(1)}}_{ij}\big(m_i^2\big)}
  {m_i^2 - m_j^2}\left(\check{\mathcal{V}}_{Sff}^{L,R\,\text{(tree)}}\right)_j
  + \left(\check{\mathcal{V}}_{Sff}^{L,R\,\text{(1)}}\big(m_i^2\big)\right)_i
  + \mathcal{O}(\text{2L})\,.\taghere[-3ex]
\end{align}
The derivative term originates in the residue, the mixing term in the
`rotation' matrix, and the 1L vertex from the expansion of
$\check{\mathcal{V}}_{Sff}^{L,R}=\check{\mathcal{V}}_{Sff}^{L,R\,\text{(tree)}}+\check{\mathcal{V}}_{Sff}^{L,R\,\text{(1)}}(s)+\mathcal{O}(\text{2L})$. We
simply recover the LSZ reduction.

\paragraph{Near-degenerate case:}
For the pole ${\cal M}_I^2$, we can write (in terms of the mixing
matrix $\mathbf{S}$ of \sect{ap:propmat} at 1L order):
\begin{align}
  g^{L,R}_{Iff} &= r_I^{1/2}\sum_{j\in D}S_{Ij}\left\{
  \left(\check{\mathcal{V}}_{Sff}^{L,R\,\text{(tree)}}\right)_j
  - \sum_{k\notin D}\frac{\hat{\Sigma}^{\text{(1)}}_{jk}\big(m_j^2\big)}{m_j^2 - m_k^2}
  \left(\check{\mathcal{V}}_{Sff}^{L,R\,\text{(tree)}}\right)_k
  + \left(\check{\mathcal{V}}_{Sff}^{L,R\,\text{(1)}}\big(m_j^2\big)\right)_j\right\}
  + \mathcal{O}(\text{2L})\,.\taghere[-3ex]
\end{align}
Instead of directly expanding the residue, it is convenient in this
case to exploit the invariance of observables (and $\mathbf{S}$) under
field renormalization and then use the on-shell scheme as
intermediary. The field counterterms are simply shifted according to
$(\delta
Z)^{\text{OS}}_{ij}=-\frac{d\hat{\Sigma}^{\text{(1)}}_{ij}}{ds}\big(m^2_{ij}\big)
+ \mathcal{O}(\text{2L})$---\mbox{the $\hat{\dotcirc}$ notation}
continues to denote renormalized quantities in the original
scheme. Correspondingly,
\begin{align}
  \left(\check{\mathcal{V}}_{Sff}^{L,R\,\text{(1)}}\big(m_j^2\big)\right)_j^{\text{OS}}=
  \left(\check{\mathcal{V}}_{Sff}^{L,R\,\text{(1)}}\big(m_j^2\big)\right)_j
  - \frac{1}{2}\,\sum\limits_{k\in D}
  \frac{d\hat{\Sigma}^{\text{(1)}}_{jk}}{ds}\big(m^2_{jk}\big)
  \left(\check{\mathcal{V}}_{Sff}^{L,R\,\text{(tree)}}\right)_k\,,\quad
  r_I^{\text{OS}} = \frac{1}{\sum\limits_{k\in D}S_{Ik}^2}\,.
\end{align}
We can define the normalized mixing matrix as
$\widetilde{S}_{Ij}\equiv \big(r_I^{\text{OS}}\big)^{1/2}\,S_{Ij}$. Finally,
we observe for $j\neq k$,
\begin{align}
  \frac{1}{2}\,\frac{d\hat{\Sigma}^{\text{(1)}}_{jk}}{ds}\big(m^2_{jk}\big) &=
  \frac{\hat{\Sigma}^{\text{(1)}}_{jk}\big(m_j^2\big)
    - \hat{\Sigma}^{\text{(1)}}_{jk}\big(m^2_{jk}\big)}{m_j^2 - m_k^2}
  + \mathcal{O}(\text{2L})\,.
\end{align}
Putting everything together, one finds
\begin{align}\label{eq:effcoup}
  g^{L,R}_{Iff}\stackrel[]{!}{=}g^{L,R\,\text{OS}}_{Iff} &=
  \sum_{i\in D}\widetilde{S}_{Ii}\,\Bigg\{\begin{aligned}[t]
  &{-}\sum_{j\in D}\frac{\hat{\Sigma}^{\text{(1)}}_{ij}\big(m_i^2\big)
    - \hat{\Sigma}^{\text{(1)}}_{ij}\big(m_{ij}^2\big)}{m_i^2 - m_j^2}
  \left(\check{\mathcal{V}}_{Sff}^{L,R\,\text{(tree)}}\right)_j
  - \sum_{k\notin D}\frac{\hat{\Sigma}^{\text{(1)}}_{ik}\big(m_i^2\big)}
       {m_i^2 - m_k^2}
  \left(\check{\mathcal{V}}_{Sff}^{L,R\,\text{(tree)}}\right)_k\\
  &{+}\left[1 - \frac{1}{2}\,
    \frac{d\hat{\Sigma}^{\text{(1)}}_{ii}}{ds}\big(m_i^2\big)\right]
  \!\left(\check{\mathcal{V}}_{Sff}^{L,R\,\text{(tree)}}\right)_i
  + \left(\check{\mathcal{V}}_{Sff}^{L,R\,\text{(1)}}\big(m_i^2\big)\right)_i\Bigg\}
  + \mathcal{O}(\text{2L})\,.\taghere\end{aligned}
\end{align}
This defines a generalized LSZ reduction for the near-degenerate case.

As shown in \citere{Domingo:2020wiy}, these loop-corrected couplings
are explicitly independent of the field renormalization and minimize
the dependence on linear gauge-fixing parameters. The properties of
the resonance can then be straightforwardly interpreted as that of a
`genuine' particle, allowing for the definition of masses and decay
widths in terms of the scattering cross-sections.

\bibliographystyle{h-physrev}
\bibliography{literature}

\end{document}